\documentclass[10pt,preprint]{AIAA}

\def\bvec{\left\{\begin{array}{c}} 
\def\evec{\end{array}\right\}}

\begin{document}

\title{Subsonic Indicial Aerodynamics for Aerofoil's Unsteady Loads via Numerical and Analytical Methods}
\author{Marco Berci\footnote{Visiting Academic, AIAA member}}
\affiliation{School of Mechanical Engineering, University of Leeds, LS2 9JT, Leeds, UK}
\author{Marcello Righi\footnote{Associate Professor, AIAA member}}
\affiliation{School of Engineering, Zurich University of Applied Sciences, 8401, Winterthur, Switzerland}

\begin{abstract}
This study deals with generating aerodynamic indicial-admittance functions for predicting the unsteady lift of two-dimensional aerofoils in subsonic flow, using approximate numerical and analytical formulations. Both a step-change in the angle of attack and a sharp-edge gust are suitably considered as small perturbations. Novel contributions concern both a systematic analysis of the computational simulations process and an effective theoretical synthesis of its outcome, providing with sound cross-validation. Good practice for generating the indicial-admittance functions via computational fluid dynamics is first investigated for several Mach numbers, angles of attack and aerofoil profiles. Convenient analytical approximations of such indicial functions are then obtained by generalising those available for incompressible flow, taking advantage of acoustic wave theory for the non-circulatory airload and Prandtl-Glauert's scalability rule for the circulatory airload. An explicit parametric formula is newly proposed for modelling the latter as function of the Mach number in the absence of shock waves, while damped harmonic terms are effectively introduced for better approximating the former. Appropriate tuning of the analytical expressions is also derived in order to mimic the numerical solutions and successfully verify the rigor of superposing circulatory and non-circulatory theoretical contributions in the light of computational fluid dynamics. Results are finally shown and critically addressed with respect to the physical and mathematical assumptions employed, within a consistent framework.
\end{abstract}

\maketitle

\section*{Nomenclature}
\noindent\begin{tabular}{@{}lcl@{}}
$a$ &=& sound speed of the reference airflow \\
$\breve{A}^W$, $\breve{B}^W$ &=& coefficients for the approximation of Wagner function \\
$\breve{A}^K$, $\breve{B}^K$ &=& coefficients for the approximation of Kussner function \\
$\hat{A}$, $\hat{B}$, $\hat{\Omega}$ &=& coefficients for the approximation of the non-circulatory lift \\
$B^1_n$, $B^{1*}_n$ &=& Bessel function of the first type and n-th order \\
$c$ &=& aerofoil chord \\
$C_L$, $\breve{C}_L$, $\hat{C}_L$ &=& total, circulatory and non-circulatory lift coefficient \\
$\breve{C}_L^W$, $\breve{C}_L^K$ &=& Wagner's and Kussner's lift-deficiency coefficient \\
$\bar{C}_L$ &=& asymptotic (steady) lift coefficient \\ 
$C_M$ &=& total pitching moment coefficient at the leading edge \\
$C_P$ &=& negative pressure coefficient \\
$\breve{C}_T$, $\breve{C}^*_T$ &=& Theodorsen function for incompressible and compressible flow \\  
$\breve{C}_S$, $\breve{C}^*_S$ &=& Sears function for incompressible and compressible flow \\  
$F$, $\tilde{F}$, &=& exact and approximate function \\
$H^2_n$, $H^{2*}_n$ &=& Hankel function of the second type and n-th order \\
$k$, $k^*$ &=& reduced frequency for incompressible and compressible flow \\
$m$ &=& number of terms for the approximation of the non-circulatory lift coefficient \\
$M$ &=& Mach number of the reference flow \\
$n_W$, $n_K$ &=& number of terms for the approximation of Wagner and Kussner functions \\
$N$ &=& number of samples for the approximate function \\
$Re$ &=& Reynolds number of the reference flow \\
$s$ &=& function samples \\
$t$ &=& time \\
$U$ &=& horizontal component of the reference airflow velocity \\
$V$, $V_G$ &=& vertical component of the reference airflow velocity due to angle of attack and gust \\
$\alpha$, $\alpha_0$ &=& angle of attack of the reference flow and zero-lift angle of the aerofoil \\
$\beta$ &=& compressibility factor of the reference flow \\
$\tau$, $\tau^*$ &=& reduced time for incompressible and compressible flow \\
$\acute{\tau}$, $\grave{\tau}$ &=& travel times of outgoing and incoming pressure waves on aerofoil \\
$\omega$ &=& frequency \\
\end{tabular} \\

\section{Introduction}
\label{sec:introduction}

Especially at its preliminary stage, aircraft's multidisciplinary design and optimisation (MDO) \cite{Alexandrov,Kesseler} requires robust and efficient methods for dynamic loads calculation \cite{Kier1,Kier2,Cavagna1} and aeroelastic stability investigation \cite{Livne}. Within this framework, indicial-admittance functions \cite{Tobak,Pike,Graham,Leishman1,Marzocca} can serve as very effective tools and reduced-order models (ROMs) \cite{Quarteroni1} for applied unsteady aerodynamics via Duhamel's convolution integral or its state-space realisation \cite{Raveh,Gennaretti}.

Considering subsonic flow perturbations \cite{Albano,Rodden} due to both a unit step in the angle of attack and a unit sharp-edged gust \cite{Wright}, few (often approximate) indicial functions \cite{Fung,Bisplinghoff} have been obtained for the lift build-up of a thin aerofoil \cite{Prandtl,Glauert1,Glauert2} in both incompressible \cite{Wagner,Theodorsen1,Kussner,Garrick,VonKarman,Sears,Jones1,Heaslet,Basu,Jones2,Peters1,Kayran,Peters2,Peters3,Gaunaa,Liu} and compressible flow \cite{Possio,Frazer,Lomax1,Mazelsky1,Mazelsky2,Mazelsky3,Lomax2,Jones3,Leishman2,Leishman3,Balakrishnan,Polyakov,Mateescu}. These functions include unsteady circulatory and non-circulatory parts \cite{Gulcat}, the physical phenomena behind which can be approached both numerically via nonlinear computational fluid dynamics (CFD) \cite{Chung} and analytically via linear potential flow theory \cite{Morino,Katz}, in order to combine the complex generality of the former approach with the solid synthesis of the latter approach and hence provide with sound cross-validation as well as thorough understanding \cite{McNamara}.

Simplified theoretical models can suitably be used to test robustness and consistency of numerical models as well as provide with essential insights on the fundamental behaviour of the physical phenomenon; they also are the best candidates for parametric sensitivity studies and affordable uncertainties quantification, especially within the MDO of complex systems. Analytical solutions for the aerodynamic indicial functions of two-dimensional airfoils in subsonic flow are only partially available and sometimes inadequate for accurate practical use. In particular, Possio's integral equation \cite{Possio} is elegantly expressed in kernel terms but has neither exact closed-form solution nor explicit expression for the aerofoil load, while Lomax's exact closed-form solution and explicit expression for the aerofoil load \cite{Lomax2} hold for the non-circulatory part only (yet including the transition to the circulatory part) and are not practical; no exact closed-form solution and explicit expression for the aerofoil load are available in the literature for the circulatory part only.

Advanced computational models can suitably be used to test hypothesis, applicability and accuracy of analytical models as well as provide with fully-detailed high-fidelity descriptions of the physical phenomenon. CFD-based generation of aerodynamic indicial functions for thin aerofoil and finite wing has been performed by several researchers in recent years \cite{Parameswaran,Silva,Cavagna2,Romanelli,DaRonch,Ghoreyshi1,Jansson,Ghoreyshi2}. Time-accurate schemes, physically consistent grid motion and deformation algorithms \cite{Cizmas} have reached a certain maturity \cite{Farhat1,Farhat2} and are available within the most popular CFD solvers developed for the aeronautical community; however, for the accurate simulation of both circulatory and non-circulatory portions of the flow response there is no fully-established best practice yet \cite{DaRonch}. In particular, an approach based on the Reynolds-Averaged Navier-Stokes (RANS) equations \cite{Anderson} would possibly provide an accurate yet computationally expensive prediction with most solvers but would also introduce an inconvenient dependence on both Reynolds number and turbulence modelling \cite{Pope} for a phenomenon which is substantially inviscid. Solutions obtained on the basis of the Euler equations \cite{Anderson} are computationally attractive but might exhibit a dependence on numerical dissipation as a result of the propagation of oscillations in the non-circulatory response \cite{Chung}.

Considering the NACA 0006, 0012, 2406 and 2412 aerofoils \cite{Jacobs,Abbott1,Ladson} in a compressible subsonic flow $0.3<M<0.6$, careful yet computationally affordable evaluation of both circulatory and non-circulatory parts of the aerodynamic response is performed in this concept demonstration study. Both a unit step-change in the angle of attack and a unit sharp-edge gust are suitably considered as perturbations. Ranging from thin symmetric to thick cambered aerofoils, the effects of domain size, spatial resolution, time stepping and integration scheme \cite{Quarteroni2} on the Euler/RANS CFD solutions are first investigated and an effective analytical synthesis is then attempted on solid physical grounds \cite{Leishman1}. A convenient methodology is proposed for approximating aerodynamic indicial functions \cite{Vepa1,Vepa2,Roger,Dunn,Dowell,Desmarais,Beddoes,Venkatesan,Peterson,Eversman,Tewari,Karpel,Poirel,Cotoi,Sedaghat,Brunton,Berci1,Berci2} of compressible subsonic flow by modifying those of incompressible flow directly, based on acoustic wave theory \cite{Pike,Lomax2} for the non-circulatory airload and Prandtl-Glauert's scalability rule \cite{Glauert2,Jones3} for the circulatory airload. In particular, an explicit parametric formula is newly proposed for modelling the latter as function of the Mach number in the absence of shock waves, while damped harmonic terms are effectively introduced alongside exponential terms for better approximating the former as well as preventing the nonlinear curve-fitting problem from being ill-conditioned, as less terms of the same analytical form become necessary. Considering the asymptotic (steady) lift coefficient as an additional free parameter, appropriate tuning of the analytical expressions is also derived in order to re-examine the rigor of superposing circulatory and noncirculatory contributions in the light of the CFD results and effectively reproduce the latter with a relatively simple approximation \cite{Wieseman,Brink,Palacios,Dimitrov,Sucipto,Berci3}, which is crucial for practical use within a consistent framework \cite{Righi}.

This paper is structured as follows: Section \ref{sec:num} describes the process followed in order to generate accurate numerical simulations via CFD, Section \ref{sec:analytical} presents the proposed analytical approximations method, a comprehensive comparison between numerical and analytical results for inviscid flow is then presented and discussed in Section \ref{sec:discussion} and conclusions are finally drawn in Section \ref{sec:conclusions}. Further details of the generalised analytical approach are provided in the Appendix for theoretical completeness.

\section{Numerical Investigations}
\label{sec:num}

The lift build-up due to a unit step in angle of attack and a unit sharp-edged gust has been calculated via time-accurate CFD solutions \cite{Chung}. All simulations have been performed with the solver Edge \cite{Eliasson1}, a well-established parallelised flow solver developed by the Swedish Defence Research Agency (FOI) for calculating two- and three-dimensional, viscous and inviscid compressible flows on unstructured grids with arbitrary elements. It can perform both steady and unsteady calculations and can couple fluid dynamics with flight mechanics and aeroelasticity; mesh adaptation functionalities and an inviscid adjoint flow solver are also included \cite{EDGE}.

\subsection{Governing Equations}
\label{subsec:equations}

All CFD simulations in this work concern subsonic compressible flow characterised by a high Reynolds number in the absence of both (strong) shock waves and (large) wake separation. The boundary layer around the aerofoil is then reasonably considered as thin and fully turbulent \cite{Schlichting} and the Euler model for inviscid flow has suitably been adopted, as it may be obtained from the Navier-Stokes equations in the theoretical limit of infinite Reynolds number (i.e., when inertial forces are much larger than viscous forces in the flow) \cite{Katz}. The RANS flow model was also successfully used in selected cases for validation purposes \cite{Righi}, with $k$-$\omega$ $EARSM$ turbulence model \cite{Wallin1,Wallin2} and standard air always assumed as ideal Newtonian gas \cite{Anderson}.

Whenever the problem involves the grid displacing/deforming at a certain rate, the effective convective velocity of the moving boundary/interface must be considered \cite{EDGE}. The correct slip/no-slip boundary condition is applied on the aerofoil surface for the Euler/RANS equations, respectively, whereas far-field conditions are imposed on the outer boundary of the computational domain \cite{Katz}. For unsteady simulations, the normal vectors at every fluid-solid interface are recalculated at each time step and all grid quantities are also recalculated whenever the problem involves grid displacement/deformation \cite{EDGE}; this study has only used the rigid motion of the computational grid, in order to simulate the aerofoil's plunge motion for the case of a unit step in the angle of attack.

\subsection{Numerical Schemes}
\label{subsec:schemes}

The central scheme proposed by Jameson-Schmidt-Turkel scheme \cite{Jameson1} has been used throughout this study to model the inviscid fluxes, for both Euler and RANS simulations. The artificial dissipation coefficients for the second- and fourth-order terms are $0.50$ and $0.02$, respectively. A few spot checks have shown no significant differences with Roe's method \cite{Roe}.

In Edge, time-accurate solutions rely on the dual-time approach based on the implicit second-order backward difference method \cite{Jameson2}, which has already been used in a number of applications \cite{Eliasson2,Peng} and may include a line-implicit approach in highly stretched grid regions \cite{Eliasson3}. Absence of significant time-step induced effects on the results is guaranteed by a careful convergence study \cite{Righi}. All simulations performed in this study use compressible schemes \cite{Chung} and can be carried out efficiently also without pre-conditioning \cite{Quarteroni2}, since the Mach number $M=\frac{U}{a}$ of the reference airflow is sufficiently high yet still subsonic.

\subsection{Computational Grids}
\label{subsec:grids}

The far-field boundaries are placed at a distance of $500$ chords from the aerofoil and the calculations have been conducted on fine meshes, with a boundary layer resolution such that the distance from the wall of the first grid point is sufficient for the RANS simulation of a realistic flow case (i.e., $Re \simeq 5\times 10^6$) without wall functions (i.e., $y^+ < 1$) \cite{Chung}. Both Euler and RANS simulations have then been conducted on the same computational grids, which have been obtained by manipulation of those made available by the Turbulence Model Benchmarking Working Group \cite{Rumsey} and feature $58300$ elements and $57800$ nodes, ensuring absence of significant grid-induced effects.

\subsection{Solution Process}
\label{subsec:solution}

For each geometry and Mach number, a steady-state flow solution has first been obtained and used as initial condition for the time-accurate simulation. The response to a unit step in angle of attack has been simulated with a vertical translation at a velocity $V$ corresponding to an angle of attack variation $\Delta\alpha=1^{\circ}$; the maneuvre starts instantaneously and is completed within the first time step. The response to a unit sharp-edged gust, instead, has been obtained by adding a travelling vertical component $V_G$ of flow velocity (still corresponding to a variation of one degree in the flow's angle of attack) in the portion of the computational domain one-chord upstream of the wing root's aerofoil. This one-chord clearance has been chosen as a suitable compromise for mitigating the mutual influence between gust shape and flow field \cite{Basu}, in order to maintain the gust edge as sharp as possible while perturbing the flow around the aerofoil as least as possible before the gust arrival \cite{Berci3}. This approach may generate reflections of the artificial gust perturbation from the boundaries of the computational domain; however, these reflections resulted barely noticeable in the asymptotic (steady) behaviour of the flow response.
 
It is worth stressing that the present study focuses on aerodynamic phenomena which are essentially inviscid \cite{Beddoes} and should ideally be independent of the Reynolds number (except for the indirect effects of the boundary layer thickness, which is not accounted for in the analytical approach anyway); therefore, the Euler-based simulations should be more significant than the RANS-based ones and produced the numerical results presented in this work.

\subsection{Sources of Error and Uncertainty}
\label{subsec:sources}

The most important sources of error are insufficient spatial and temporal resolutions. 

An excessively coarse mesh leads to inaccurate circulation around the aerofoil and the relevance of this error is strictly related to the numerical dissipation introduced by the integration scheme. Also, poor grid resolution normal to the aerofoil wall causes an inaccurate reconstruction of the boundary layer as well as pressure oscillations in the impulsive part of the flow response, especially when solving the Euler equations for unsteady simulations.

An insufficient size of the computational domain may also cause spurious oscillations, due to acoustic waves reflection or unsatisfied far-field boundary conditions.

An excessively large time step leads to under-resolution of the impulsive and transitory parts of the flow response; however, insufficient convergence of the subiterations within each time step causes oscillations at the end of the transitory part.

Finally, by affecting the boundary layer thickness and introducing viscous losses, an inappropriate or mistuned turbulence model may lead to an inaccurate asymptotic (steady) value of the aerofoil lift.

\section{Analytical Approximations}
\label{sec:analytical}

The approximation of the compressible indicial aerodynamic functions is here obtained from the corresponding incompressible one \cite{Berci1}, by means of Prandtl-Glauert's transformation \cite{Glauert2,Jones3} for the circulatory part $\breve{C}_L$ and acoustic wave theory \cite{Pike,Lomax2} for the non-circulatory part $\hat{C}_L$, which are then linearly superposed as $C_L = \breve{C}_L + \hat{C}_L$ \cite{Leishman1}. In particular, the circulatory contribution is significant along the entire indicial functions and accounts for the decaying effect of the wake's downwash on the aerofoil's lift build-up \cite{Jones2}, whereas the non-circulatory contribution is significant at the start of indicial functions only and consists of an initial impulsive-like reaction (where piston theory holds for any aerofoil shape \cite{Lighthill,Ashley}) followed by a relatively short transitory region (where the characteristic lines of the wave equation intersect), which represents the most complex part of the aerodynamic response \cite{Pike}. With respect to the lift development of a thin aerofoil due to both a unit step in the angle of attack and a unit sharp-edged gust, exact analytical solutions are available for both the non-circulatory contribution \cite{Lomax2} and the circulatory contribution in incompressible flow \cite{Wagner,Theodorsen1,Kussner,Garrick,VonKarman,Sears} (see Appendix), whereas few approximate analytical solutions are available for the circulatory contribution in compressible flow \cite{Lomax1,Mazelsky1,Mazelsky2,Mazelsky3,Lomax2,Jones3,Leishman2,Leishman3}, the exact solution being given in complex functional form \cite{Possio,Frazer,Balakrishnan,Polyakov}. However, in order to compare numerical and analytical results thoroughly, the analytical models still require appropriate tuning so to match the limit behaviour of the CFD simulations \cite{Wieseman,Brink,Dimitrov,Palacios,Sucipto,Berci3} in both circulatory and non-circulatory parts of the flow response.

\subsection{Circulatory Part}

Using the Prandtl-Glauert factor $\beta=\sqrt{1-M^2}$ to scale the reduced time $\tau=\frac{2*U}{c}t$ \cite{Jones3}, the aerofoil's circulatory lift development due to a unit step in the angle of attack may be written as:
\begin{equation}
\breve{C}_L^\alpha = \bar{C}_L \left[ 1 - 2\left(1-\frac{\pi}{\bar{C}_L} \right) \sum_{j=1}^{n_W} \breve{A}_j^W e^{-\breve{B}_j^W \beta^2 \tau} \right],   
\qquad   \sum_{j=1}^{n_W} \breve{A}_j^W=\frac{1}{2},
\end{equation}
whereas, according to the ``frozen gust'' approach \cite{Chiang,Hoblit}, the circulatory lift development due to a unit sharp-edged gust may be written as:
\begin{equation}
\breve{C}_L^G = \bar{C}_L \left( 1 - \sum_{j=1}^{n_K} \breve{A}_j^K e^{-\breve{B}_j^K \beta^2 \tau} \right),   \qquad   \sum_{j=1}^{n_K} \breve{A}_j^K=1,
\end{equation}
where $\bar{C}_L$ is generally taken from steady CFD simulations directly and contains most of the nonlinear flow effects \cite{McNamara,Berci3}. $\breve{A^W}$, $\breve{B^W}$ and $\breve{A^K}$, $\breve{B^K}$ are the coefficients for the exponential approximation of Wagner's \cite{Wagner} and Kussner's \cite{Kussner} functions for incompressible flow and coincide with those for the rational approximation of Theodorsen's \cite{Theodorsen1} and Sears' \cite{VonKarman,Sears} functions in the reduced-frequency domain (see Appendix), respectively; Table \ref{tab:2D} reports all $\breve{A}$ and $\breve{B}$ coefficients with $n_W=3$ and $n_K=5$, as obtained via constrained nonlinear optimisation \cite{Holmes,Vanderplaats} by best-fitting the exact curves \cite{Berci1,Tiffany} (see Appendix). Note that $\bar{C}_L\approx\frac{2\pi}{\beta}$ for thin aerofoils \cite{Glauert2} but the initial values of the circulatory lift coefficients still coincide with those of incompressible flow (i.e., $\breve{C}_{L0}^\alpha=\pi$ and $\breve{C}_{L0}^G=0$), since the information about the compressible nature of the singular perturbation travels with some delay due to the sound speed \cite{Mazelsky1,Mazelsky2,Mazelsky3,Lomax2}; indeed, all approximations for incompressible flow are consistently resumed with $M=0$ and $\beta=1$.

\begin{table}[h!]
\begin{center}
\begin{tabular}{lcccccccccccccccccccc}
        & $\breve{A}_1$ & $\breve{A}_2$ & $\breve{A}_3$ & $\breve{A}_4$ & $\breve{A}_5$ & $\breve{A}_6$ &         $\breve{B}_1$ & $\breve{B}_2$ & $\breve{B}_3$ & $\breve{B}_4$ & $\breve{B}_5$ & $\breve{B}_6$ \\  \hline 
Wagner  & $0.0271$ & $0.1262$ & $0.2541$ & $0.0926$ & $-$ & $-$ & $0.0095$ & $0.0615$ & $0.2077$ & $0.6630$ & $-$ & $-$ \\
Kussner & $0.3694$ & $0.0550$ & $0.2654$ & $0.1829$ & $0.0861$ & $0.0412$ & $0.3733$ & $0.0179$ & $0.1096$ & $1.6003$ & $10.428$ & $170.93$
\end{tabular}
\end{center}
\caption{Coefficients of the optimal approximation of Wagner's and Kussner's functions for flat aerofoil in incompressible flow \cite{Berci1,Berci2}.}
\label{tab:2D}
\end{table}

\subsubsection{Simplest (Closed) Form}

The simplest approximation of the circulatory lift development due to a unit step in the angle of attack employs a single exponential term (i.e., $n_\alpha=1$), where no optimisation is necessary but the constraints are still satisfied and imposing the exact limit behaviours (see Appendix) directly results in $\breve{A}_1=\frac{1}{2}$ and $\breve{B}_1=\frac{1}{4}$. For the case of a unit sharp-edged gust, the penetration of the latter must also be accounted for and the simplest approximation of the circulatory lift development hence employs a second exponential term (i.e., $n_G=2$) with $\breve{A}_2=\frac{1}{2}$ and $\breve{B}_2=\frac{3}{4}$. Such approximations are computationally convenient for crude lift estimations and hence suitable for investigating complex multidisciplinary applications with heavy state-space representation only \cite{Berci1}.

\subsection{Non-Circulatory Part}

The aerofoil's non-circulatory lift contribution is known either analytically via acoustic wave theory \cite{Pike} or numerically via CFD \cite{Parameswaran} and extend to a complex transitory region, where acoustic waves interact before eventually joining the decaying non-circulatory solution with the developing circulatory solution \cite{Lomax2}; thus, these contributions may be approximated with a series of damped oscillatory terms as:
\begin{equation}
\hat{C}_L^\alpha = \sum_{j=1}^{m_\alpha} \hat{A}_j^\alpha e^{-\hat{B}_j^\alpha \beta^2 \tau} \cos \left( \hat{\Omega}_j^\alpha \beta^2 \tau \right),   \qquad
\hat{C}_L^G = \sum_{j=1}^{m_G} \hat{A}_j^G e^{-\hat{B}_j^G \beta^2 \tau} \cos \left( \hat{\Omega}_j^G \beta^2 \tau \right),
\end{equation}
which are best-fitted to the difference between exact non-circulatory solution and approximate circulatory contribution in a range including the transitory region, in order to obtain the coefficients $\hat{A}$, $\hat{B}$ and $\hat{\Omega}$. Note that damped oscillatory terms own an exact Laplace transform in rational form \cite{Roger,Righi} and their time evolution may hence be reproduced by an equivalent state-space system \cite{Leishman4,Meyer} when added aerodynamic states are suitably defined; moreover, they reduce to simple exponential terms when $\hat{\Omega}=0$ in the absence of any frequency content. In the case of a unit step in the angle of attack, the unknown variables of the nonlinear optimisation problem are constrained as:
\begin{equation}
\sum_{j=1}^{m_\alpha} \hat{A}_j^\alpha = \frac{4}{M} - \pi,   \qquad
\sum_{j=1}^{m_\alpha} \hat{A}_j^\alpha \hat{B}_j^\alpha = 2\left( \bar{C}_L-\pi \right) \sum_{j=1}^{n_W} \breve{A}_j^W \breve{B}_j^W + \frac{2\left(1-M\right)}{\beta^2 M^2},
\end{equation}
where it is common to satisfy only the initial condition of piston theory \cite{Lighthill,Ashley} (see Appendix) using just one exponential term for practical applications \cite{Leishman1,Beddoes}; however, this is typically conservative (especially for low Mach numbers) and a more realistic physical representation is obtained when the series of damped oscillators is best-fitted to the difference between non-circulatory and circulatory contributions in the entire validity range of the former, by minimising the approximation error via constrained nonlinear optimisation \cite{Berci1,Tiffany} (see Appendix). In the case of a unit sharp-edged gust, at least two exponential terms are formally necessary to satisfy the initial conditions of piston theory \cite{Lighthill,Ashley} (see Appendix) and the unknown variables of the nonlinear optimisation problem are constrained as:
\begin{equation}
\sum_{j=1}^{m_G} \hat{A}_j^G = 0,   \qquad
\sum_{j=1}^{m_G} \hat{A}_j^G \hat{B}_j^G = \bar{C}_L \sum_{j=1}^{n_K} \breve{A}_j^K \breve{B}_j^K - \frac{2}{\beta^2 \sqrt{M}};
\end{equation}
a small transitory region joining non-circulatory and circulatory parts also exists and may appropriately be modelled by few additional terms, although not strictly necessary for typical practical applications \cite{Leishman1,Beddoes}. The flow condition being identically unperturbed ahead of the aerofoil \cite{Lomax2}, note that the initial value of both circulatory and non-circulatory contributions coincide in both tuned and untuned cases; then the lift develops with a different rate.

\subsubsection{Simplest (Closed) Form}

The simplest approximation of the non-circulatory lift development due to a unit step in the angle of attack also employs a single exponential term (i.e., $m_\alpha=1$ and $\hat{\Omega}_1^\alpha=0$), where no optimisation is necessary but the constraints are still satisfied and imposing the exact limit behaviour of piston theory (see Appendix) directly results in:
\begin{equation}
\hat{A}_1^\alpha = \frac{4}{M} - \pi,   \qquad
\hat{B}_1^\alpha = \frac{2}{\hat{A}_1^\alpha} \left[\left( \bar{C}_L-\pi \right) \sum_{j=1}^{n_W} \breve{A}_j^W \breve{B}_j^W + \frac{1-M}{\beta^2 M^2}\right].
\end{equation}

For the case of a unit sharp-edged gust, the penetration of the latter must also be accounted for and the last two exponential terms in Table \ref{tab:2D} (which are just devoted to reproducing the initial radical behaviour of the circulatory contribution \cite{Garrick,VonKarman,Sears} and then decay very rapidly within the duration of the transitory region) are hence effectively replaced by a single exponential term (i.e., $\hat{\Omega}_1^G=0$) satisfying the exact limit behaviour of piston theory directly; this may formally be achieved with $m_G=3$ by imposing:
\begin{equation}
\hat{A}_1^G = -\bar{C}_L \sum_{j=n_K-1}^{n_K} \breve{A}_j^K,   \qquad
\hat{B}_1^G = \frac{1}{\hat{A}_1^G} \left(\bar{C}_L \sum_{j=1}^{n_K-2} \breve{A}_j^K \breve{B}_j^K - \frac{2}{\beta^2 \sqrt{M}}\right),
\end{equation}
along with $\hat{A}_2=\bar{C}_L \breve{A}_5^K$, $\hat{B}_2=\breve{B}_5^K$, $\hat{\Omega}_2=0$ and $\hat{A}_3=\bar{C}_L \breve{A}_6^K$, $\hat{B}_3=\breve{B}_6^K$, $\hat{\Omega}_3=0$, so that the last two exponential terms in Table \ref{tab:2D} for the approximation of the circulatory contribution are explicitly cancelled out. Thus, note that the approximation of the full aerodynamic indicial function effectively employs $n=5$ exponential terms in both unit step in the angle of attack and unit sharp-edged gust cases.

Of course, when the simplest approximation of the circulatory lift development due to a unit sharp-edged gust is employed in the first place then only its last exponential term may effectively be replaced and formally cancelled out with $m_G=2$ and $\hat{A}_2=\bar{C}_L \breve{A}_2$, $\hat{B}_2=\breve{B}_2$, $\hat{\Omega}_2=0$; therefore, the approximation of the full indicial aerodynamic function would effectively employ $n=2$ exponential terms for the case of a unit step in the angle of attack and $n=3$ exponential terms for the case of a unit sharp-edged gust.

\section{Results Discussion and Validation}
\label{sec:discussion}

The two-dimensional lift due to a unit step in the angle of attack (AoA) and a unit sharp-edged gust (SEG) has been obtained for four NACA aerofoils: the 0006 (thin and symmetric), the 0012 (thick and symmetric), the 2406 (thin and cambered) and the 2412 (thick and cambered) aerofoils. Note that the first NACA digit gives the aerofoil's maximum camber in percentage of the chord, whereas the second digit gives the distance of maximum camber from the leading edge in tens of percents of the chord and the last two digits give half of the maximum thickness (always located at $30\%$ of the chord) in percentage of the chord; finally, the leading-edge radius is about $110\%$ of the square of the maximum thickness \cite{Jacobs,Abbott1,Ladson}.

Several angles of attack and Mach numbers are considered, thus encompassing suitable combinations of flow incidence and compressibility aerofoil as well as camber and thickness. All these parameters are generally relevant for linearised flow regimes \cite{Ghoreyshi2,Motta} and aerodynamic indicial functions are independent of them for linear flow regimes only \cite{Leishman1} (i.e., for thin airfoil at small incidence in subsonic flow); extreme cases with very thick aerofoil, high angle of attack or transonic Mach number typically include shock waves \cite{Anderson} (as well as boundary layer separation, in the case of RANS simulations) and are outside the scope of the present work. In the CFD simulations, the small perturbation suitably imposed to the vertical component of the reference flow velocity (i.e., the variation imposed to the angle of attack or the ``frozen'' gust speed) gives a marginal contribution to the overall magnitude of the latter and is indeed coherently neglected in the analytical model \cite{Bisplinghoff}. Computational and theoretical results are hence compared and all differences critically assessed based on solid physical and mathematical grounds, in order to provide with a clear and thorough concept demonstration of the combined analytical-numerical strategy hereby proposed for the derivation of aerodynamic indicial functions, including nonlinear effects.

\subsection{Steady Solutions: Static Load}

The initial unperturbed condition corresponding to steady compressible flow is first investigated and the critical Mach number is estimated for each test case, based on the pressure distribution for steady incompressible flow, in order to ensure that the aerodynamic flow can actually be considered as subsonic and no significant shock is present (even when the unsteady perturbations are eventually included). The test cases calculated within this study are then summarised in Table \ref{tab:testcases}: all aerofoils are analysed at their own zero-lift angle of attack $\alpha=\alpha_0$, so to compare the results for different aerofoil shapes giving the same steady lift coefficient $C_L \approx 0$; however, the symmetric aerofoils 0006 and 0012 are also analysed at the opposite zero-lift angle of attack $\alpha=-\alpha_0$ of the corresponding cambered aerofoils 2406 and 2412 (having $\alpha_0 \approx -2.1^{\circ}$ \cite{Jacobs}), so to compare the results for different aerofoil shapes at the same effective angle of attack.

\begin{table}[h!]
\begin{center}
\begin{tabular}{lcccc}
NACA & $\alpha$ & Mach & Perturbations & CFD \\ \hline 
0006 & $0^{\circ}$, $2.1^{\circ}$ & 0.3, 0.4, 0.5, 0.6 & AoA, SEG & Euler \\
0012 & $0^{\circ}$, $2.1^{\circ}$ & 0.3, 0.4, 0.5, 0.6 & AoA, SEG & Euler \\
2406 & $-2.1^{\circ}$ & 0.3, 0.4, 0.5, 0.6 & AoA, SEG & Euler \\
2412 & $-2.1^{\circ}$ & 0.3, 0.4, 0.5, 0.6 & AoA, SEG & Euler \\
\end{tabular}
\end{center}
\caption{Test cases analysed.}
\label{tab:testcases}
\end{table}

\begin{figure}[h!]
\begin{center}
\includegraphics[width=67mm]{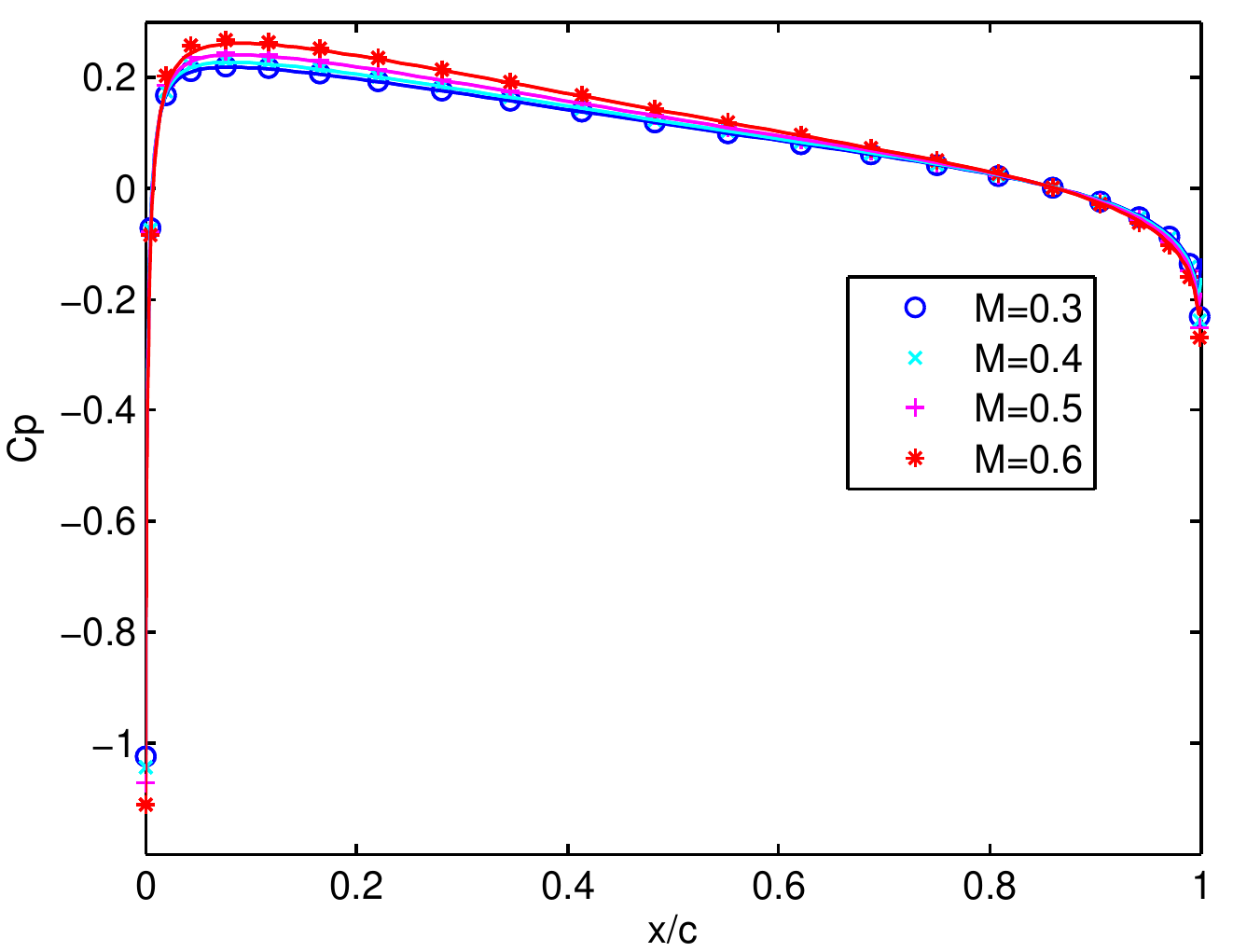}
\includegraphics[width=67mm]{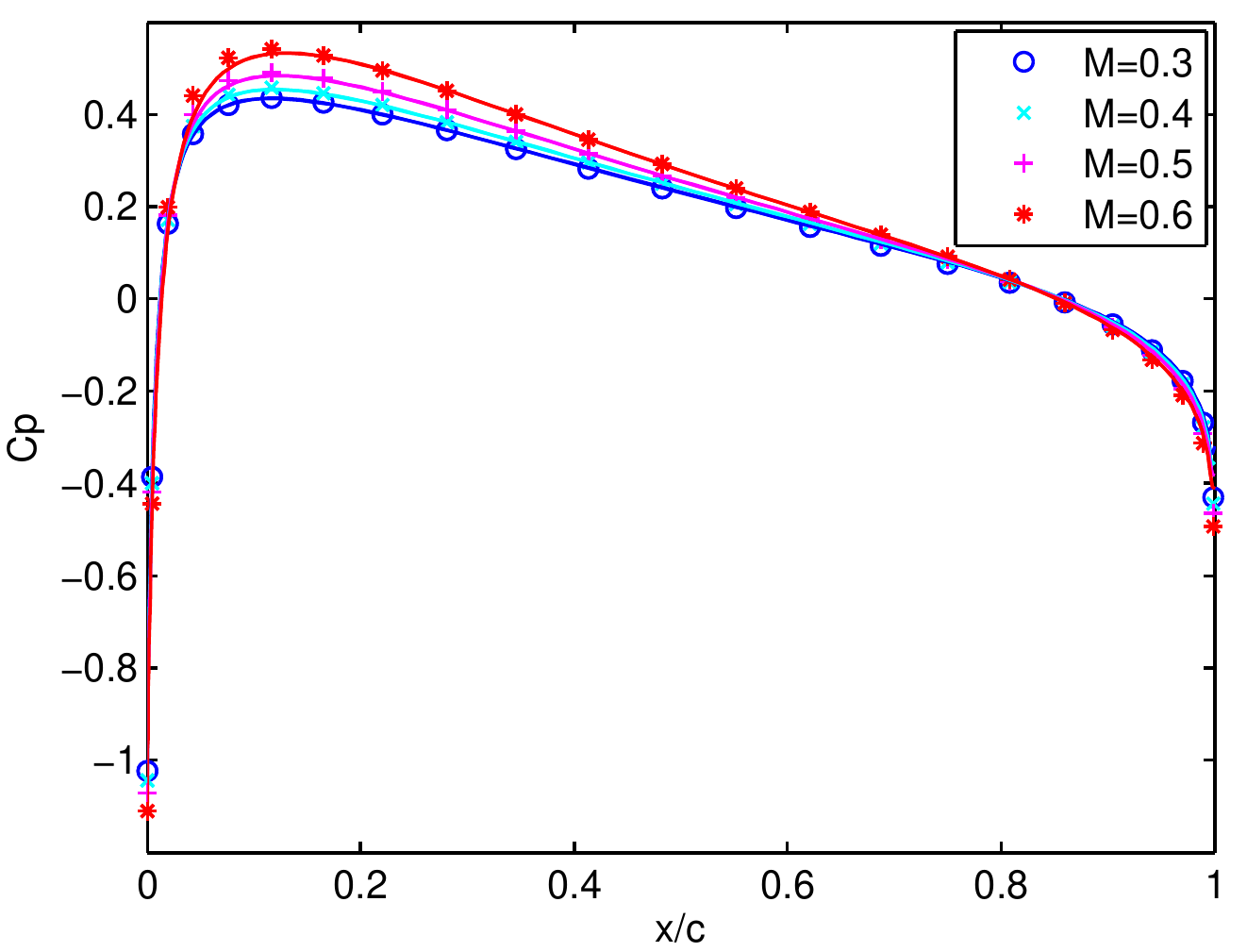}
\end{center}
\caption{Negative pressure coefficient distribution for $\alpha=0^{\circ}$: NACA 0006 (left) and 0012 (right) aerofoils; lines = Euler CFD, symbols = analytical solution.}
\label{fig:AOA0}
\end{figure}

\begin{figure}[h!]
\begin{center}
\includegraphics[width=67mm]{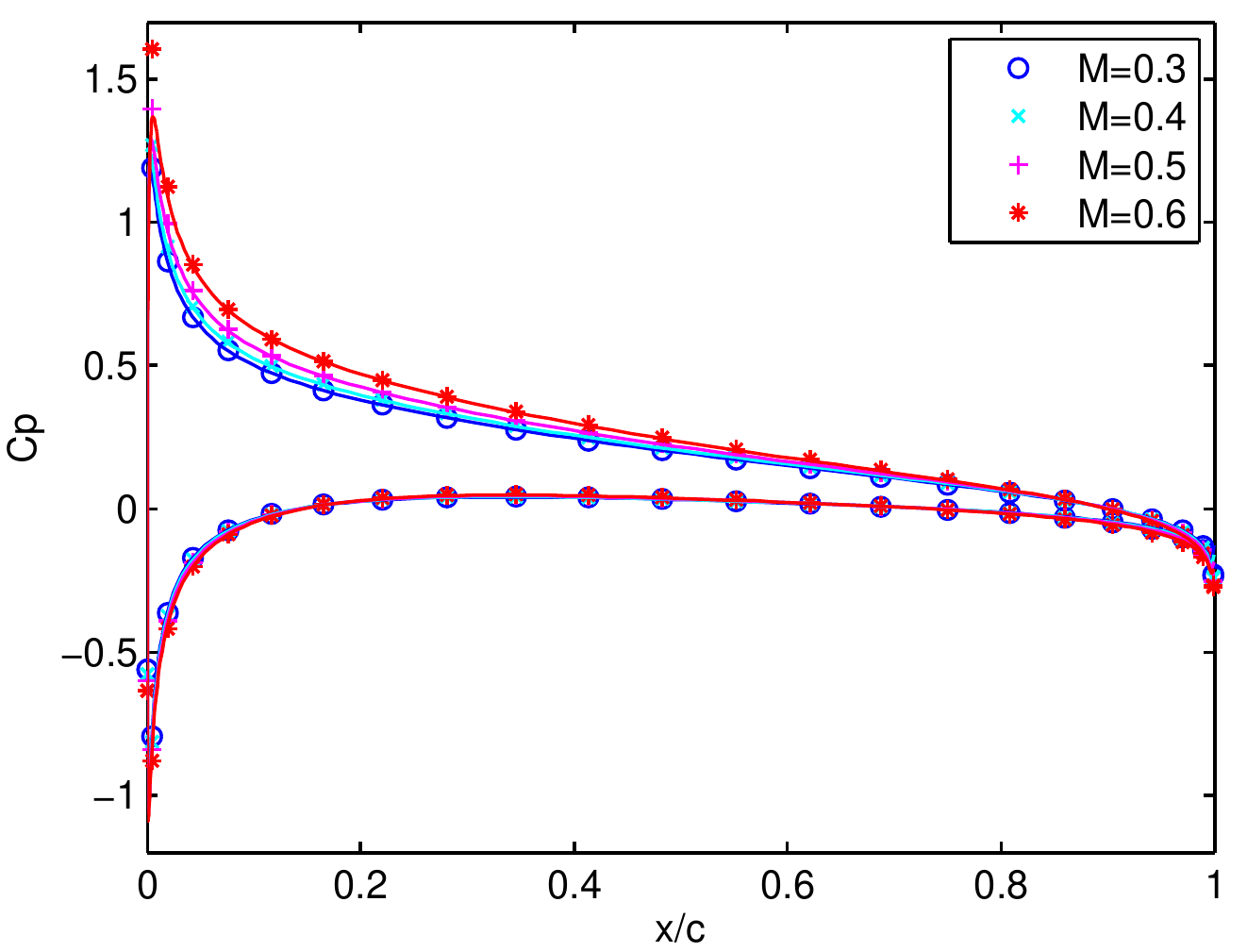}
\includegraphics[width=67mm]{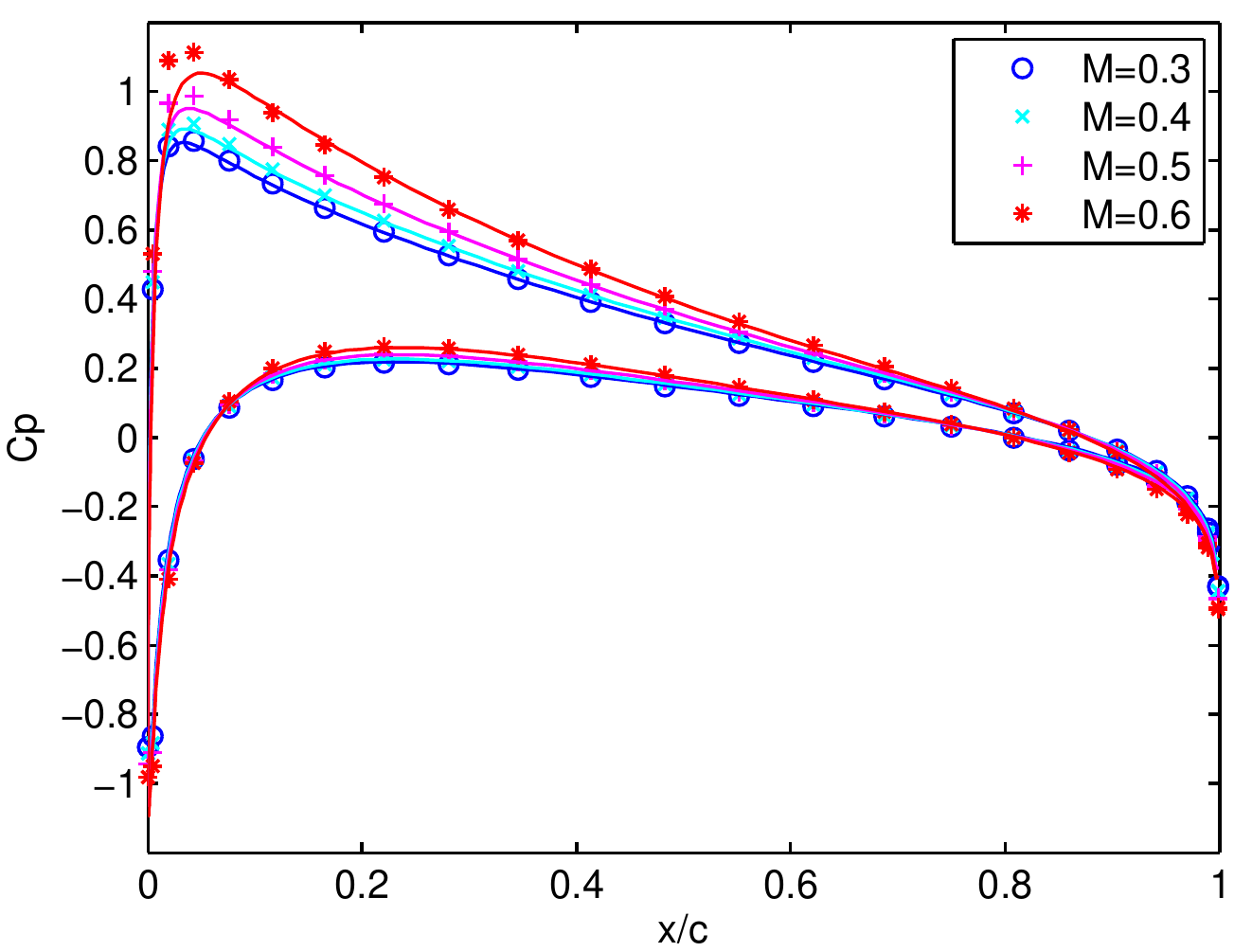}
\end{center}
\caption{Negative pressure coefficient distribution for $\alpha=2.1^{\circ}$: NACA 0006 (left) and 0012 (right) aerofoils; lines = Euler CFD, symbols = analytical solution.}
\label{fig:AOA2}
\end{figure}

\begin{figure}[h!]
\begin{center}
\includegraphics[width=67mm]{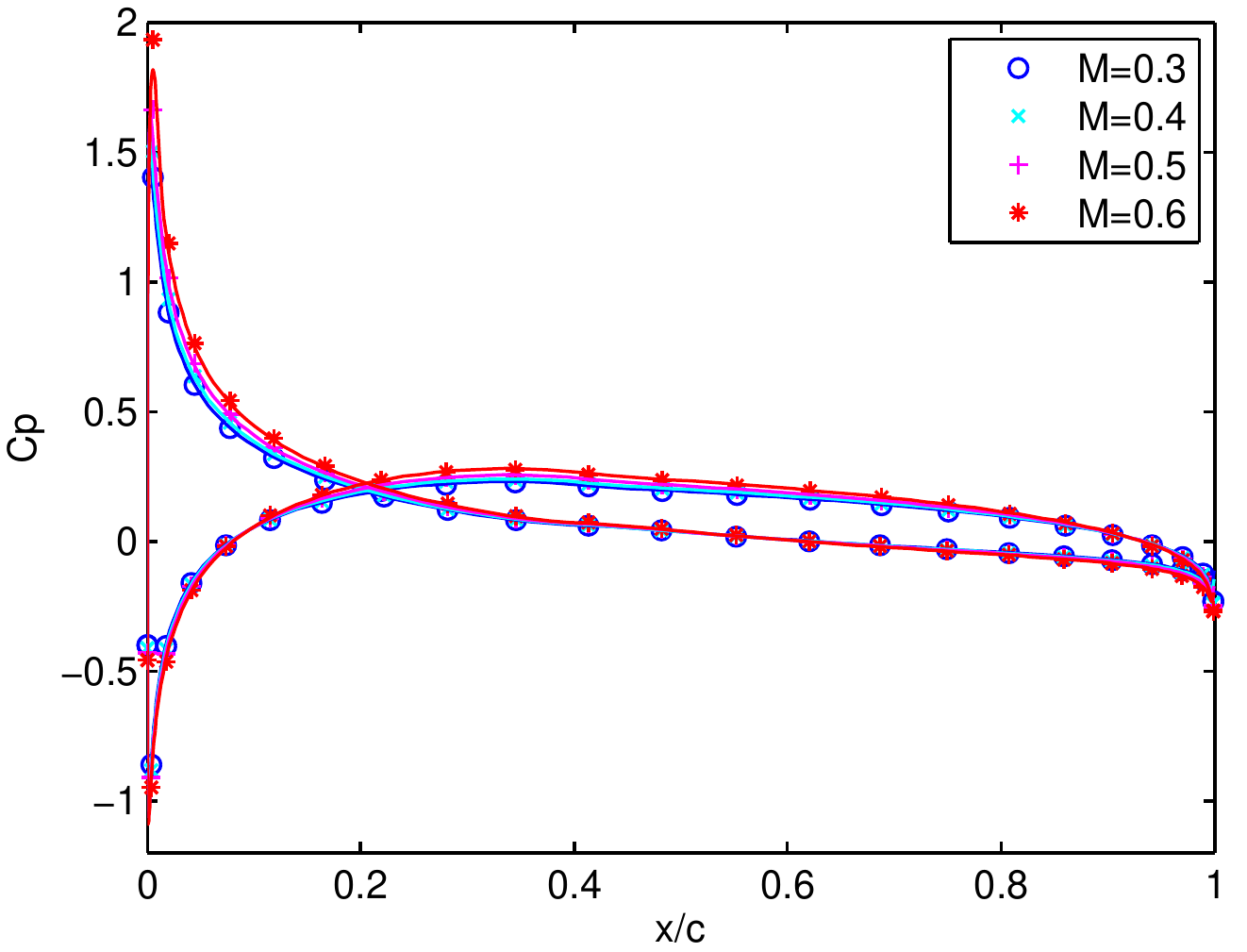}
\includegraphics[width=67mm]{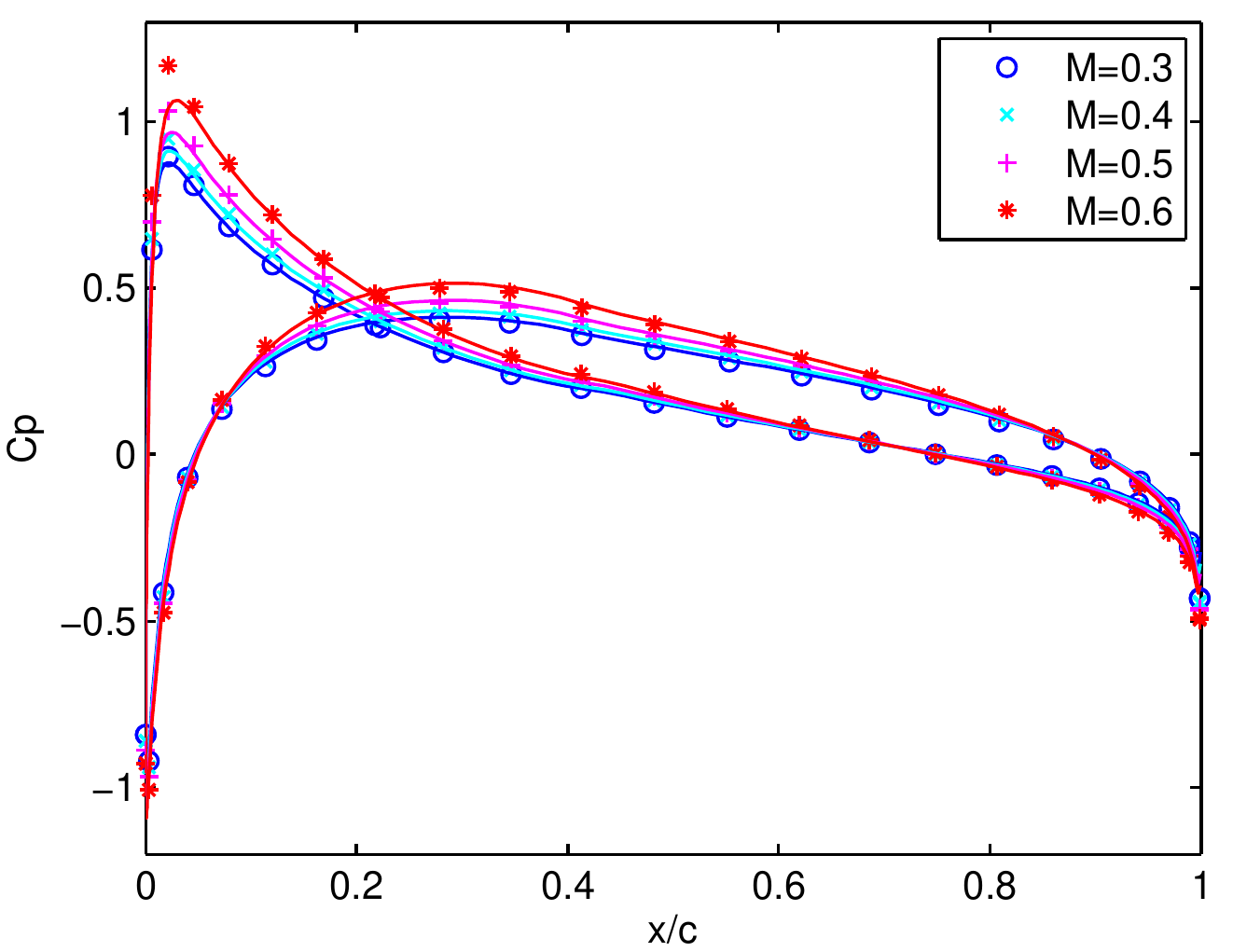}
\end{center}
\caption{Negative pressure coefficient distribution for $\alpha=-2.1^{\circ}$: NACA 2406 (left) and 2412 (right) aerofoils; lines = Euler CFD, symbols = analytical solution.}
\label{fig:AOAn2}
\end{figure}

Figures \ref{fig:AOA0} and \ref{fig:AOA2} compare the negative pressure coefficient $C_P$ distributions for the symmetric NACA 0006 and 0012 aerofoils at their reference angles of attack, as obtained numerically via Euler CFD \cite{Anderson} and analytically via Theodorsen theory \cite{Theodorsen2,Theodorsen3} (which satisfies Kutta-Joukowski condition \cite{Kutta,Joukowski} and is based on conformal mapping \cite{Imai,Karamcheti}, thus including second-order effects \cite{Barger,Mateescu2}) with Karman-Tsien compressibility correction \cite{Spreiter}; Figure \ref{fig:AOAn2} then compares the negative pressure coefficient distributions for the cambered NACA 2406 and 2412 aerofoils at their (steady) effective angles of attack. Excellent agreement is always found, especially in the case of the thinner NACA 0006 and 2406 aerofoils flying at the lower Mach numbers, which better suite the terms and conditions for the applicability of the analytical compressibility correction \cite{Glauert2}.

\begin{figure}[h!]
\begin{center}
\includegraphics[width=67mm]{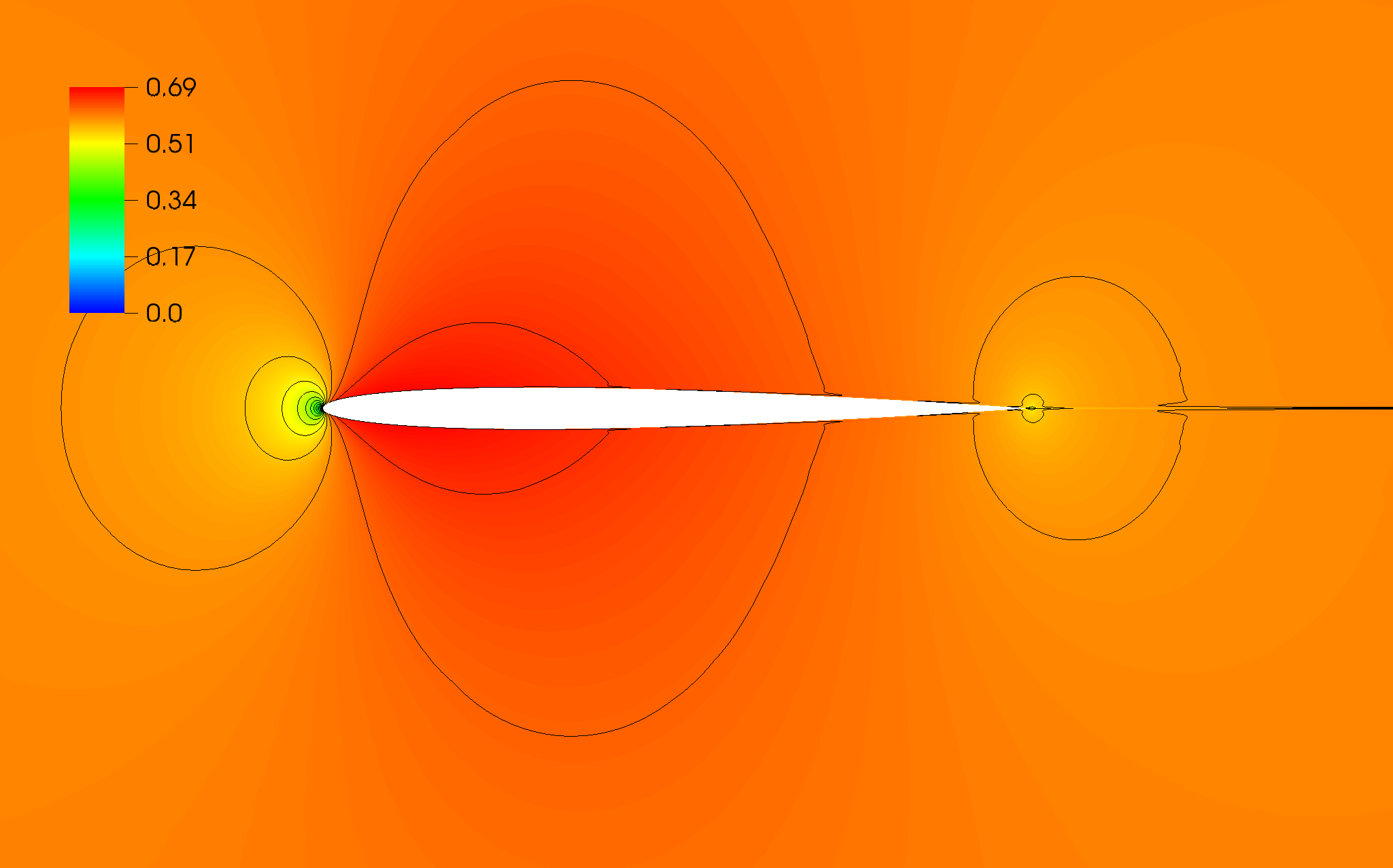}
\includegraphics[width=67mm]{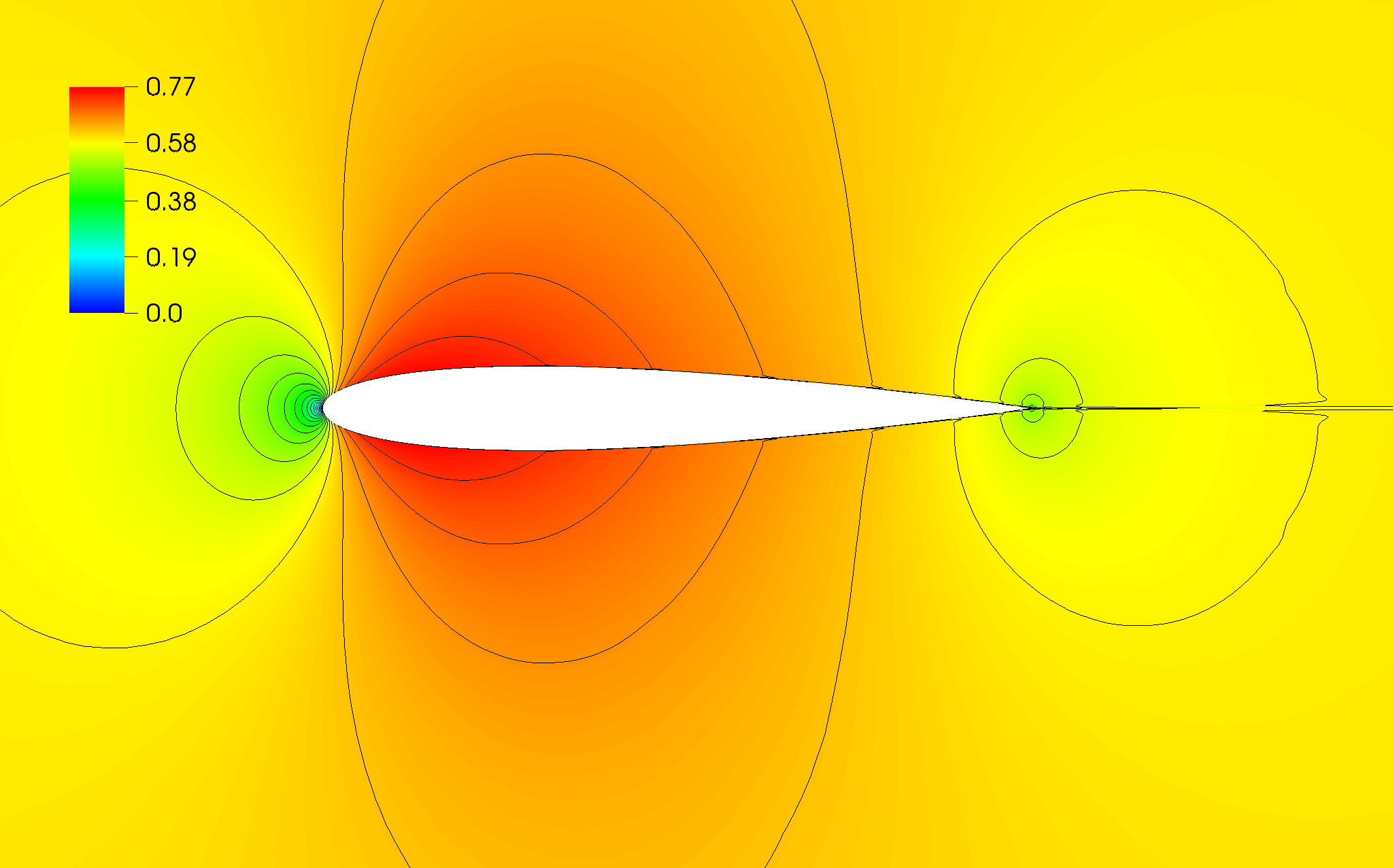}
\end{center}
\caption{Euler Mach field for $M=0.6$ and $\alpha=0^{\circ}$ via Euler CFD: NACA 0006 (left) and 0012 (right) aerofoils.}
\label{fig:mach0}
\end{figure}

\begin{figure}[h!]
\begin{center}
\includegraphics[width=67mm]{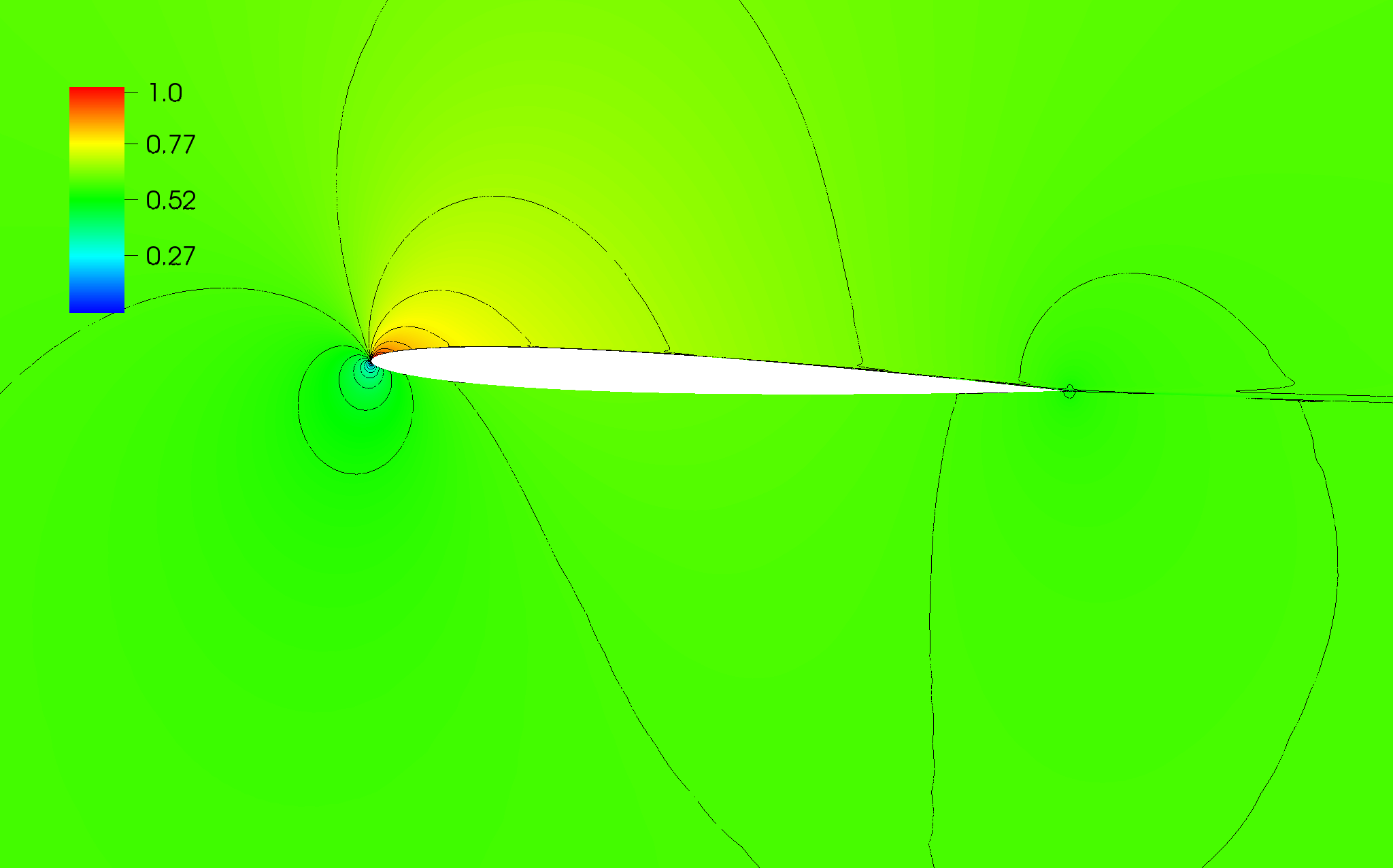}
\includegraphics[width=67mm]{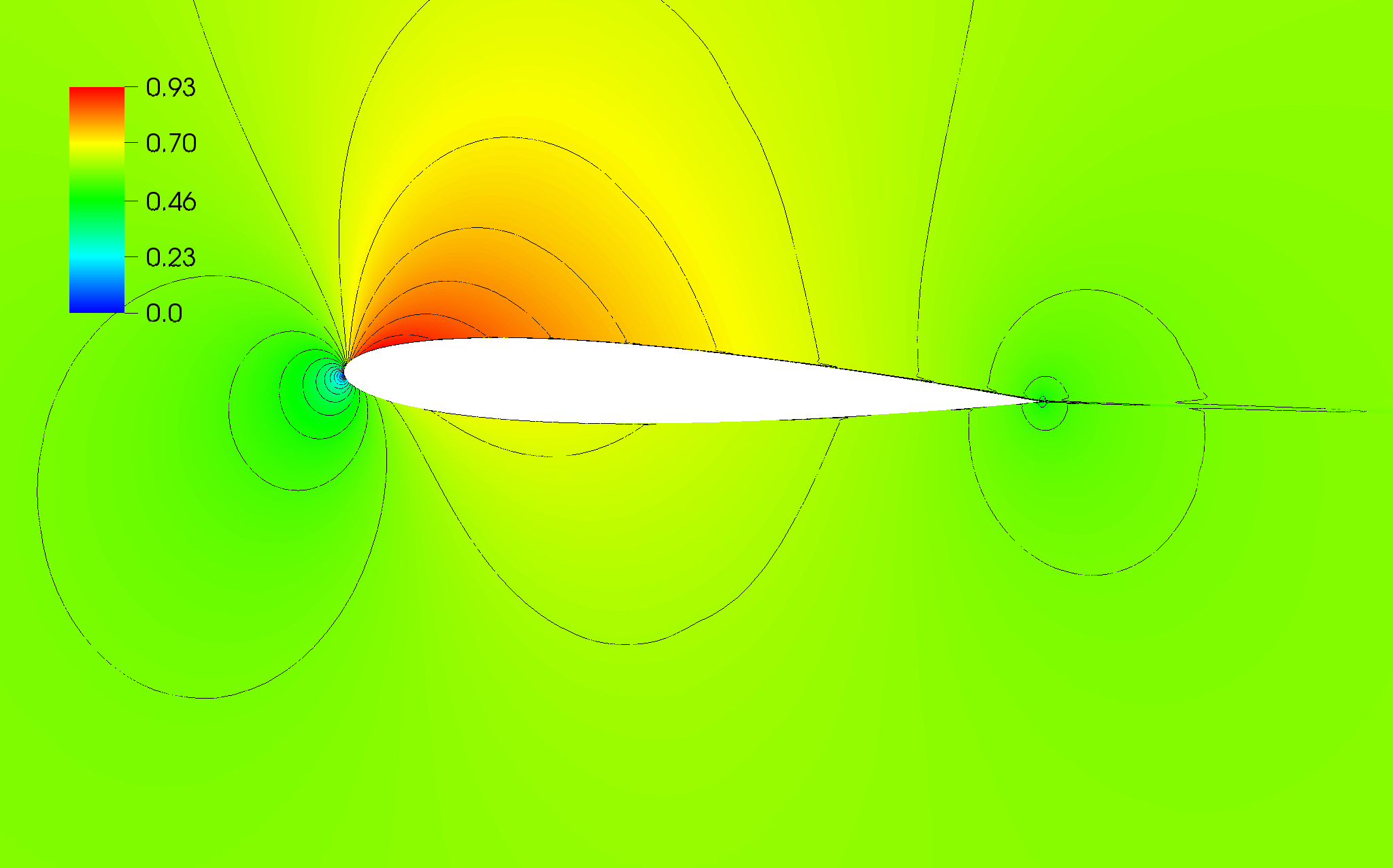}
\end{center}
\caption{Euler Mach field for $M=0.6$ and $\alpha=2.1^{\circ}$ via Euler CFD: NACA 0006 (left) and 0012 (right) aerofoils.}
\label{fig:mach2}
\end{figure}

\begin{figure}[h!]
\begin{center}
\includegraphics[width=67mm]{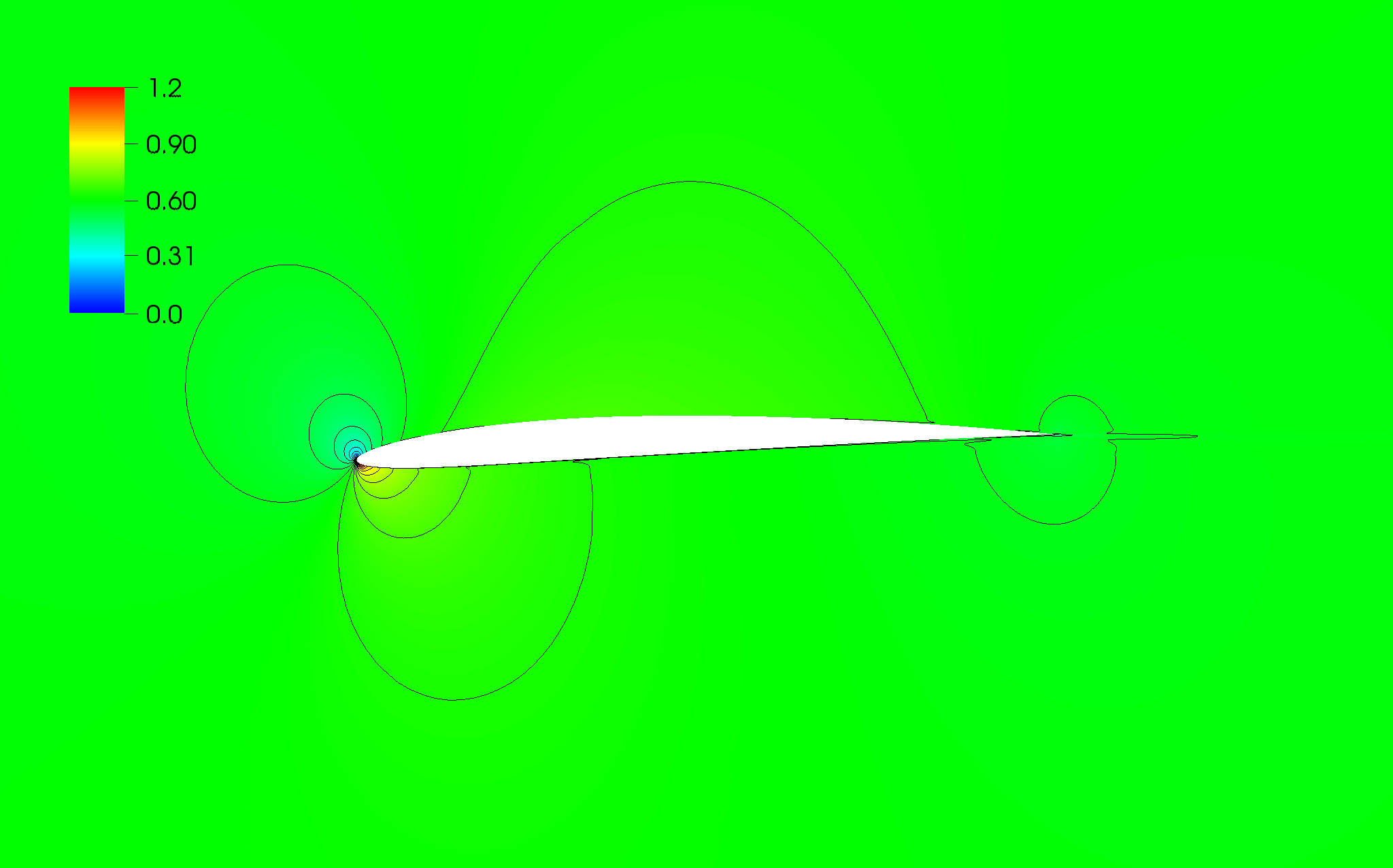}
\includegraphics[width=67mm]{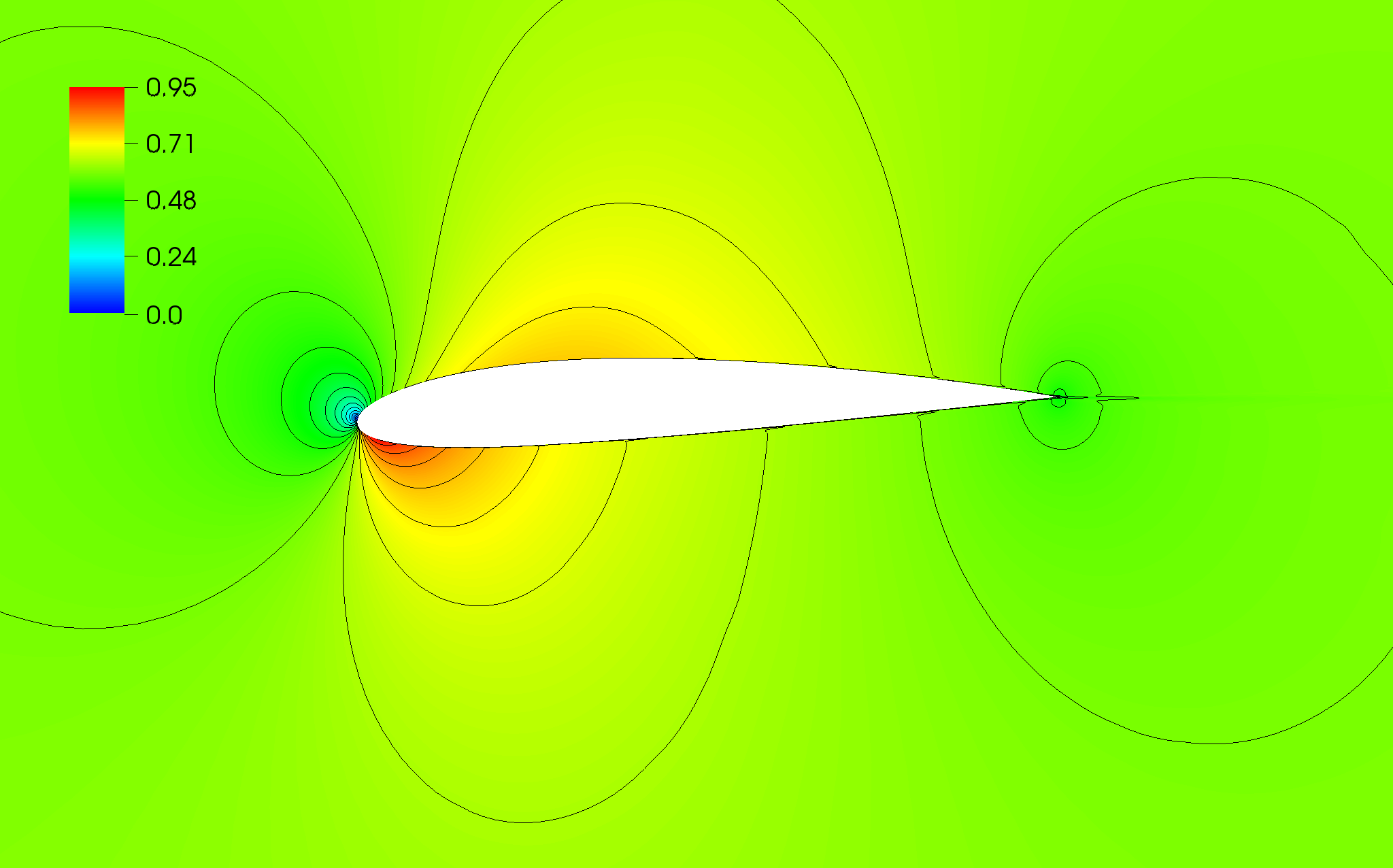}
\end{center}
\caption{Euler Mach field for $M=0.6$ and $\alpha=-2.1^{\circ}$ via Euler CFD: NACA 2406 (left) and 2412 (right) aerofoils.}
\label{fig:machn2}
\end{figure}

Figures \ref{fig:mach0} to \ref{fig:machn2} show the Mach field of the steady reference solution for all aerofoils flying at $M=0.6$. Note that when the unsteady flow perturbation $\Delta\alpha=1^{\circ}$ is also included the critical Mach number drops below $M=0.6$ for both NACA 0006 and 0012 aerofoils at $\alpha=2.1^{\circ}$, a sonic bubble indeed appearing in the pressure field calculated via CFD (see Figure \ref{fig:mach2}); therefore, such cases will not be investigated any further, as outside the applicability and scope of this work. Moreover, it is also possible to note that the high peak in the local Mach number causes the small discrepancy between numerical and analytical pressure coefficients at the aerofoil's nose, as all scalability rules hereby employed for subsonic flow (both steady and unsteady) do break down in transonic conditions \cite{Leishman1,Bendiksen}.

Numerical results based on RANS CFD have also been computed for the sake of generality and further cross-validation of the proposed computational method \cite{Righi} and consistent agreement with those based on Euler CFD was always found, especially in the case of the thinner NACA 0006 and 2406 aerofoils flying at the lower angles of attack (which prevent significant wake separation at the aerofoils' trailing edge). However, slightly lower Mach peaks and smaller transonic regions occurred, due to physical and numerical dissipation in the boundary layer with no-slip condition on the aerofoil.

\subsection{Unsteady Solutions: Dynamic Load}

Having obtained and verified the steady flow solution to be used as the initial condition, time-accurate simulations have then been performed to compute the unsteady flow response to both a unit step in angle of attack and a unit sharp-edged gust in terms of lift coefficient development, for each aerofoil geometry and Mach number; a thin aerofoil is also considered in the first place, for the sake of both a theoretical understanding and a direct validation of the proposed analytical model.

Note that, although the aerofoil is also pitching in most dynamic conditions, the unsteady lift due to a unit step-change in the pitch rate is not presented as its analysis follows the same principles of the one for the plunge motion and would hence add little substantial value to the present work; in fact, the aerodynamic indicial function for a unit step-change in the pitch rate is typically assumed just identical to that for a unit step-change in the angle of attack \cite{Leishman1,Wagner,Theodorsen1,Kussner,Sears}.

\subsubsection{Thin Aerofoil}

\begin{figure}[h!]
\begin{center}
\includegraphics[width=67mm]{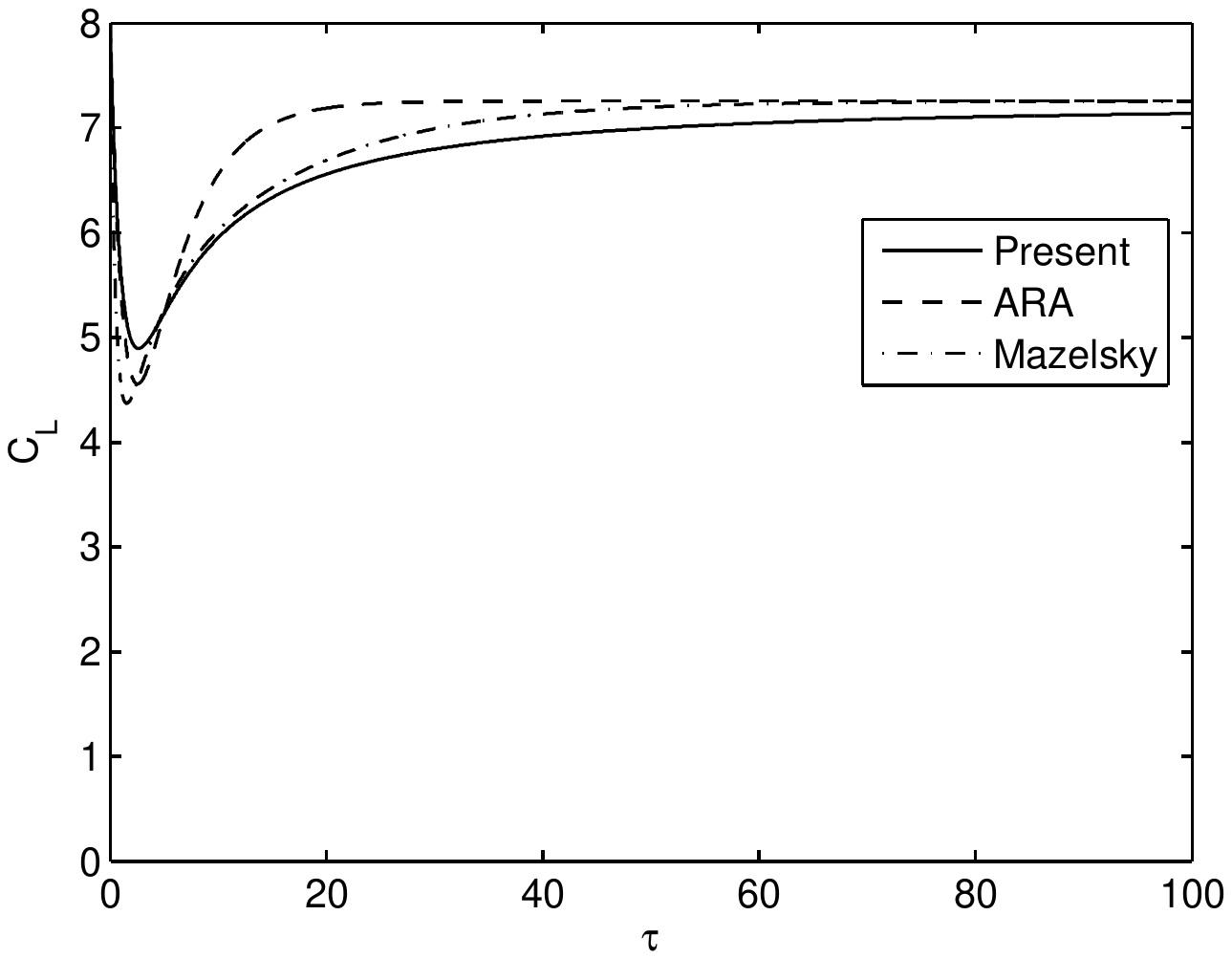}
\includegraphics[width=67mm]{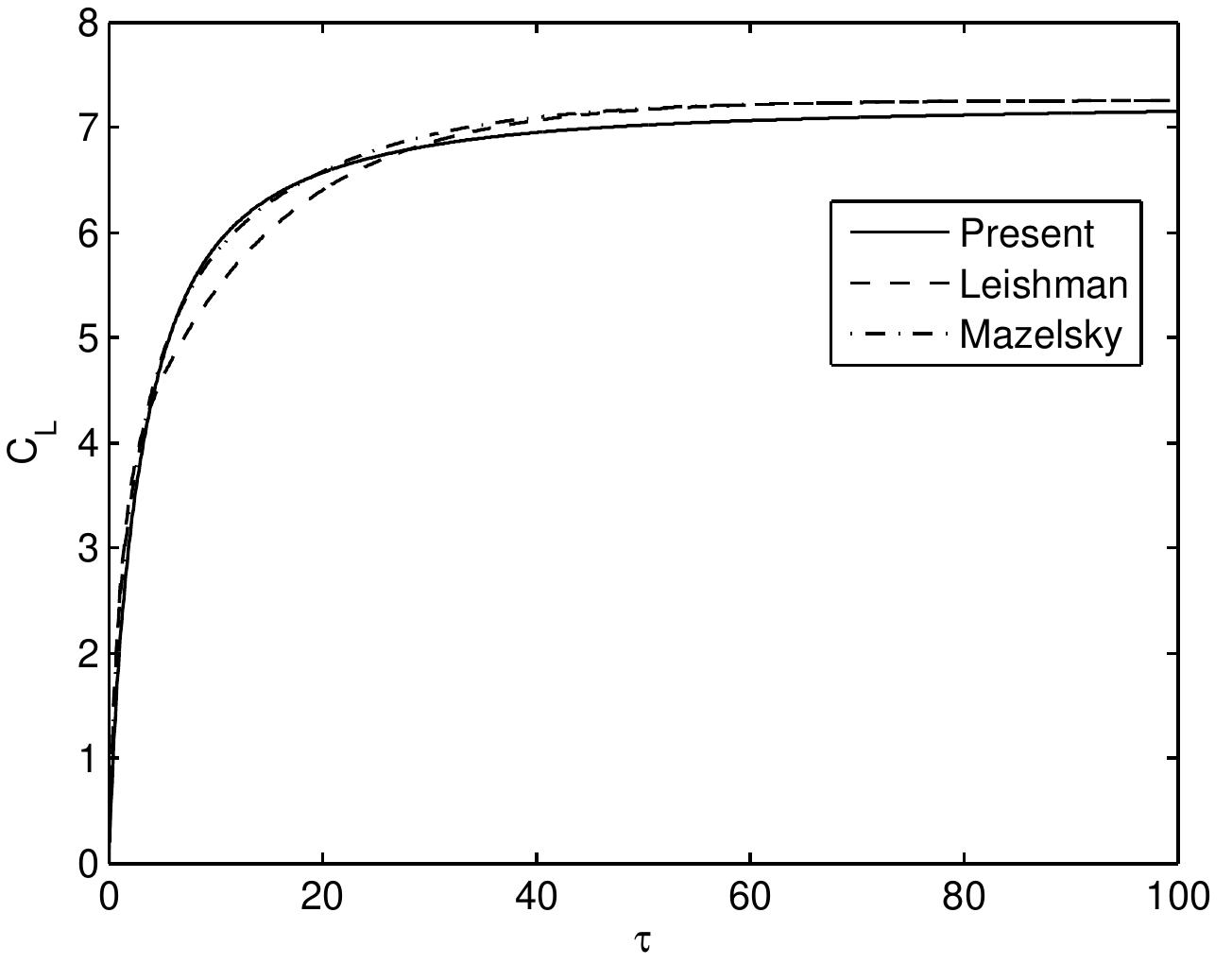}
\end{center}
\caption{Lift coefficient of a thin aerofoil at $M=0.5$: approximate analytical response to unit AoA (left) and unit SEG (right).}
\label{fig:simplest}
\end{figure}

\begin{figure}[h!]
\begin{center}
\includegraphics[width=67mm]{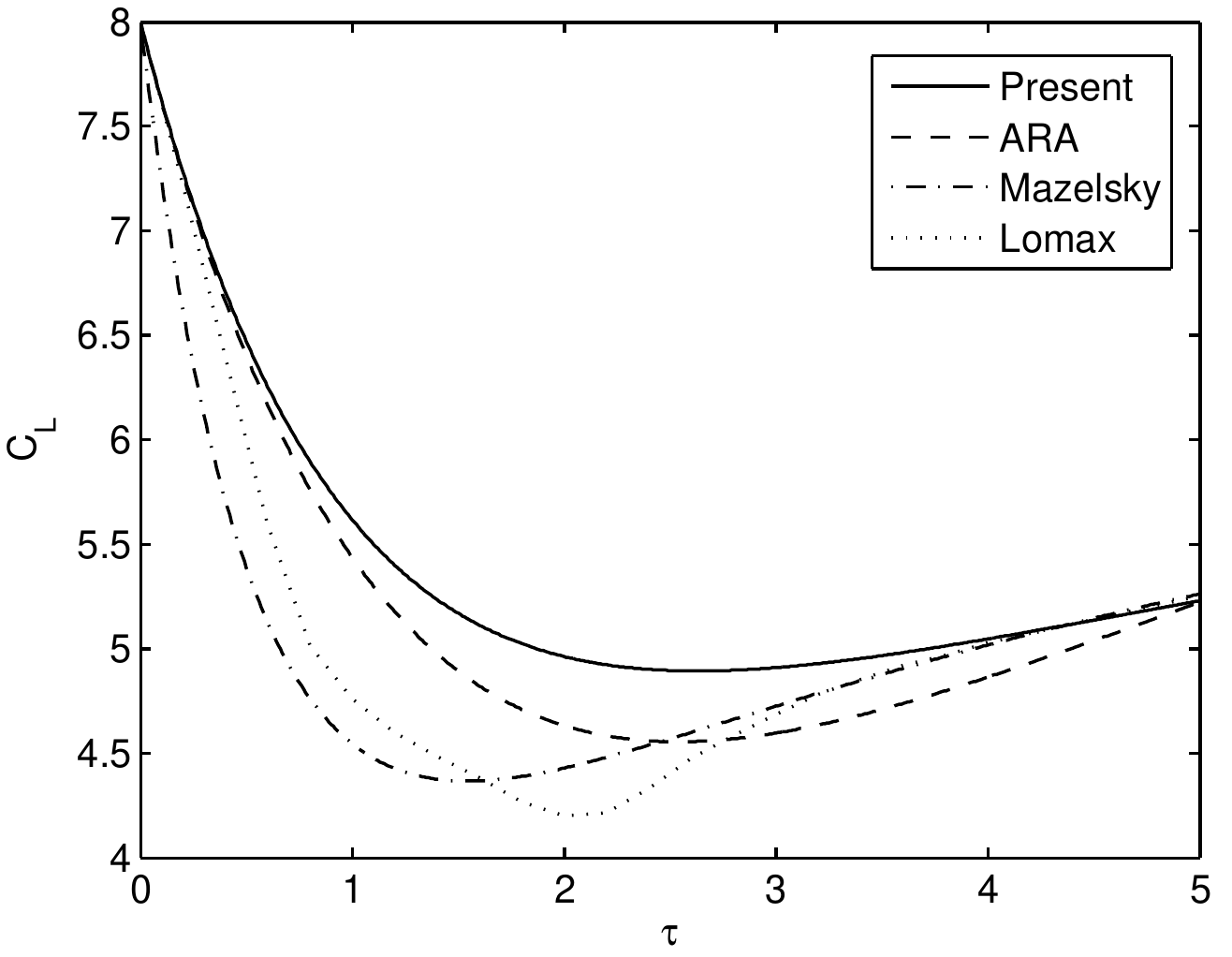}
\includegraphics[width=67mm]{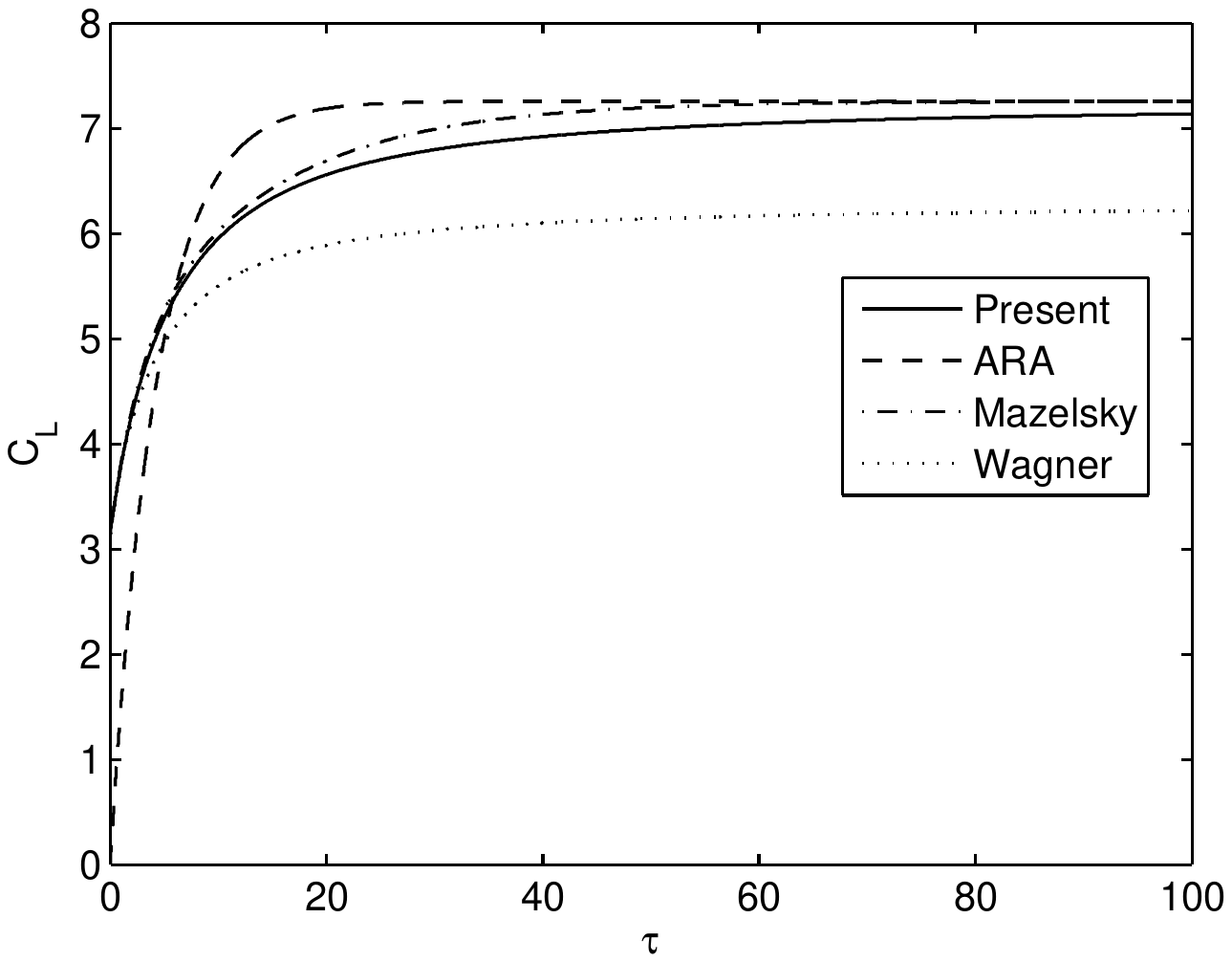}
\end{center}
\caption{Lift coefficient of a thin aerofoil at $M=0.5$ due to unit AoA: approximate non-circulatory (left) and circulatory (right) analytical response.}
\label{fig:simplestAoA}
\end{figure}

\begin{figure}[h!]
\begin{center}
\includegraphics[width=67mm]{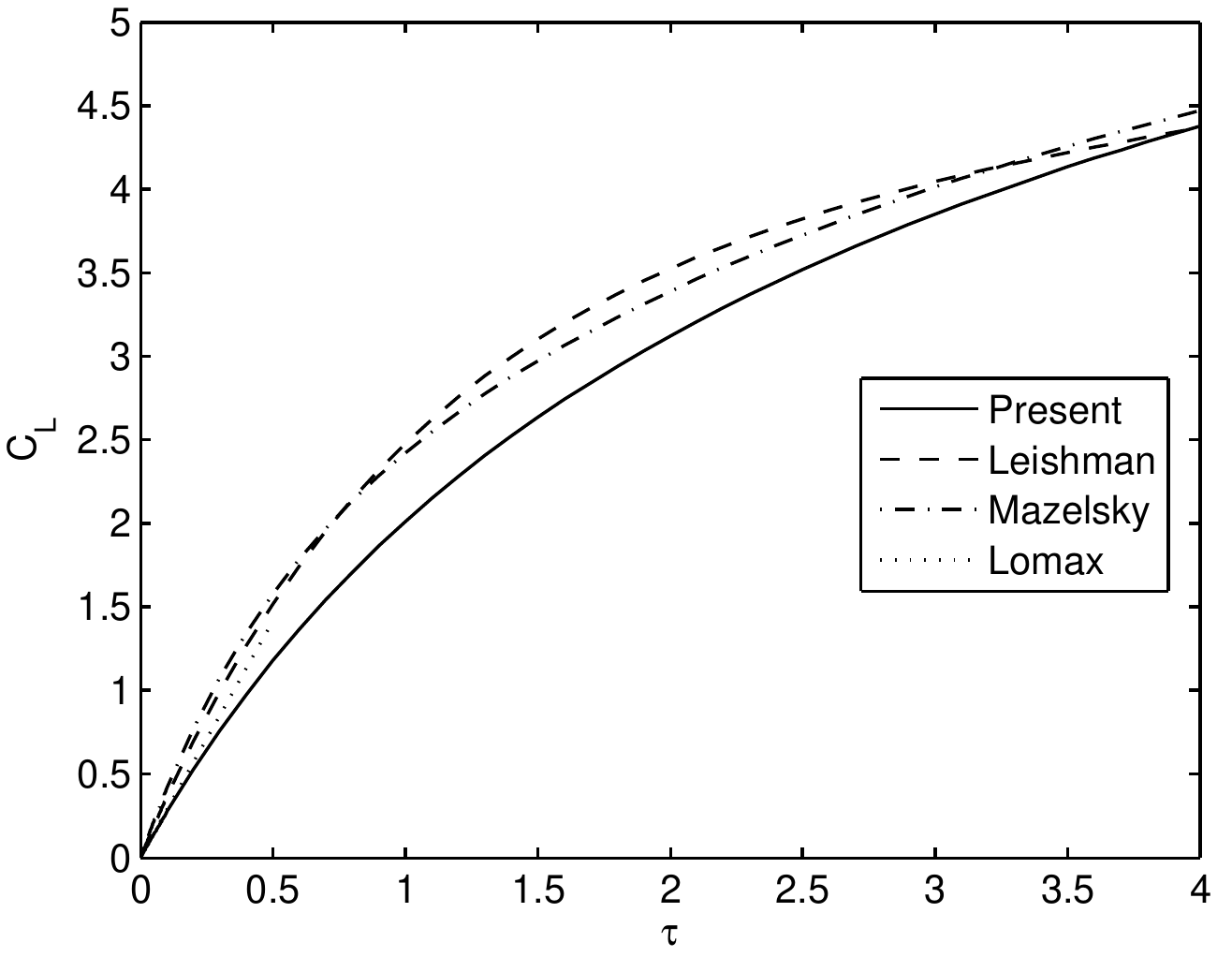}
\includegraphics[width=67mm]{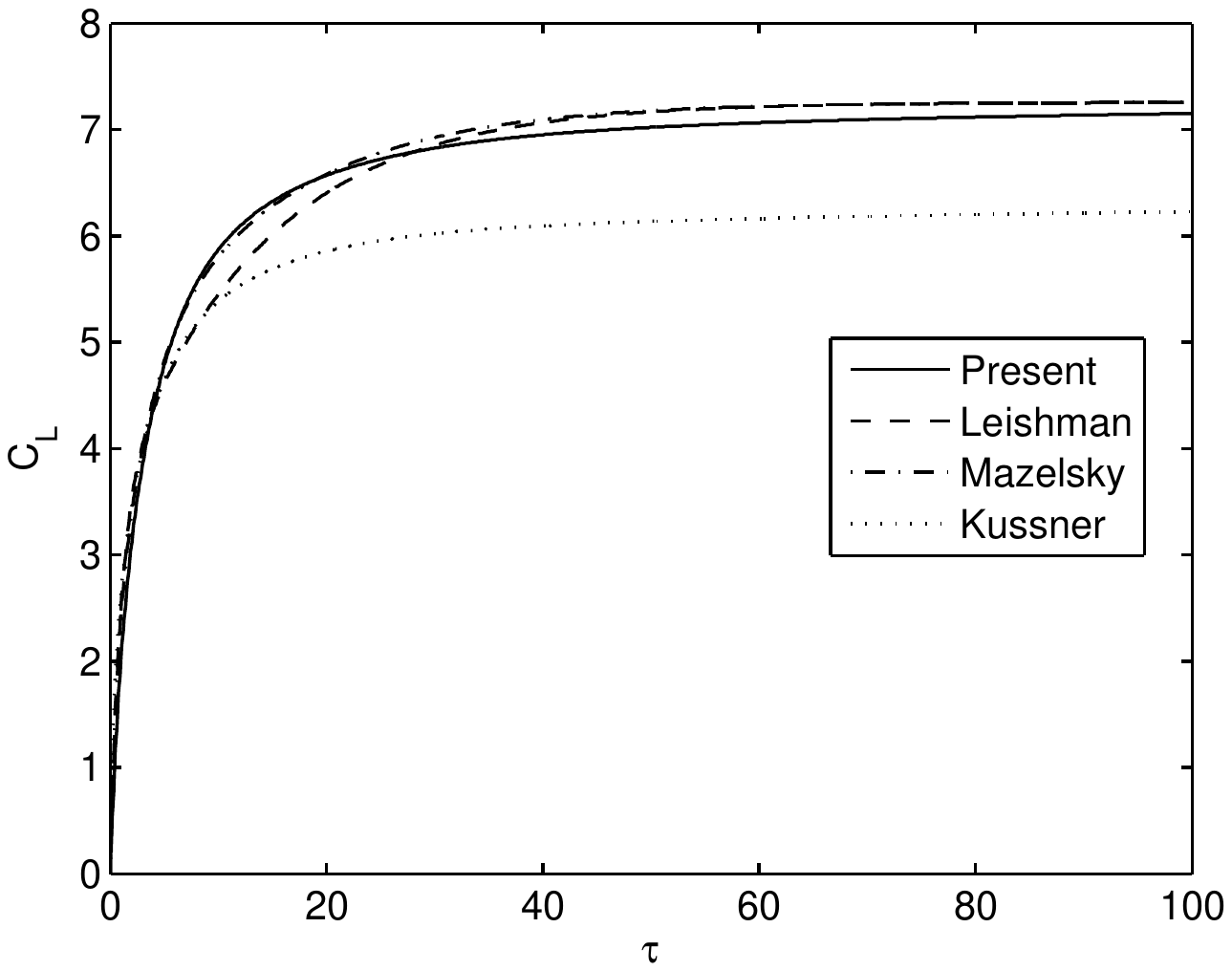}
\end{center}
\caption{Lift coefficient of a thin aerofoil at $M=0.5$ due to unit SEG: approximate non-circulatory (left) and circulatory (right) analytical response.}
\label{fig:simplestSEG}
\end{figure}

Figure \ref{fig:simplest} compares the present analytical solution with the few results available in the literature for the lift build-up due to a unit step in the angle of attack (left) and a unit sharp-edged gust (right) for a thin aerofoil at $M=0.5$; in both cases, the simplest form (i.e., a single effective exponential term) is employed for approximating the entire non-circulatory part. Figures \ref{fig:simplestAoA} and \ref{fig:simplestSEG} then show the non-circulatory (left) and circulatory (right) parts of the lift build-up; Lomax's exact solution \cite{Lomax2} is also depicted for direct validation of the non-circulatory lift build-up, while Wagner's \cite{Wagner} and Kussner's \cite{Kussner} functions are shown as the exact references for the circulatory lift build-up of incompressible potential flow.

For the the unit AoA case, both present and ARA's approximations \cite{Leishman1} agree well with piston theory \cite{Lighthill,Ashley} in the impulsive region at the start of the flow response; ARA's approximation is then closer to Mazelsky's results \cite{Mazelsky3} in the transitory region (where the latter is closer to Lomax's solution), whereas the present approximation is closer to Mazelsky's results in the circulatory region. Note that ARA's approximation of the circulatory contribution starts at $\breve{C}_{L0}^\alpha=0$ rather than $\breve{C}_{L0}^\alpha=\pi$ and predicts a totally different build-up of the circulatory lift, which does not resume Wagner's \cite{Wagner} function in the incompressible limit.

For the unit SEG case, instead, all approximations agree well with piston theory in the impulsive region at the start of the flow response, Leishman's \cite{Leishman1} and Mazelsky's \cite{Mazelsky3} results being closer (without possessing the exact limit behaviour though) also in the transitory region; then the present approximation stays close to Mazelsky's results for the rest of the flow response. Note that Leishman's approximation of the circulatory contribution does not resume Kussner's \cite{Kussner} function accurately in the incompressible limit.

It is worth stressing that both ARA's and Mazelsky's approximations were determined from the (Laplace-transform based) reciprocal relations between aerodynamic indicial functions for a unit step in the angle of attack in the reduced-time domain and oscillatory lift coefficients in the reduced-frequency domain \cite{Bisplinghoff,Leishman1}. The latter were always obtained from prescribed harmonic plunging of the aerofoil at a limited set of individual reduced frequencies but were collected experimentally (under nominally attached flow conditions) by ARA whereas numerically (from different sources and methods) by Mazelsky, who subsequently calculated the aerodynamic indicial functions for a unit sharp-edge gust from its reciprocal relations with those for a unit step in the angle of attack \cite{Garrick,Mazelsky3}. Opposite to the proposed formulation, all approximate indicial functions provided by Mazelsky were separately derived (with related specific sets of coefficients) at selected Mach numbers, which is not ideal for routine application within the preliminary MDO of flexible subsonic wings.

\begin{figure}[h!]
\begin{center}
\includegraphics[width=67mm]{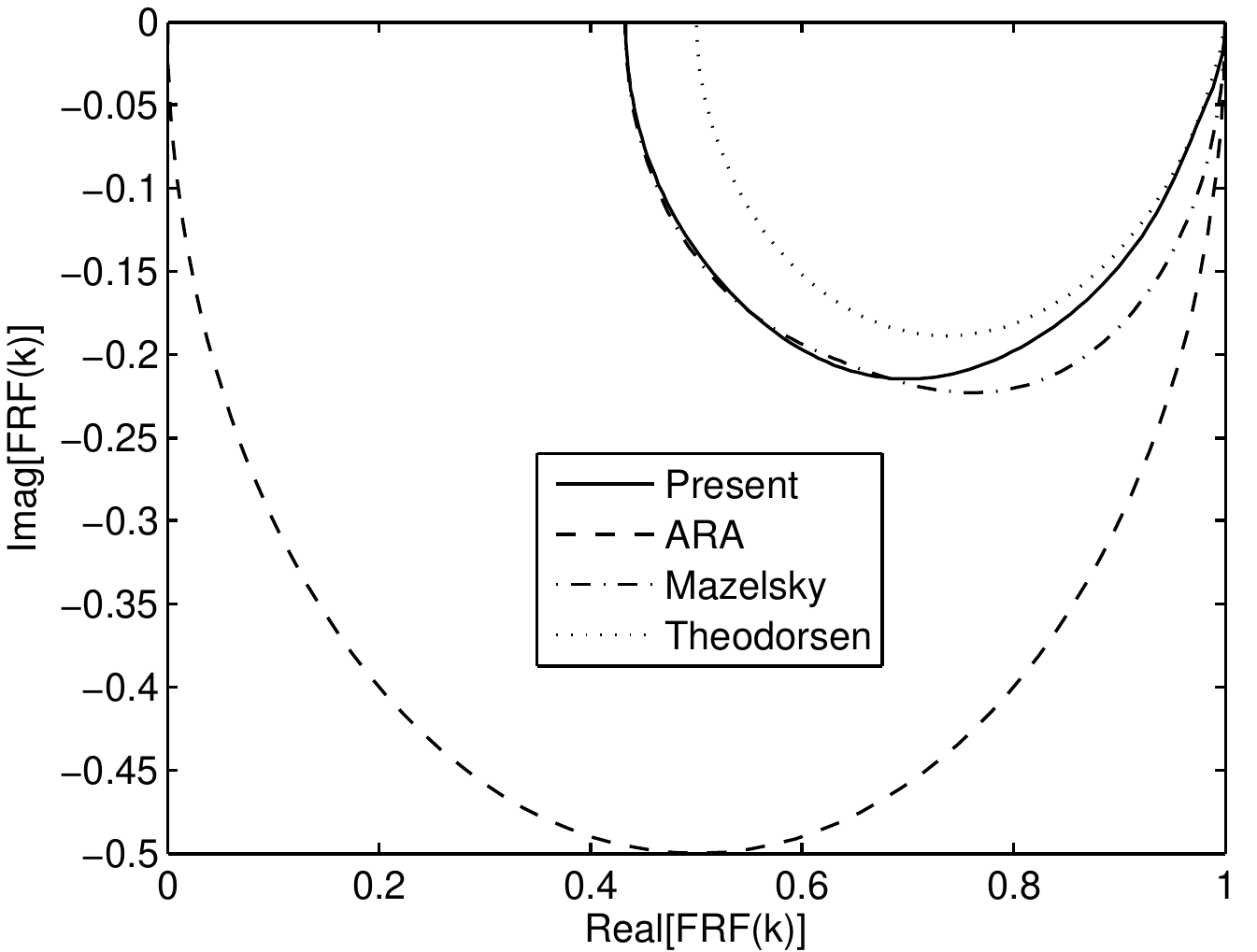}
\includegraphics[width=67mm]{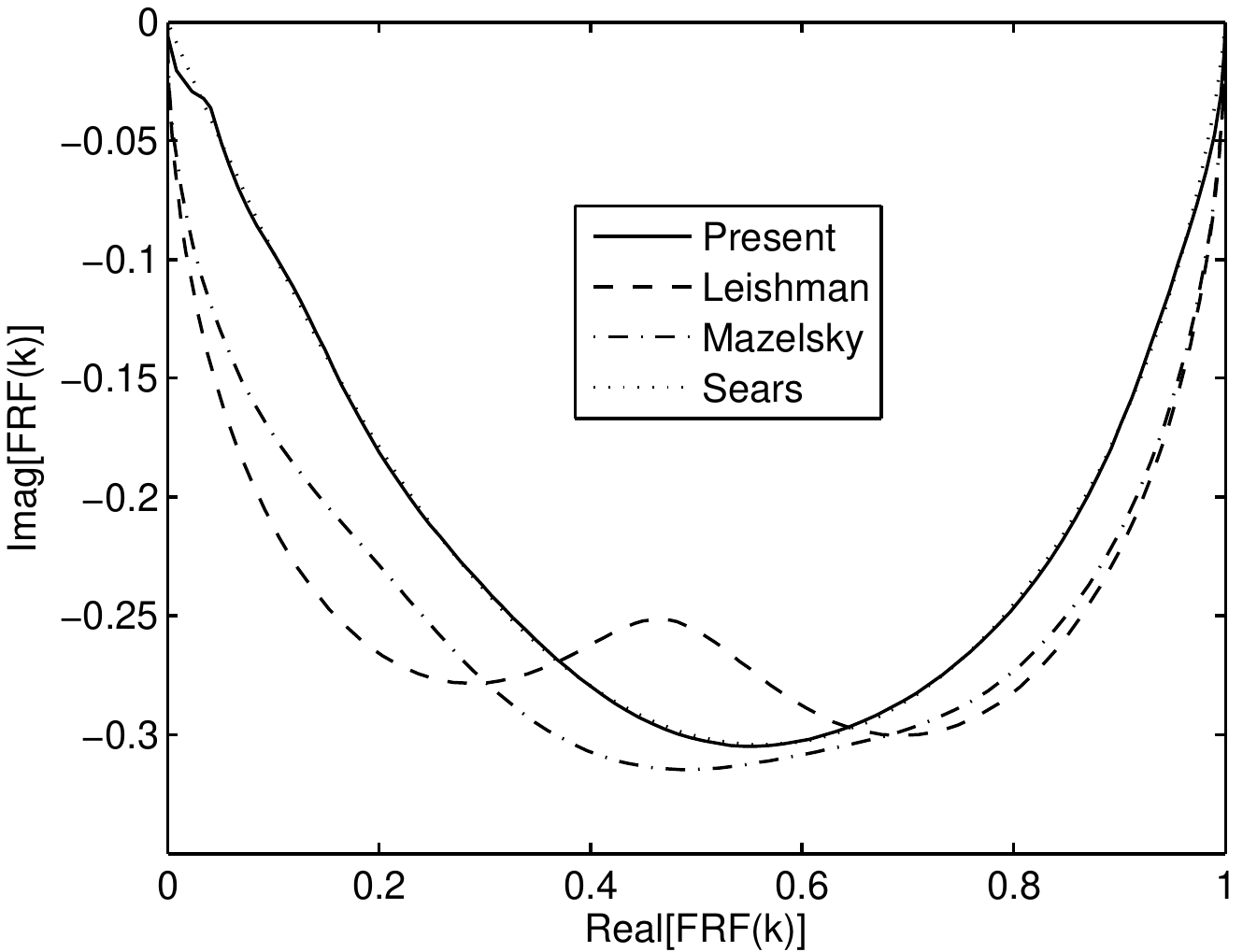}
\end{center}
\caption{Frequency response function for the circulatory contribution of a thin aerofoil at $M=0.5$: approximate analytical response to unit AoA (left) and unit SEG (right).}
\label{fig:compaFRF}
\end{figure}

Figure \ref{fig:compaFRF} presents the frequency response functions (FRF) of the circulatory lift build-up due to unit AoA (left) and unit SEG (right); Theodorsen's \cite{Theodorsen1} and Sears' \cite{VonKarman,Sears} functions are also shown as exact references for incompressible flow. Due to Laplace transform \cite{Quarteroni2}, any discrepancy in approximating the indicial-admittance function translates into an discrepancy in approximating the relative frequency response function and vice-versa (see Appendix) \cite{Dowell}; therefore, ARA's approximation \cite{Leishman1} predicts a totally different FRF for the circulatory lift build-up. In particular, a good accuracy in approximating the latter at high reduced frequencies (i.e., highly-unsteady flow) translates into a good accuracy in approximating the former at low reduced time (i.e., transitory response), while a good accuracy at low reduced frequencies (i.e., quasi-steady flow) translates into a good accuracy at high reduced time (i.e., asymptotic response) \cite{Berci1}.

\begin{figure}[h!]
\begin{center}
\includegraphics[width=67mm]{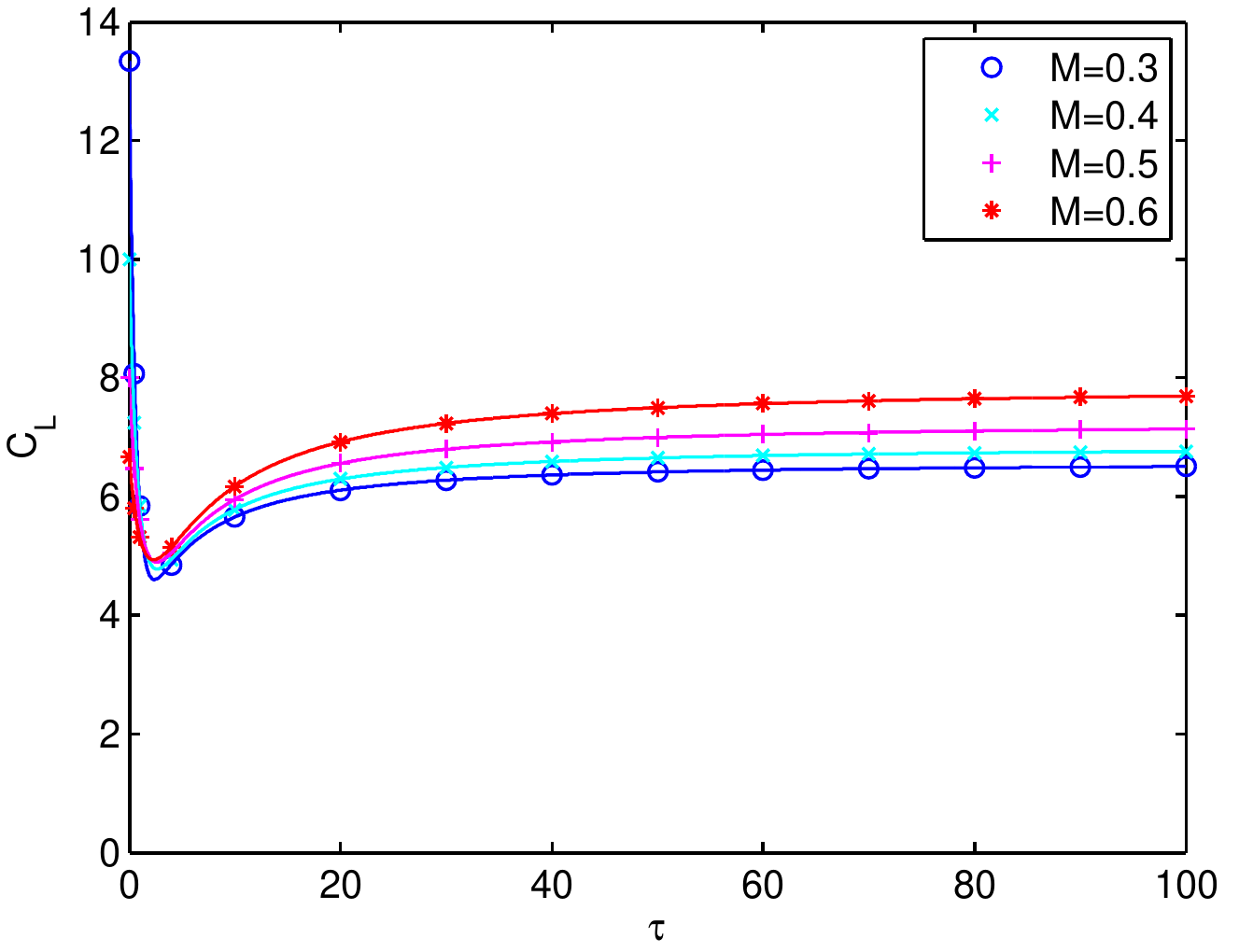}
\includegraphics[width=67mm]{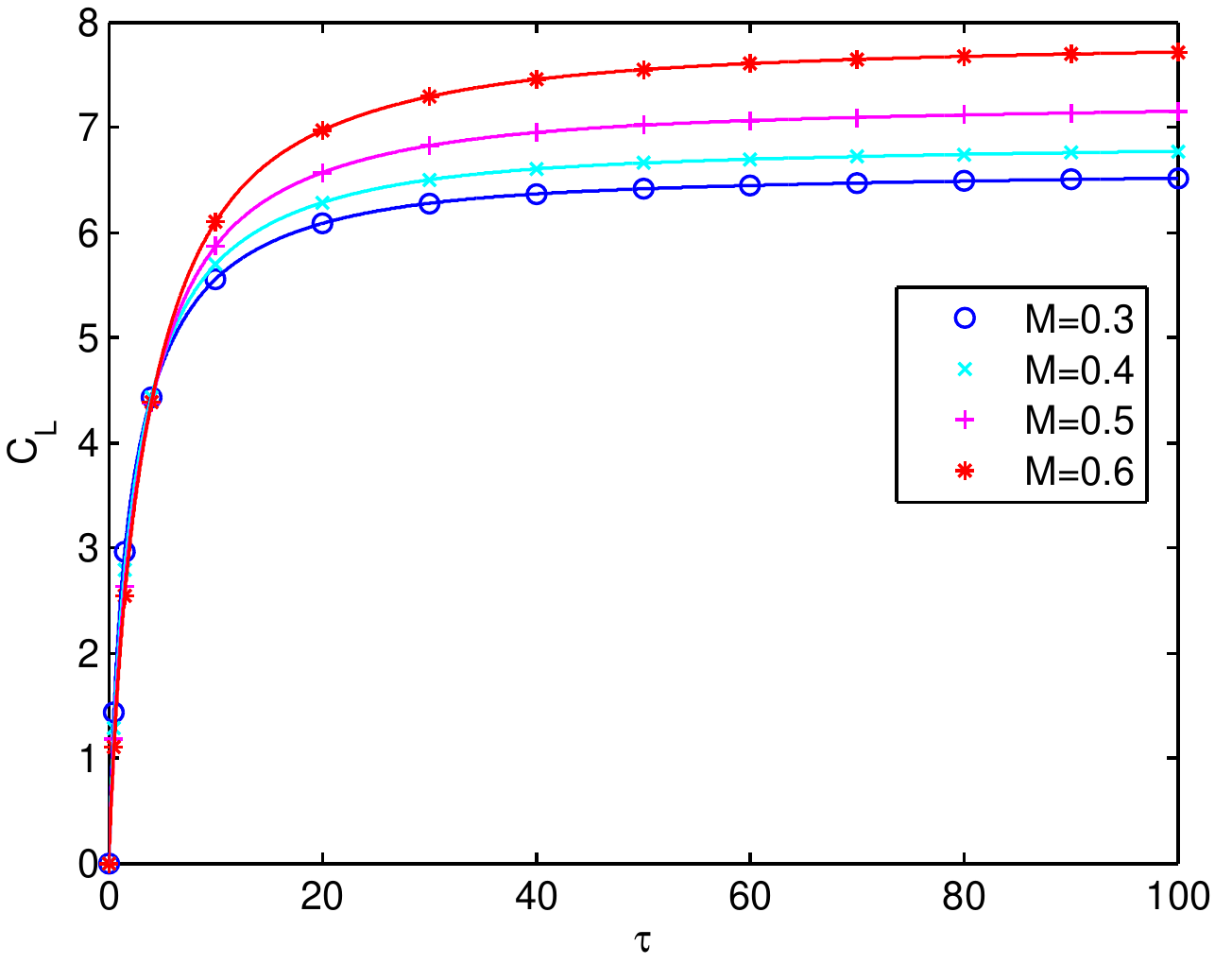}
\end{center}
\caption{Lift coefficient of a thin aerofoil: response to unit AoA (left) and unit SEG (right) with simplest analytical approximation of the non-circulatory contribution.}
\label{fig:allsimplest}
\end{figure}

Figure \ref{fig:allsimplest} then shows the unsteady lift coefficient due to unit AoA (left) and unit SEG (right) for a thin aerofoil at $M=0.3$, $M=0.4$, $M=0.5$ and $M=0.6$, using the simplest approximation form (i.e., a single effective exponential term) for the non-circulatory contribution. All curves agree exactly with piston theory \cite{Lighthill,Ashley} at the impulsive start of the non-circulatory response and asymptotically approach Prandtl-Glauert's theoretical (steady) value \cite{Glauert2} at the end of the circulatory response. As expected, the lower the Mach number the higher the impulsive lift whereas the opposite is true for the circulatory lift, with relative differences exacerbated towards the lower $M\rightarrow0$ and upper $M\rightarrow1$ Mach limits; the transitory region comes out rather similar for all curves, instead, as they cross each other thereby. The frequency response functions of the circulatory lift build-up are fully shown in the Appendix.

In both AoA and SEG cases, the approximation of the non-circulatory response may finally be improved when a series of damped oscillatory terms is best-fitted to the difference between non-circulatory and circulatory responses in the entire validity range of the former \cite{Righi}, by minimising the approximation error via nonlinear optimisation \cite{Berci1,Tiffany} (see Appendix). Figure \ref{fig:HOTR} shows the unsteady lift coefficient due to unit AoA (left) and unit SEG (right) in the transitory region for a thin aerofoil at all subsonic Mach numbers considered, where either a single damped oscillatory term or a couple of exponential terms are effectively employed; the applicable coefficients are shown in Table \ref{tab:HO2}. Note that, in the SEG case only, the last three exponential terms in Table \ref{tab:2D} for the approximation of the circulatory contribution have explicitly been cancelled out and essentially replaced by either the single damped oscillatory term or the couple of exponential terms implicitly satisfying the exact limit behaviour of piston theory. The latter is now well reproduced but the net improvement in approximating the entire non-circulatory response is remarkable in the acoustic-waves driven transitory region too, at least in terms of correct overall trend \cite{Lomax2}. Note that the approximation employing a couple of exponential terms is slightly smoother than the one using a single damped oscillatory term, the former approximation also being more conservative in the case AoA cases while the opposite is true in the SEG case; however, both approximations decay and join the circulatory response at roughly the same time (which depends on the Mach number: the higher the latter, the longer the non-circulatory response \cite{Pike}). Moreover, the reduced instants where the curves cross are also similar for both approximation types in both AoA and SEG cases.

\begin{figure}[h!]
\begin{center}
\includegraphics[width=67mm]{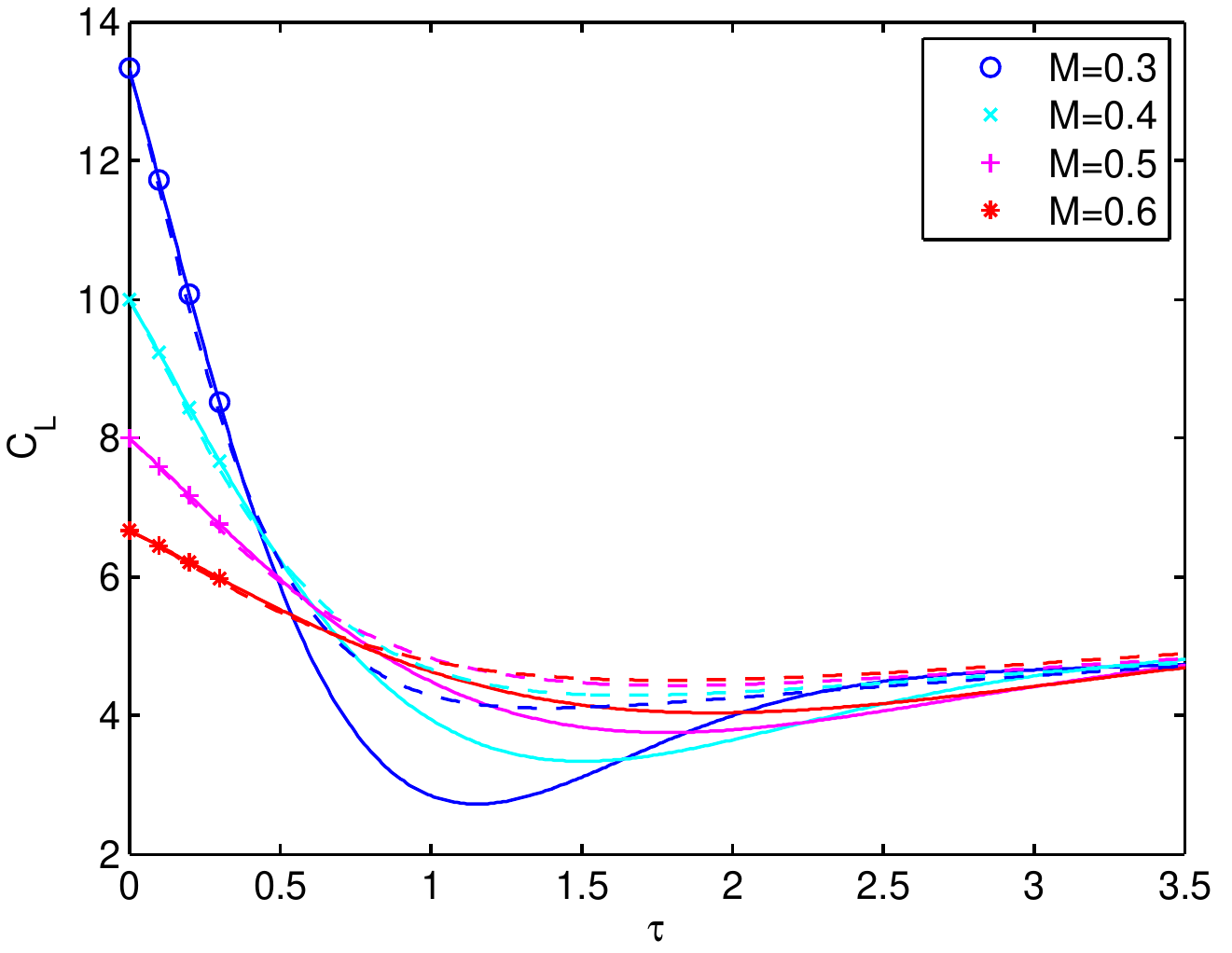}
\includegraphics[width=67mm]{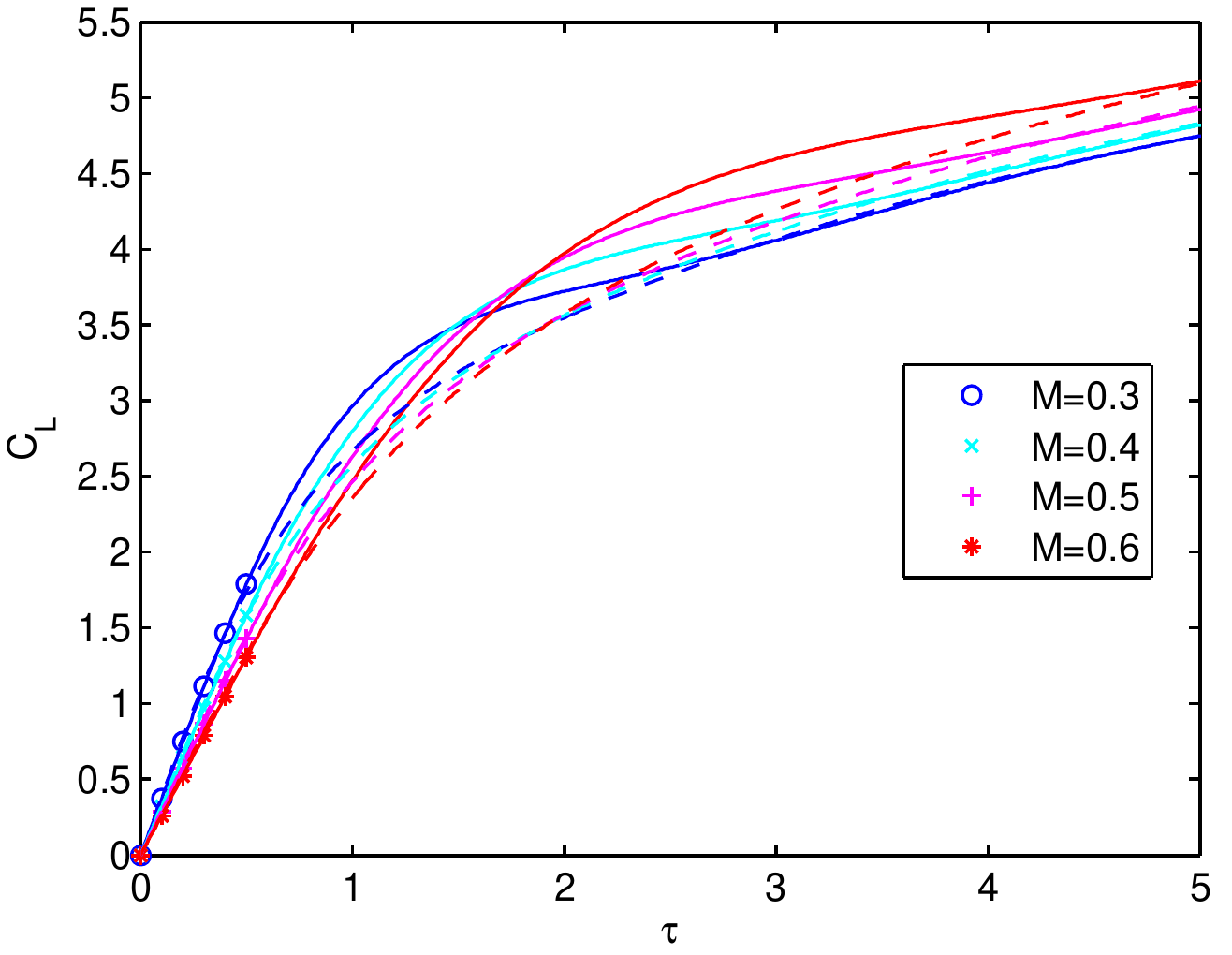}
\end{center}
\caption{Lift coefficient of a thin aerofoil in the transitory region: approximate analytical response to unit AoA (left) and unit SEG (right); symbols = piston theory, continuous lines = single damped oscillatory term, dashed lines = couple of exponential terms.}
\label{fig:HOTR}
\end{figure}                

\begin{table}[h!]
\begin{center}
\begin{tabular}{lcccc}
$M$ & $0.3$ & $0.4$ & $0.5$ & $0.6$ \\ \hline 
$\hat{A}_1^{\alpha}$ & $10.192$ & $6.8584$ & $4.8584$ & $3.5251$ \\
$\hat{B}_1^{\alpha}$ & $1.7598$ & $1.4342$ & $1.3047$ & $1.3117$ \\
$\hat{\Omega}_1^{\alpha}$ & $2.2471$ & $1.7793$ & $1.5711$ & $1.5374$ \\
$\hat{A}_1^G$ & $-2.0432$ & $-2.1266$ & $-2.2506$ & $-2.4363$ \\
$\hat{B}_1^G$ & $1.4224$ & $1.2288$ & $1.1342$ & $1.1144$ \\
$\hat{\Omega}_1^G$ & $1.8191$ & $1.5910$ & $1.4698$ & $1.4297$ \\
\vspace*{7mm}
\end{tabular}
\hspace*{10mm}
\begin{tabular}{lcccc}
$M$ & $0.3$ & $0.4$ & $0.5$ & $0.6$ \\ \hline 
$\hat{A}_1^{\alpha}$ & $19.637$ & $11.112$ & $6.9200$ & $4.6276$ \\
$\hat{A}_2^{\alpha}$ & $-9.4453$ & $-4.2535$ & $-2.0616$ & $-1.1025$ \\
$\hat{B}_1^{\alpha}$ & $3.9519$ & $2.9579$ & $2.4813$ & $2.3563$ \\
$\hat{B}_2^{\alpha}$ & $6.3172$ & $5.4147$ & $5.2541$ & $5.6960$ \\
$\hat{A}_1^G$ & $0.5203$ & $0.5539$ & $0.5456$ & $0.5095$ \\
$\hat{A}_2^G$ & $-2.5634$ & $-2.6805$ & $-2.7961$ & $-2.9458$ \\
$\hat{B}_1^G$ & $7.1144$ & $6.1526$ & $5.7815$ & $5.8689$ \\
$\hat{B}_2^G$ & $2.5777$ & $2.2462$ & $2.0409$ & $1.9368$ \\
\end{tabular}
\end{center}
\caption{Coefficients for the optimal analytical approximation of thin aerofoil's non-circulatory response to unit AoA and unit SEG for compressible flow: single damped oscillatory term (left) and couple of exponential terms (right).}
\label{tab:HO2}
\end{table}

\subsubsection{NACA Aerofoils}

Up to this point, no information has been used from the CFD results. In order to make proper comparisons between numerical and analytical results, the latter have suitably been tuned so to give the (steady) asymptotic value of the circulatory lift build-up of the former \cite{Wieseman,Brink,Palacios,Dimitrov,Sucipto,Berci3}. Moreover, two damped oscillatory terms are now effectively employed for best-fitting the difference between CFD results and circulatory analytical response, so that also the transitory region may accurately be reproduced \cite{Righi}. Note that the initial step of the circulatory lift build-up is kept as the theoretical value $\breve{C}_L^\alpha=\pi$, since aerofoil-specific circulatory effects (e.g., due to shape thickness and boundary layer) take time to develop \cite{Lomax2}. Still, in the SEG case only, the last three exponential terms in Table \ref{tab:2D} for the approximation of the circulatory contribution have explicitly been cancelled out and essentially replaced by either the two damped oscillatory terms implicitly satisfying the exact limit behaviour of piston theory \cite{Lighthill,Ashley}.

\begin{figure}[h!]
\begin{center}
\includegraphics[width=67mm]{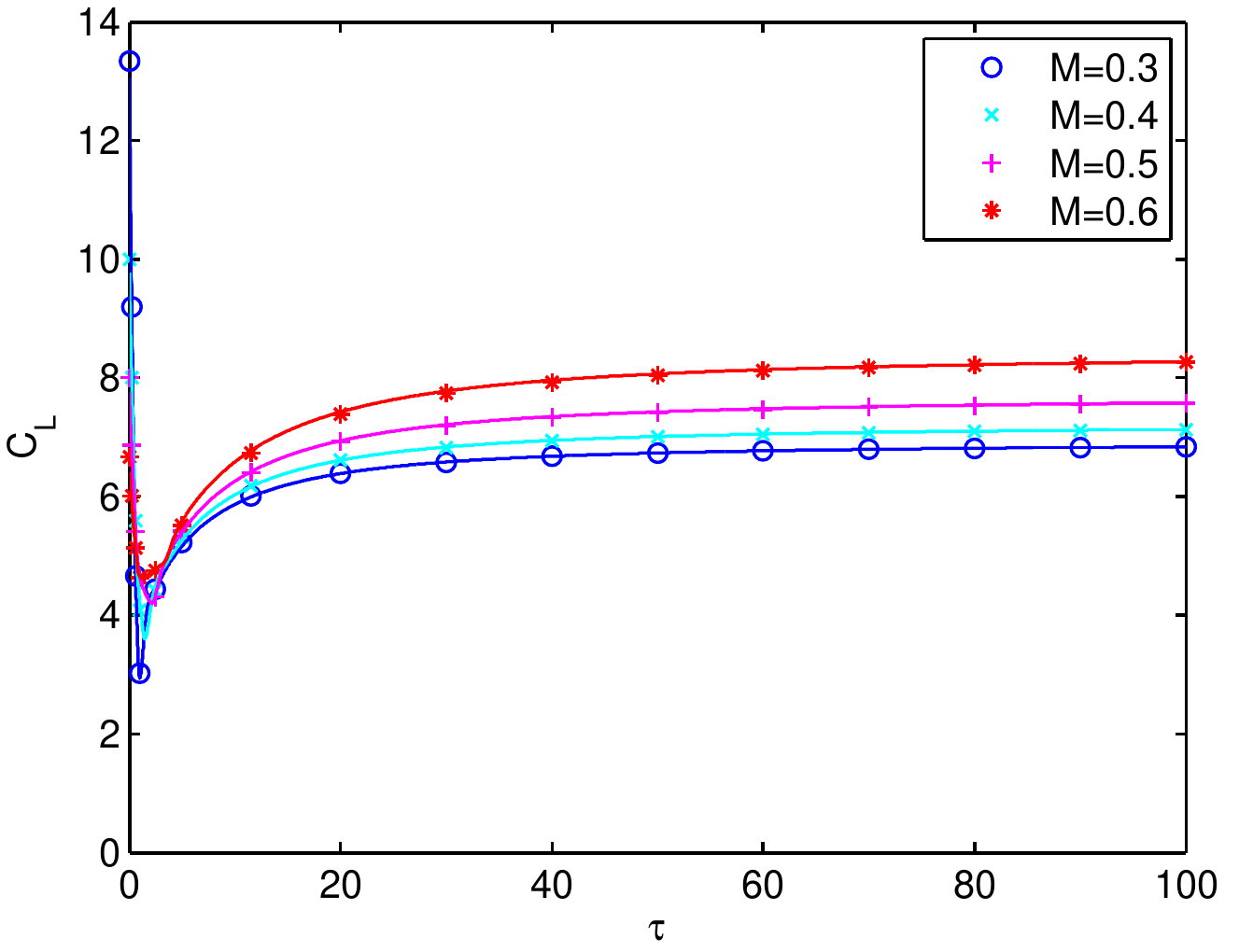}
\includegraphics[width=67mm]{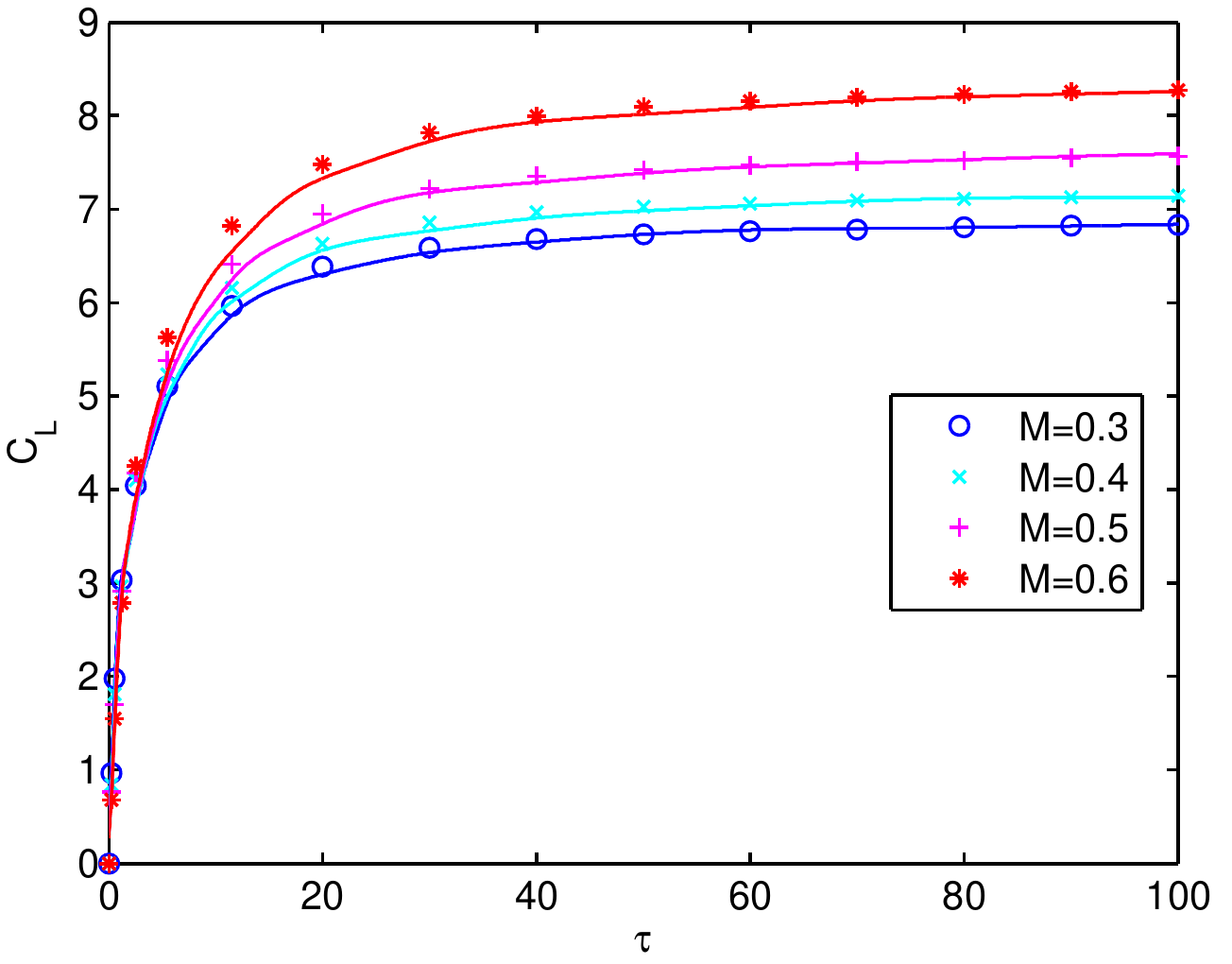}
\end{center}
\caption{NACA 0006 lift coefficient for $\alpha=0^{\circ}$: tuned response to unit AoA (left) and unit SEG (right); lines = Euler CFD, symbols = analytical approximation.}
\label{fig:tuned0006}
\end{figure}

Figure \ref{fig:tuned0006} shows the unsteady lift coefficient due to unit AoA (left) and unit SEG (right) for the NACA 0006 aerofoil at $\alpha=0^{\circ}$, where the analytical results have been tuned so to mimic the Euler CFD results for each subsonic Mach number considered. Except for marginal oscillations in the numerical gust response (also due to the travelling gust profile being calculated as part of the CFD solution at each time step and hence slightly modified by the aerofoil's presence \cite{Basu}), a striking agreement is found between tuned analytical approximations and CFD results in both AoA and SEG cases for all Mach numbers, as the aerofoil is thin and symmetric. The same considerations relative to the results for the response to unit AoA apply also to the results for the response to unit SEG; the flow perturbation being unitary, the (steady) asymptotic values of the circulatory lift build-up are also correctly the same for each Mach number in all corresponding (tuned) analytical and numerical cases.

\begin{figure}[h!]
\begin{center}
\includegraphics[width=67mm]{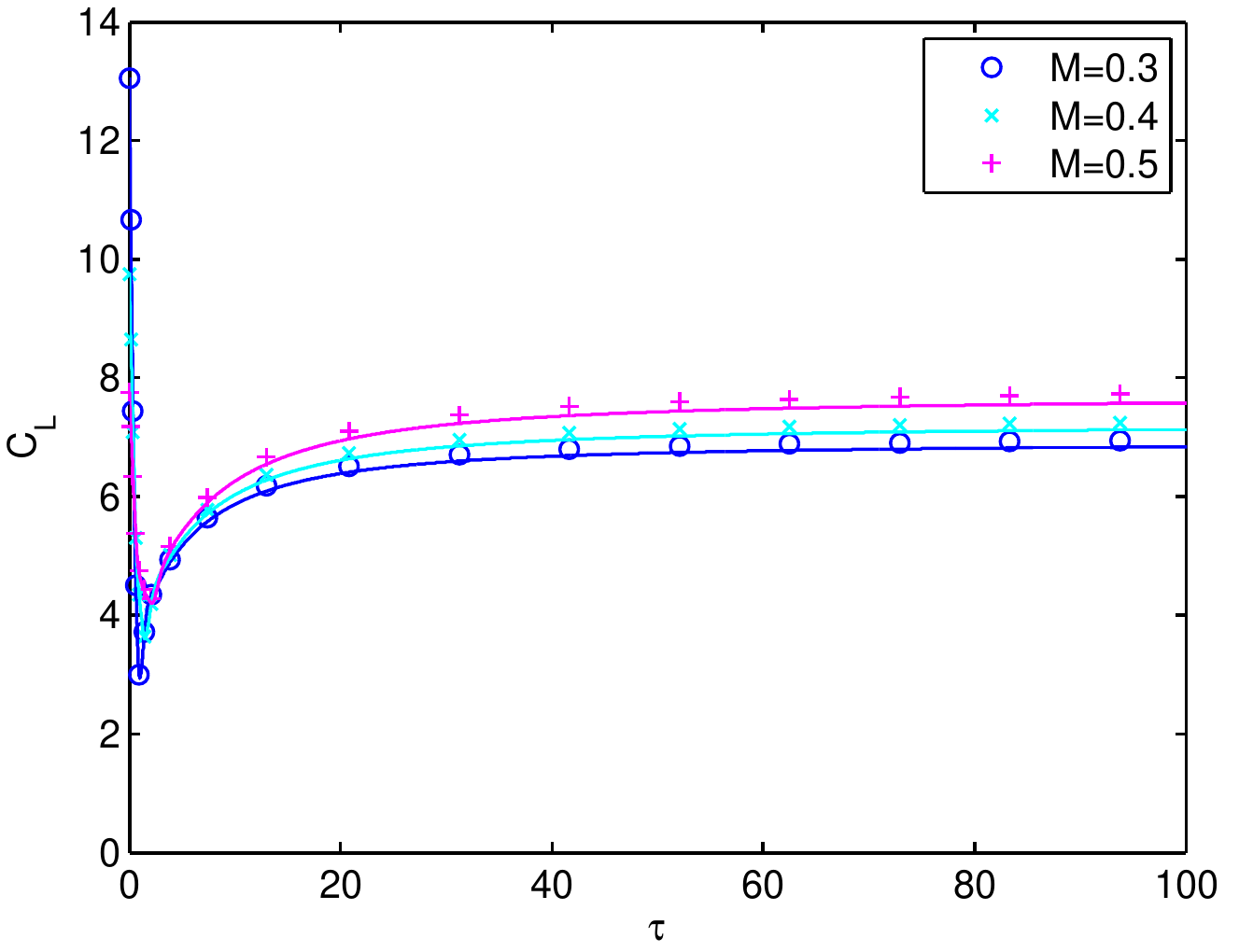}
\includegraphics[width=67mm]{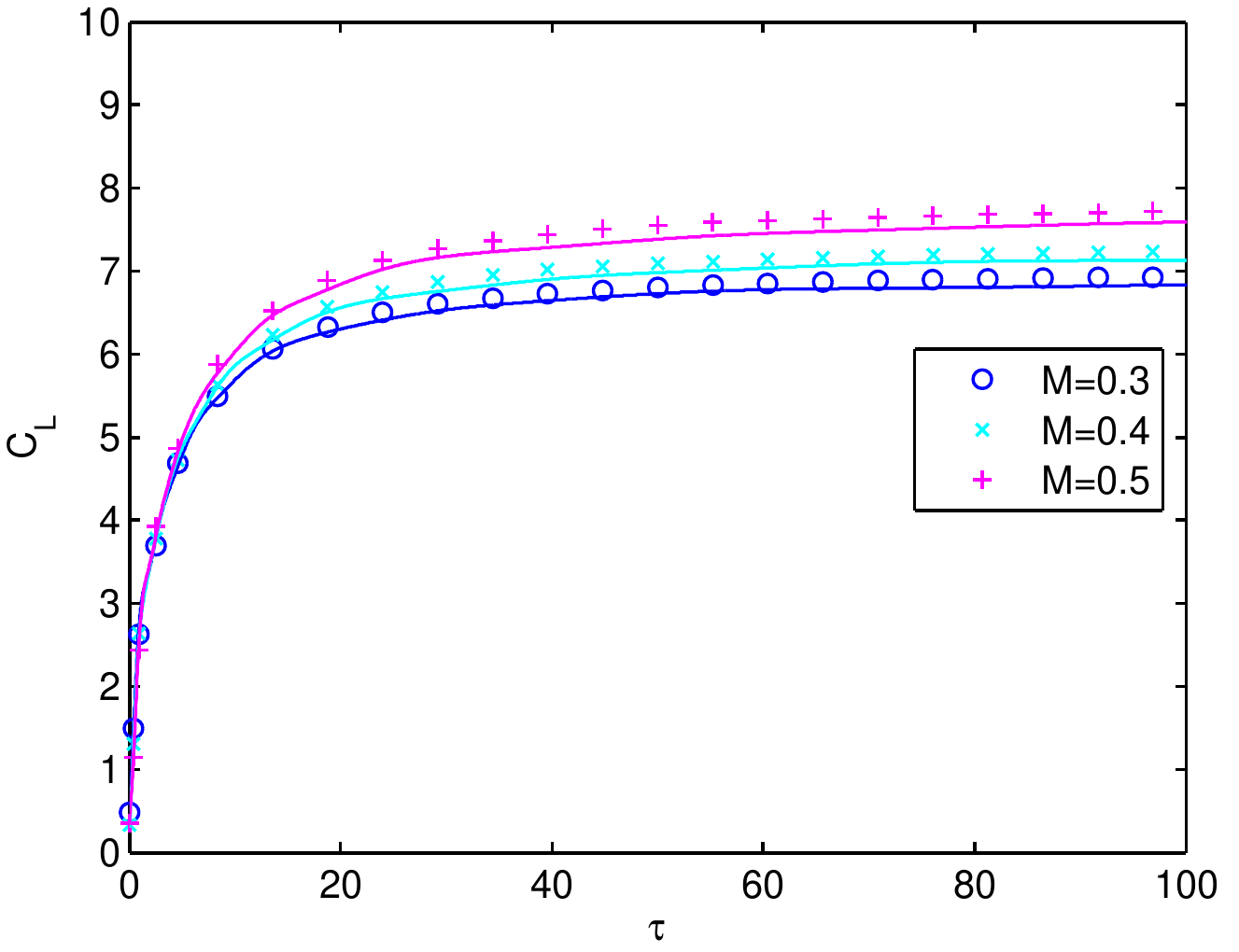}
\end{center}
\caption{NACA 0006 lift coefficient via Euler CFD: response to unit AoA (left) and unit SEG (right); lines = solution at $\alpha=0^{\circ}$, symbols = solution at $\alpha=2.1^{\circ}$.}
\label{fig:EURA0006}
\end{figure}

\begin{figure}[h!]
\begin{center}
\includegraphics[width=67mm]{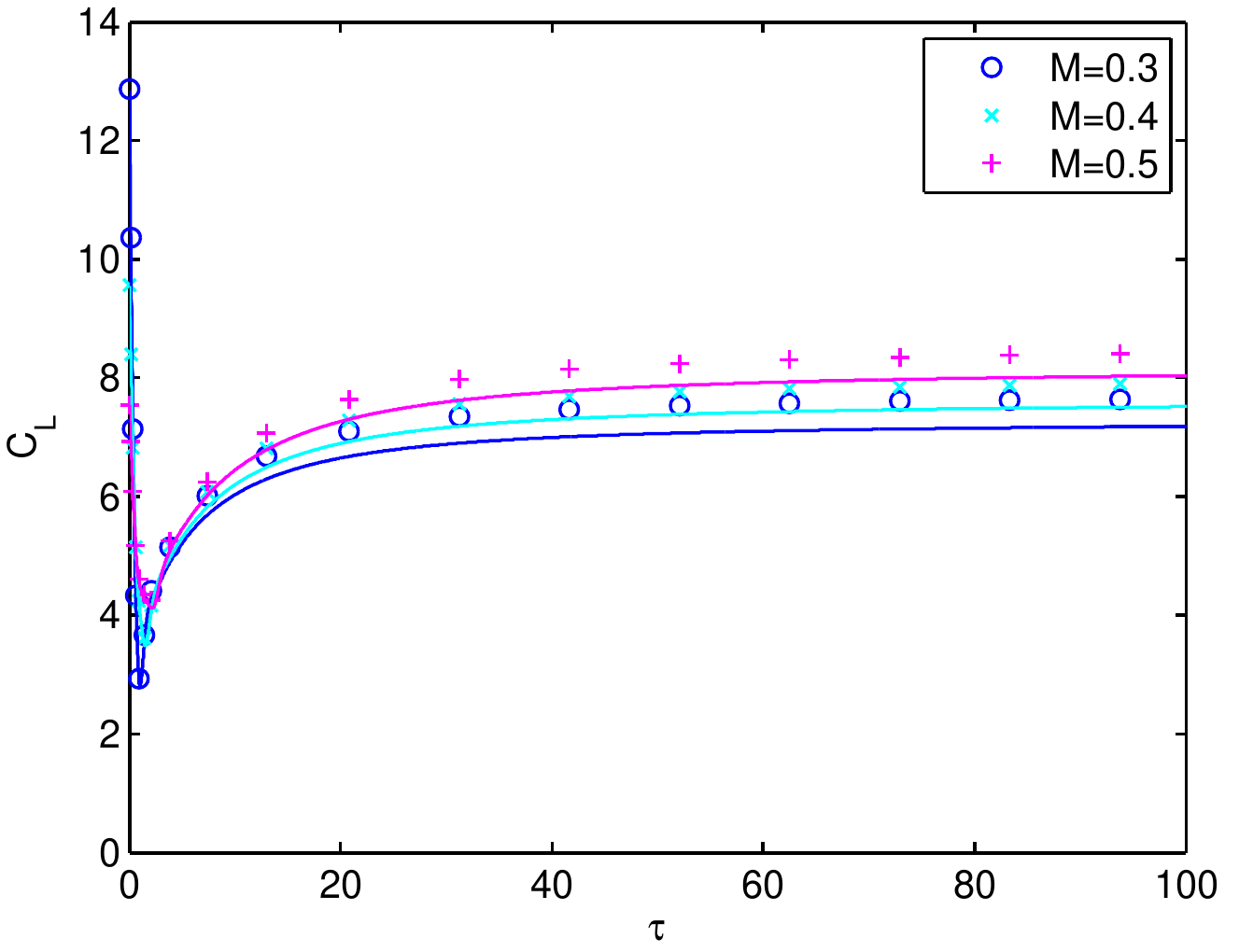}
\includegraphics[width=67mm]{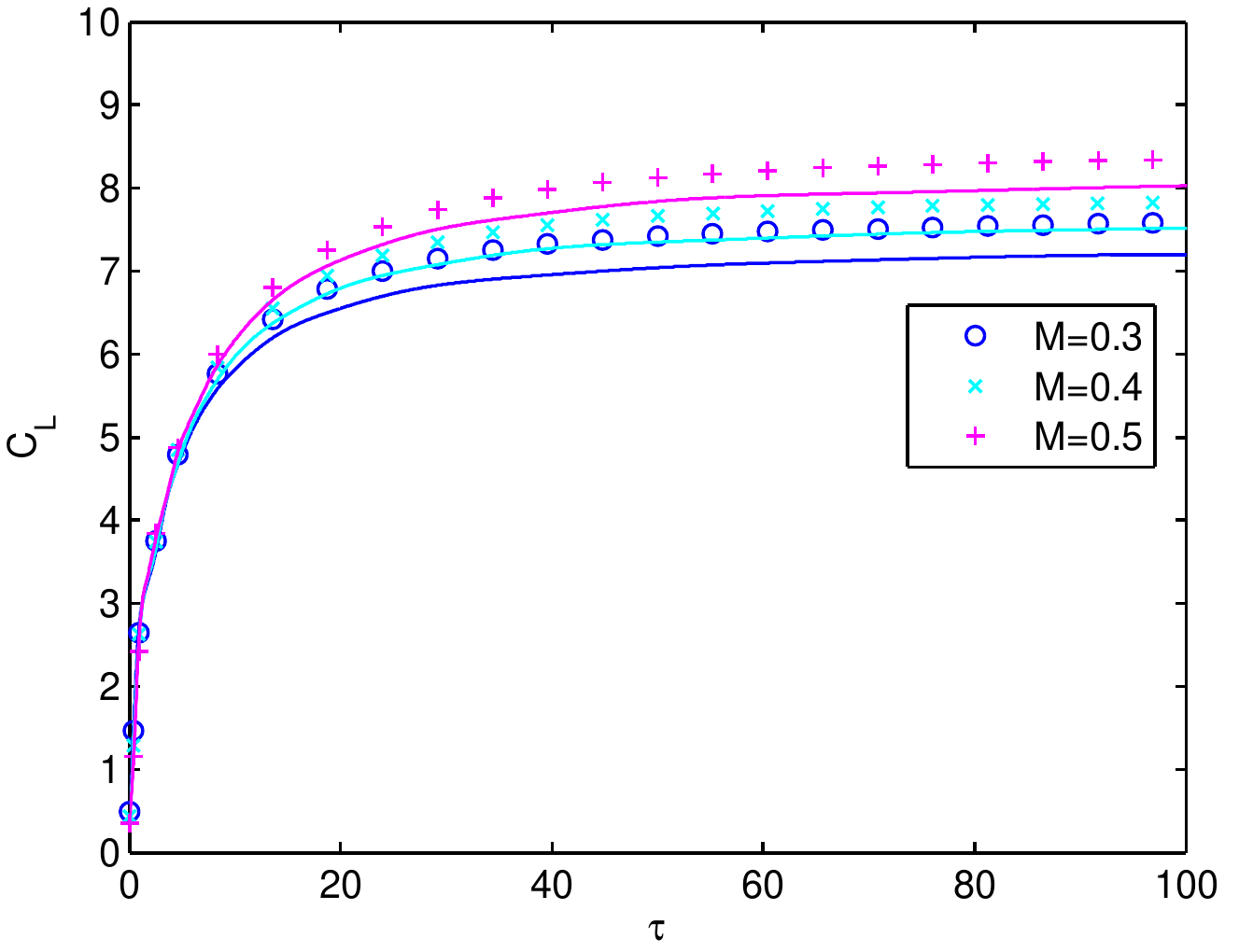}
\end{center}
\caption{NACA 0012 lift coefficient via Euler CFD: response to unit AoA (left) and unit SEG (right); lines = solution at $\alpha=0^{\circ}$, symbols = solution at $\alpha=2.1^{\circ}$.}
\label{fig:EURA0012}
\end{figure}

Figures \ref{fig:EURA0006} and \ref{fig:EURA0012} then compare the lift development for the NACA 0006 and 0012 aerofoils at both $\alpha=0^{\circ}$ and $\alpha=2.1^{\circ}$, as calculated by Euler CFD analysis for all subsonic Mach number considered: the non-circulatory build-up being almost identical (especially in the initial impulsive region), the non-circulatory build-up exhibits the same trend but for the (steady) asymptotic value, which differs due to nonlinear AoA-induced effects \cite{Ghoreyshi2} especially at the higher Mach numbers, as the flow lines have higher local curvature and hence acceleration at the aerofoil's nose for $\alpha=2.1^{\circ}$.

\begin{figure}[h!]
\begin{center}
\includegraphics[width=67mm]{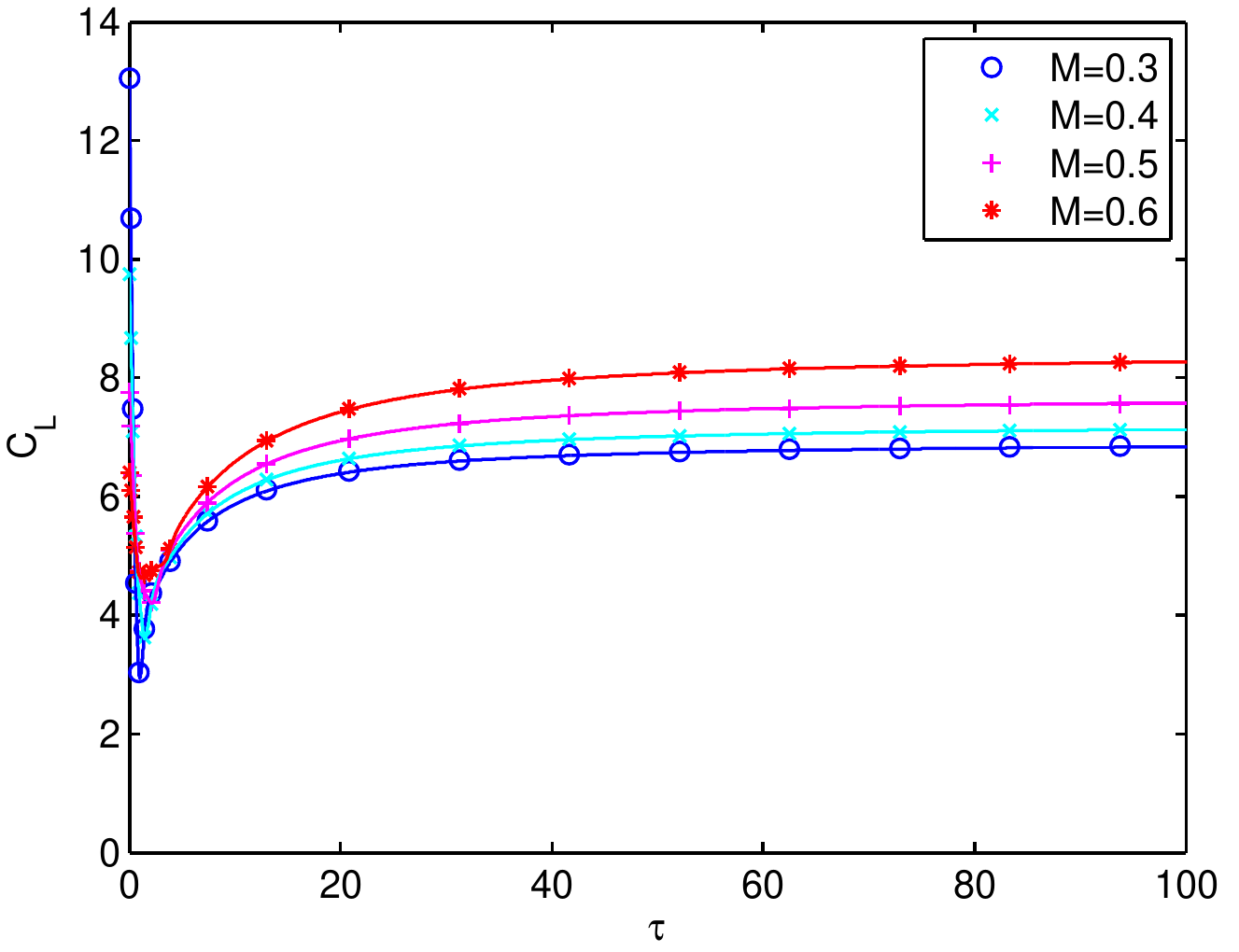}
\includegraphics[width=67mm]{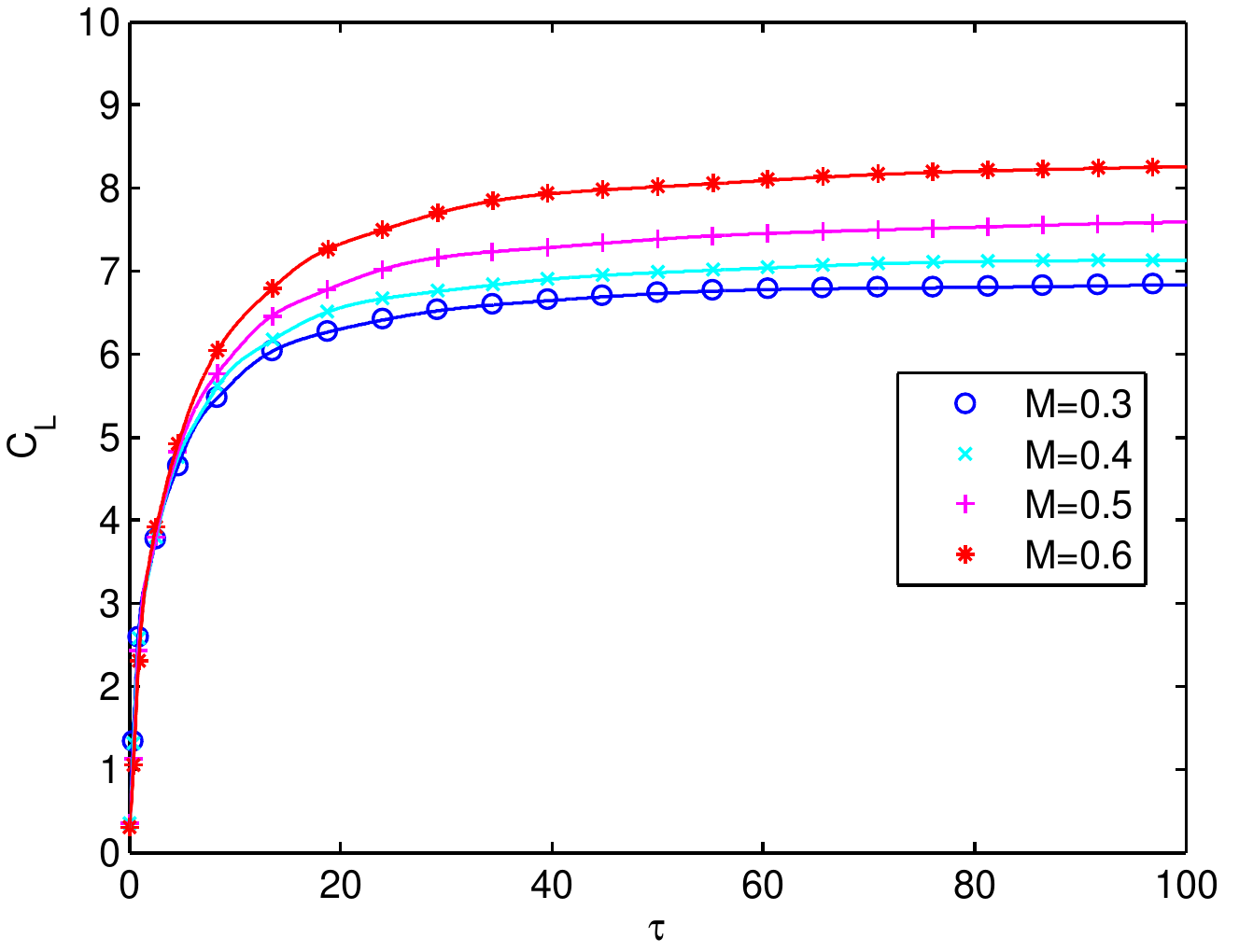}
\end{center}
\caption{NACA 2406 lift coefficient via Euler CFD: response to unit AoA (left) and unit SEG (right); lines = NACA 0006 at $\alpha=0^{\circ}$, symbols = NACA 2406 at $\alpha=-2.1^{\circ}$.}
\label{fig:EURA2406}
\end{figure}

\begin{figure}[h!]
\begin{center}
\includegraphics[width=67mm]{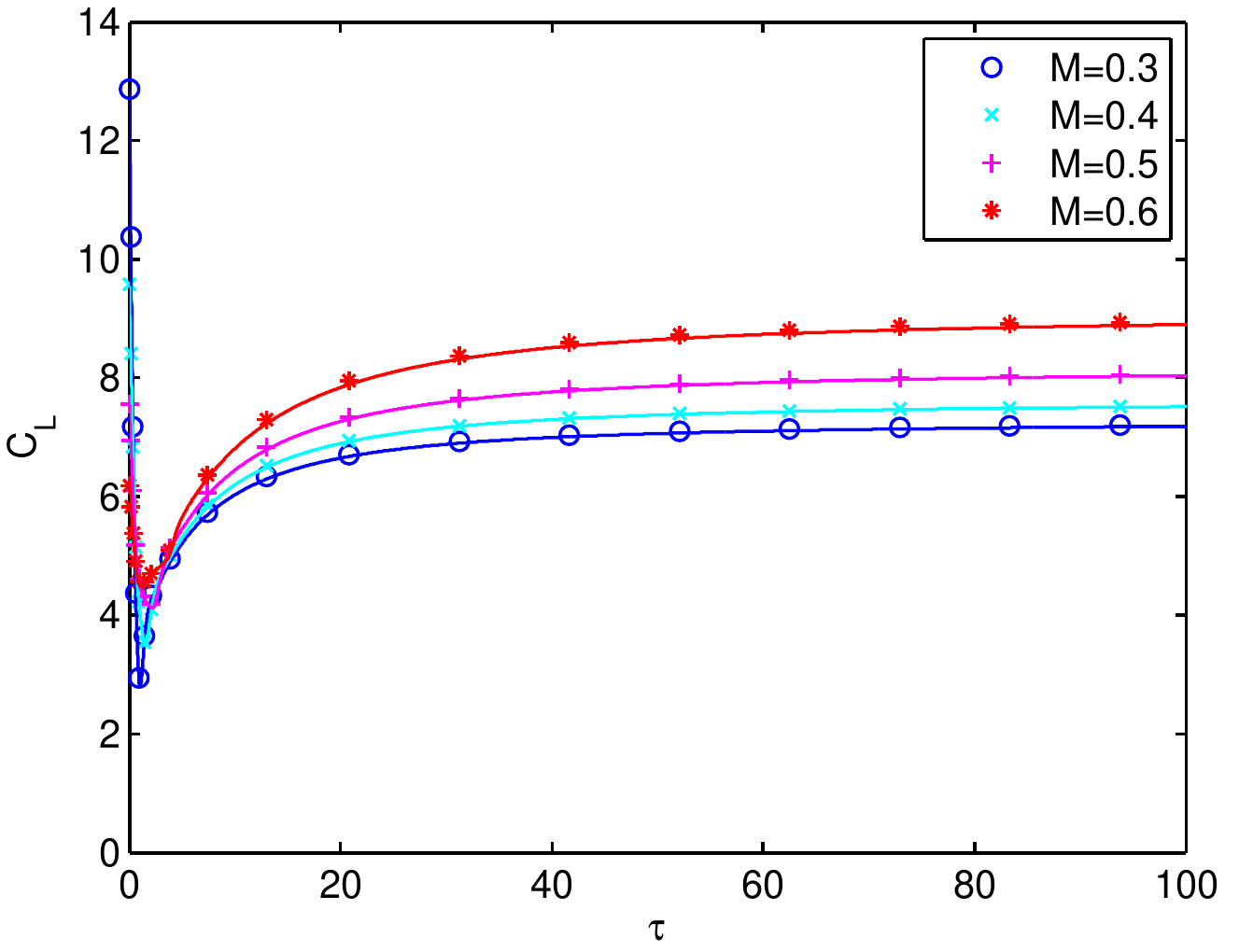}
\includegraphics[width=67mm]{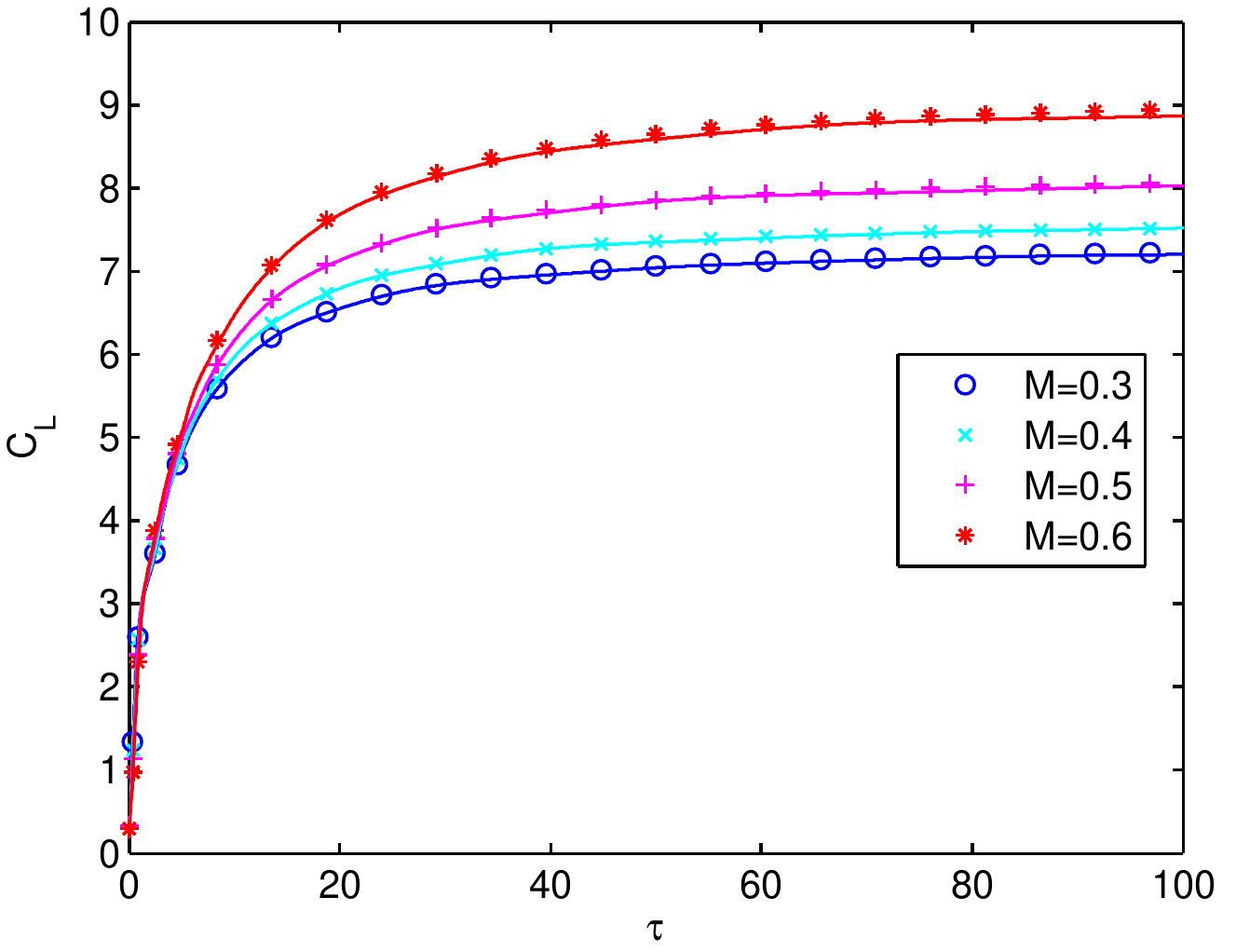}
\end{center}
\caption{NACA 2412 lift coefficient via Euler CFD: response to unit AoA (left) and unit SEG (right); lines = NACA 0012 at $\alpha=0^{\circ}$, symbols = NACA 2412 at $\alpha=-2.1^{\circ}$.}
\label{fig:EURA2412}
\end{figure}

Figures \ref{fig:EURA2406} and \ref{fig:EURA2412} compare the lift development for the NACA 0006 and 0012 aerofoils at $\alpha=0^{\circ}$ with that for the NACA 2406 and 2412 aerofoils at $\alpha=-2.1^{\circ}$, as calculated by Euler CFD analysis for all subsonic Mach number considered: the reference angle of attack being the zero-lift one $\alpha_0$ in all cases, the entire lift build-up is essentially identical in every case.

\begin{figure}[h!]
\begin{center}
\includegraphics[width=67mm]{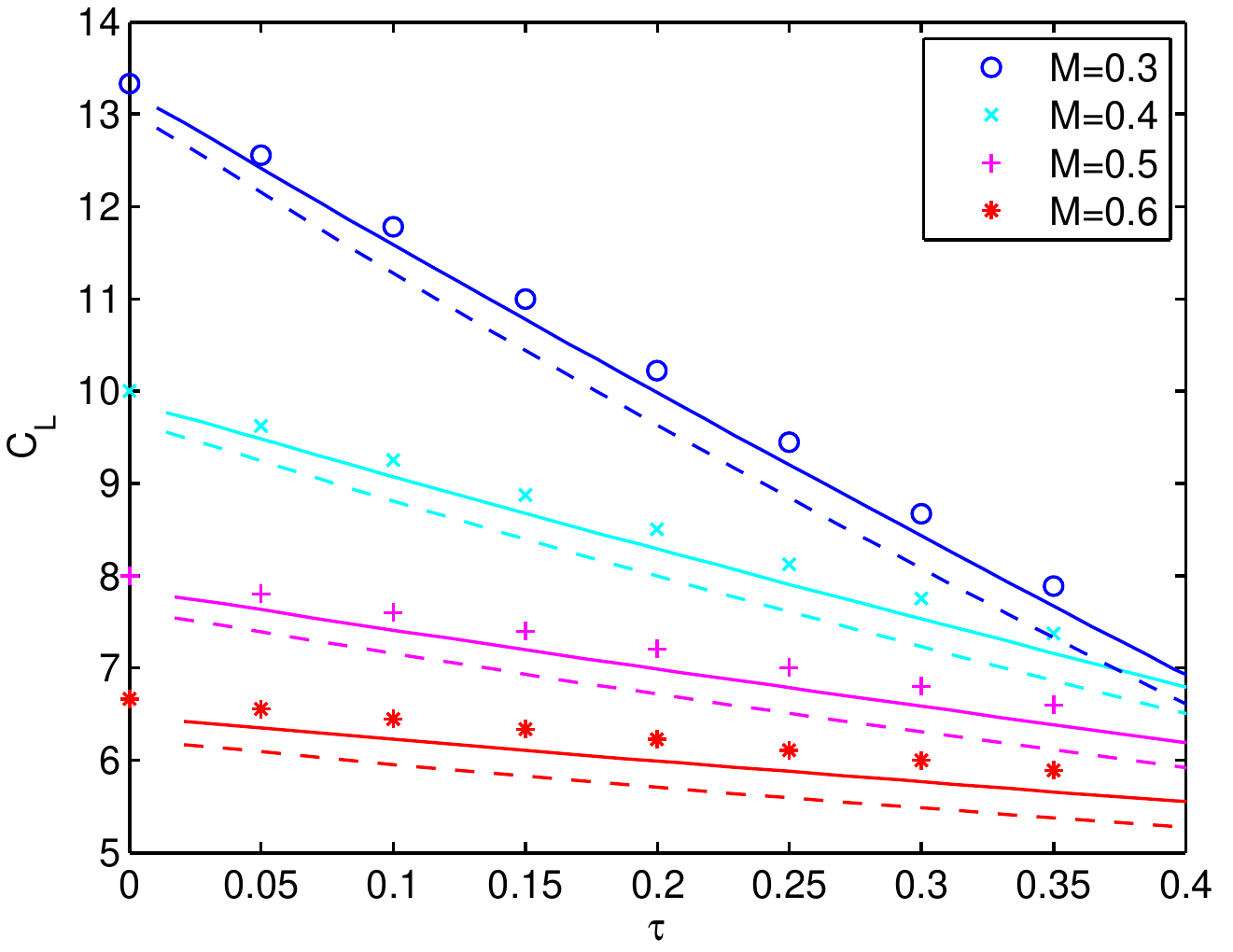}
\includegraphics[width=67mm]{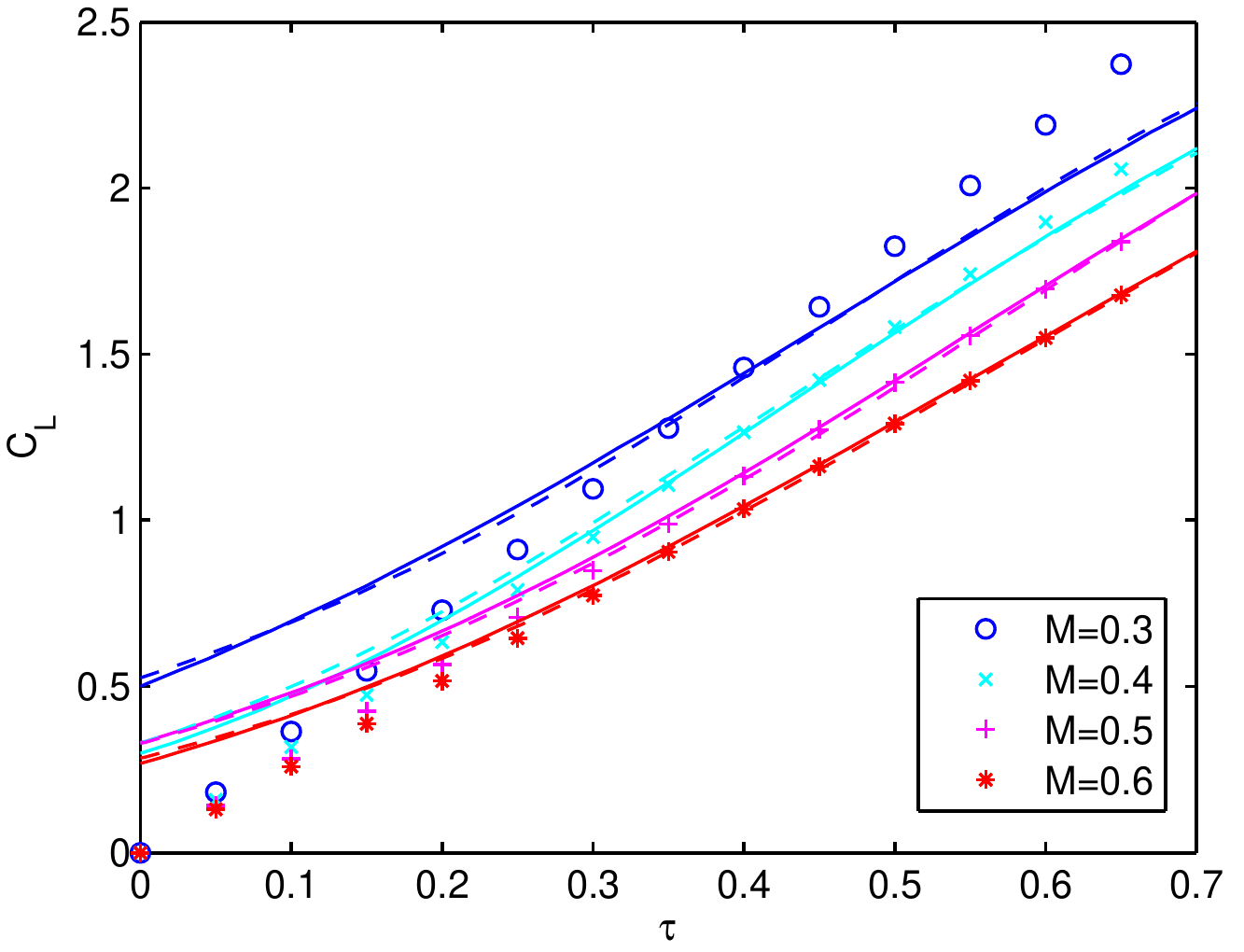}
\end{center}
\caption{NACA 0006 and 0012 lift coefficient for $\alpha=0^{\circ}$: impulsive response to unit AoA (left) and unit SEG (right); continuous lines = Euler CFD for NACA 0006, dashed lines = Euler CFD for NACA 0012, symbols = piston theory.}
\label{fig:EURANR}
\end{figure}

Figure \ref{fig:EURANR} focuses on the initial lift development for the NACA 0006 and 0012 aerofoils at $\alpha=0^{\circ}$ in the impulsive region, as calculated by Euler CFD analysis for all subsonic Mach number considered: for the AoA case, excellent agreement with piston theory \cite{Lighthill,Ashley} is always found. For the SEG case, the flow being subsonic and the travelling vertical gust being introduced slightly ahead of the aerofoil, the influence of the former is felt by the latter in advance \cite{Basu,Berci3}; therefore, the CFD results do not start at the theoretical value $C_L^G=0$ when the gust hits the aerofoil's leading edge and the good agreement with piston theory becomes evident only at a slightly later time, especially at the higher Mach numbers (for which the validity region of piston theory is larger \cite{Pike}).

\begin{figure}[h!]
\begin{center}
\includegraphics[width=67mm]{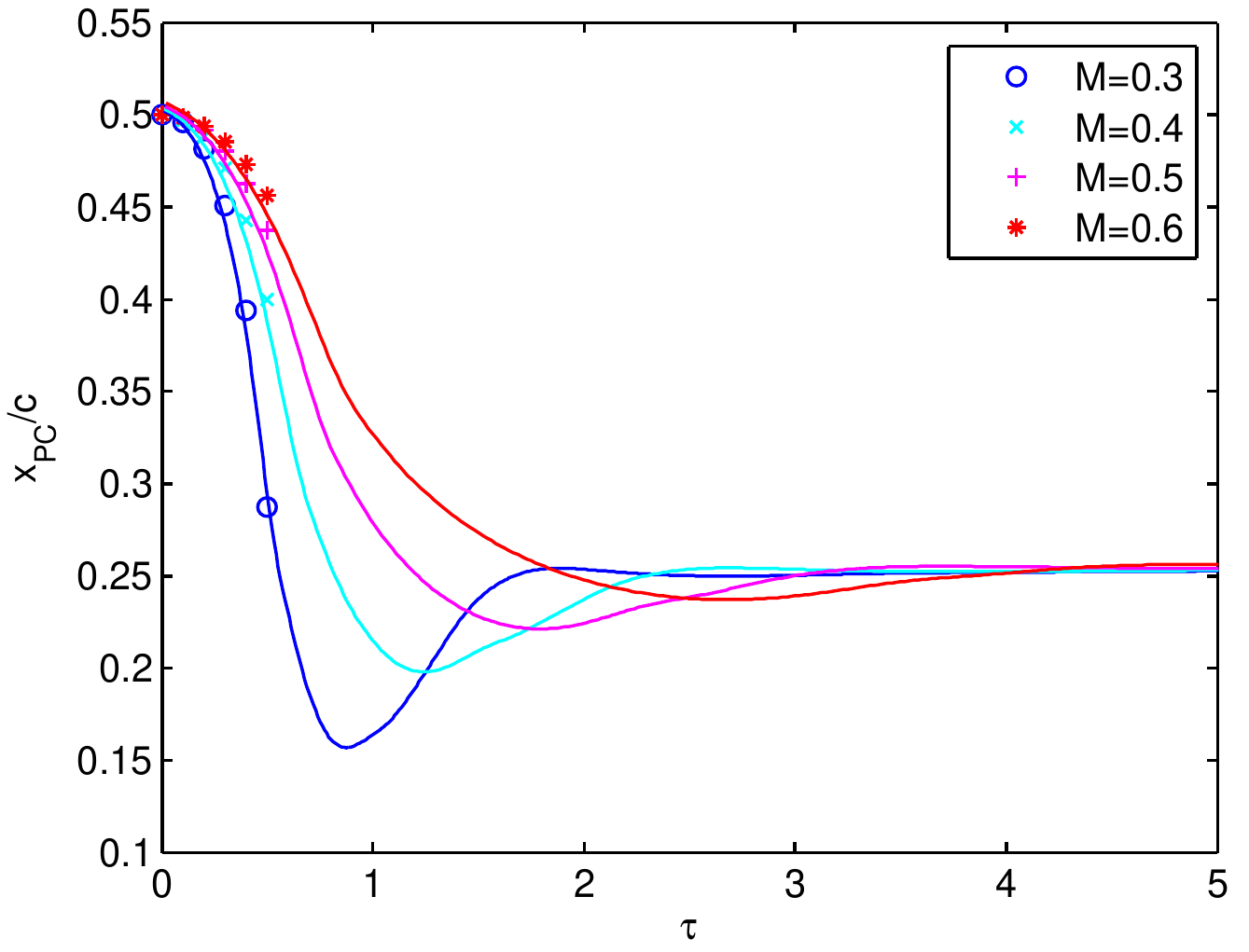}
\includegraphics[width=67mm]{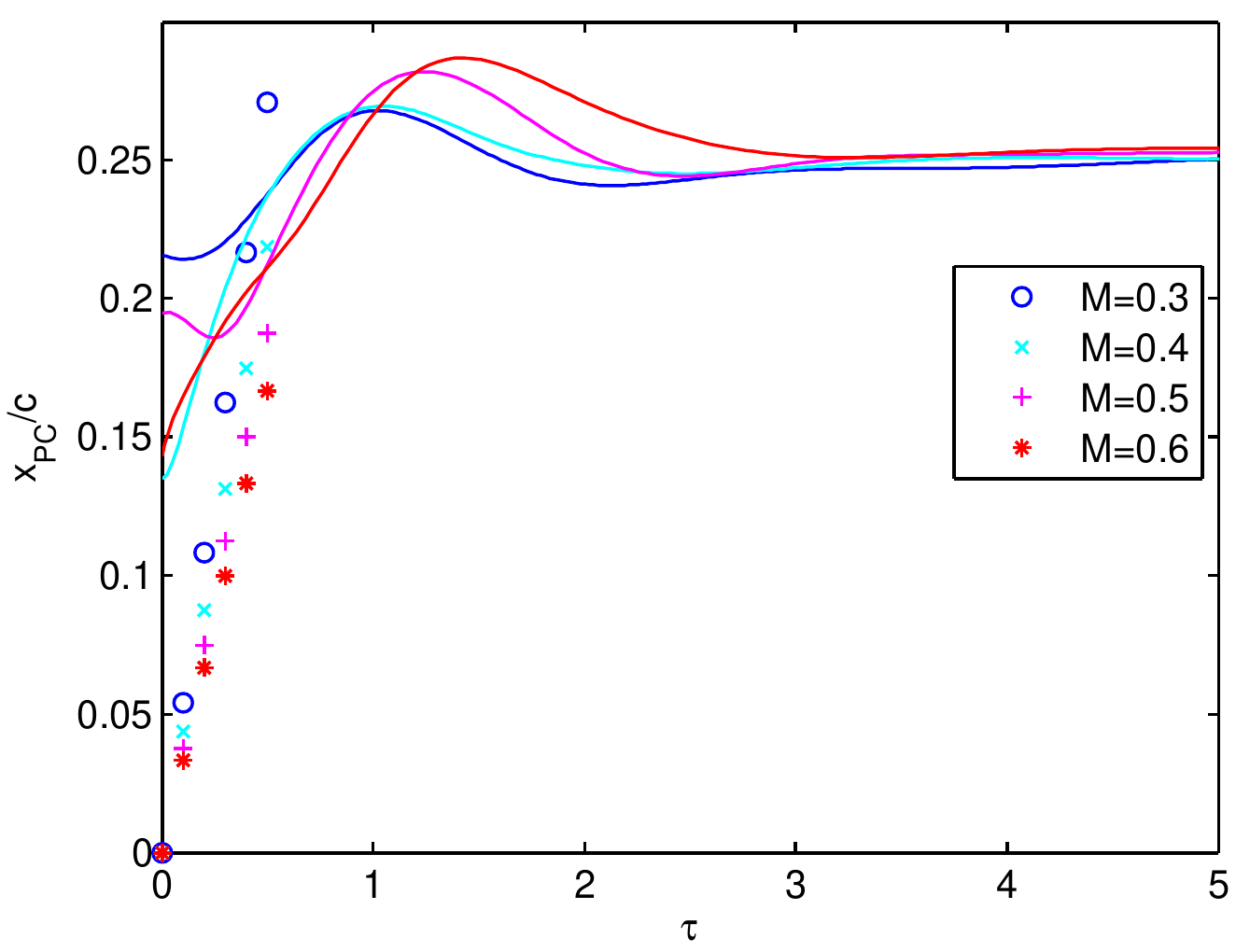}
\end{center}
\caption{Center of pressure's position for the NACA 0006 aerofoil at $\alpha=0^{\circ}$: response to unit AoA (left) and unit SEG (right); lines = Euler CFD, symbols = piston theory.}
\label{fig:CP_CFD}
\end{figure}

Figure \ref{fig:CP_CFD} compares the position of the center of pressure during the AoA and SEG responses for the symmetric NACA 0006 aerofoil at its reference angle of attack $\alpha=0^{\circ}$, as obtained numerically via Euler CFD by dividing the instantaneous pitching moment at the aerofoil's leading edge by the instantaneous lift; Lomax's solution \cite{Lomax2} is also shown as the exact reference. Perfect agreement is confirmed in the impulsive region, then a consistent variation of the center of pressure's position is still found in the transitory region but soon converges to a slightly different asymptotic (steady) value in the circulatory region. In the case of the flow response to unit AoA step, the center of pressure correctly starts at the aerofoil's mid-chord according to piston theory \cite{Lighthill,Ashley} and then eventually approaches the aerofoil's quarter-chord for all considered Mach numbers (since suitably far from transonic conditions) according to thin aerofoil theory \cite{Prandtl,Glauert2}; this last point is also true in the case of the flow response to unit SEG, where the center of pressure initially oscillates around the asymptotic (steady) position and exact agreement with piston theory is not possible as the aerofoil's presence is felt by the gust before its actual arrival (i.e., the present CFD simulations are more realistic and do not follow the classic "`frozen"' approach \cite{Chiang,Hoblit}, where the load builds up from scratch).

\begin{figure}[h!]
\begin{center}
\includegraphics[width=67mm]{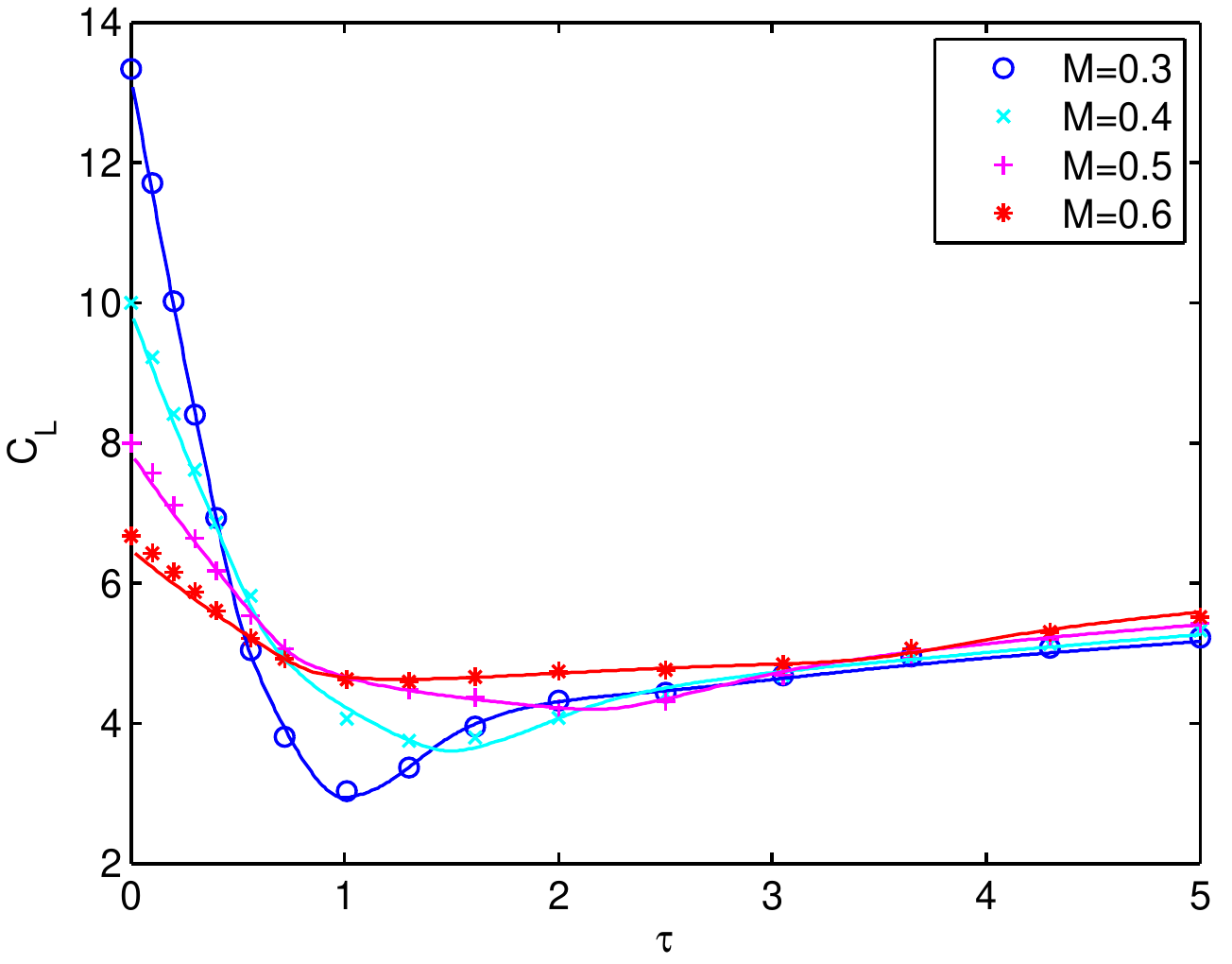}
\includegraphics[width=67mm]{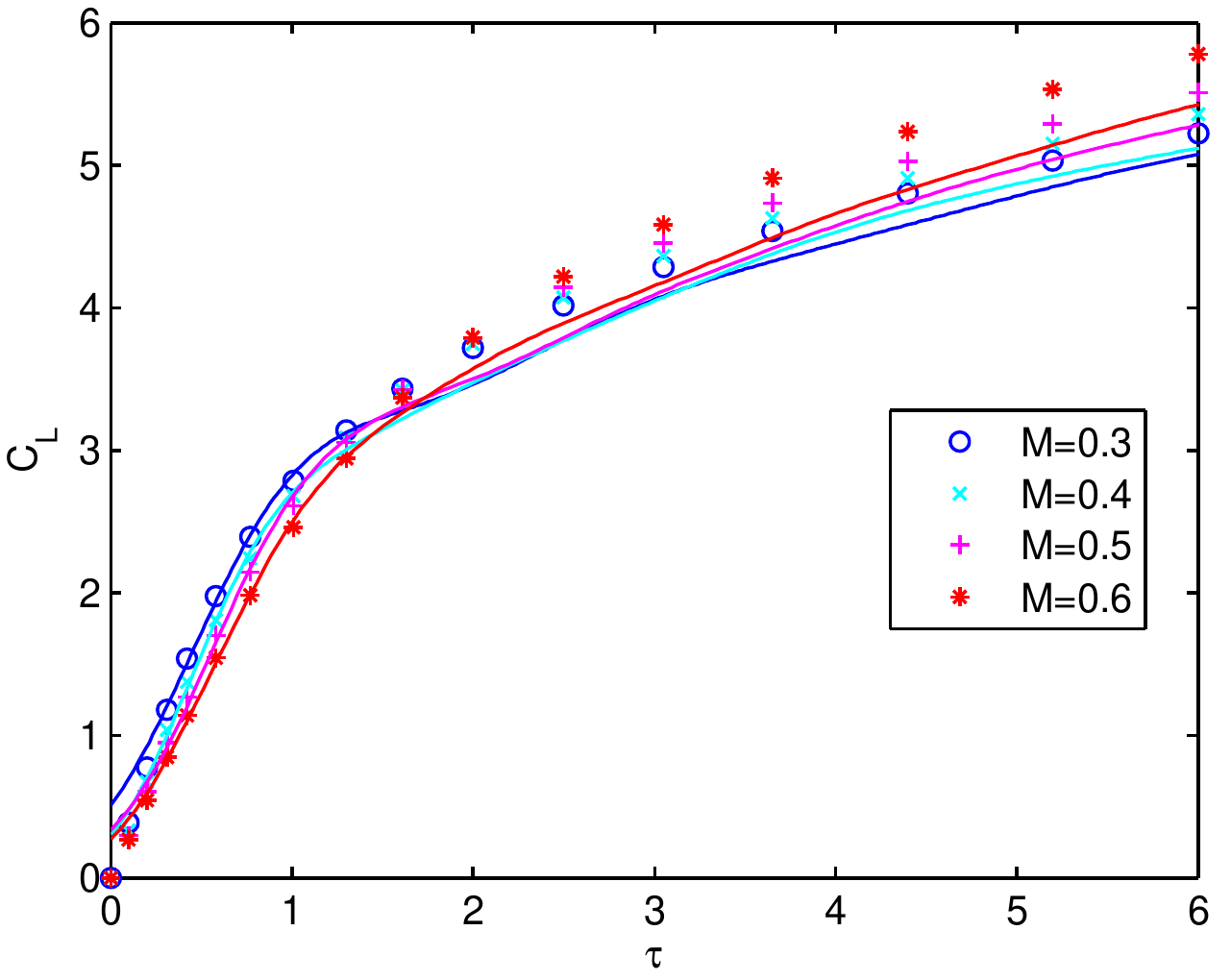}
\end{center}
\caption{NACA 0006 lift coefficient in the transitory region for $\alpha=0^{\circ}$: tuned response to unit AoA (left) and unit SEG (right); lines = Euler CFD, symbols = analytical approximation.}
\label{fig:tuned0006NR}
\end{figure}

Figure \ref{fig:tuned0006NR} compares the lift development for the NACA 0006 aerofoil at $\alpha=0^{\circ}$ in the entire non-circulatory region, where the analytical results have been tuned so to mimic the Euler CFD results for each subsonic Mach number considered: for the AoA case, excellent agreement is always found (this is also true with respect to Lomax's exact solution \cite{Lomax2}, which provides further validation). For the SEG case, excellent agreement is still found in the initial impulsive region (but for the CFD results not starting at the theoretical value $C_L^G=0$ of the ``frozen'' approach \cite{Chiang,Hoblit}), whereas the optimal approximation of the transitory region may further be improved by employing more damped oscillatory terms, although this is not deemed strictly necessary as the present accuracy can safely be considered good enough already for practical applications \cite{Leishman1,Beddoes}.

To summarise, it appears that the acoustic waves generation (in the impulsive region) and their subsequent interactions (in the transitory region) are well reproduced by the CFD results in the entire non-circulatory part of the flow response, where the analytical solution behaves correctly only if either the complex exact solution derived by Lomax \cite{Lomax2} or the efficient tuned approximation proposed in this work are employed, for both unit AoA and unit SEG cases. Theoretical analytical approximations assuming a thin aerofoil and CFD results then exhibit the same trend for the circulatory lift build-up but for the (steady) asymptotic value, which differs due to aerofoil-specific thickness effects \cite{Motta} especially at the higher Mach numbers; however, the very same circulatory lift build-up is readily obtained as soon as the analytical approximation is suitably tuned so to reach the (steady) asymptotic value predicted by the CFD simulations.

Still, numerical results based on RANS CFD have also been computed for further cross-validation \cite{Righi} and perfect agreement was always found in the impulsive region, then a consistent trend of the load build-up is still observed in both transitory and circulatory regions but converges to a different asymptotic (steady) value, which scales the lift growth rate; yet, differences in the latter are indeed small, especially in the case of the thinner aerofoils at the lower angles of attack.

\section{Conclusions}
\label{sec:conclusions}

In this study, aerodynamic indicial-admittance functions for the unsteady lift of wing aerofoils in subsonic flow have been generated via CFD and theoretical formulations, providing with sound cross-validation. Both a step-change in the angle of attack and a sharp-edge gust were suitably considered as small perturbations and good practice for calculating such functions was then demonstrated for a comprehensive range of Mach numbers, angles of attack and aerofoils profiles.

Many CFD solvers in the aerospace community can accurately solve time-accurate problems involving grid motion too and CFD may be regarded a reliable and accurate tool to generate aerodynamic indicial functions for the unsteady loads of any generic aircraft configurations; nevertheless, many factors may prevent the users from obtaining physically correct results free from spurious effects due to grid size and resolution, numerical integration scheme, turbulence model or Reynolds number and thorough validation is always needed.

A powerful combination of CFD and theoretical solutions has hence been exploited in order to devise a systematic and robust process for calculating the unsteady lift of wing aerofoils in subsonic flow. This healthy combination provides with both the complex generality of the numerical approach and the solid synthesis of the analytical approach (effectively supporting the conventional exercises on grid and methods convergence) and thus allows a thorough understanding and interpretation of the underlying unsteady phenomena, for both non-circulatory and circulatory contributions in the absence of shock waves.

In particular, an explicit parametric formula was newly proposed for the circulatory lift build-up as function of the Mach number and showed excellent agreement with CFD results in all cases. Damped harmonic terms were also effectively introduced to better approximate the non-circulatory lift, accurately reproducing both the initial impulsive phase (with a single term) and the subsequent transitory phase (with an additional term, as the fluid mechanics is thereby dominated by a complex interaction of acoustic waves) while beneficially preventing the nonlinear curve-fitting problem from resulting ill-conditioned, as less approximation terms become required.

Considering the asymptotic (steady) lift coefficient as free parameter, appropriate tuning of the analytical expressions has also been derived in order to mimic the CFD results and make proper comparisons between the two approaches, where excellent agreement has been found. All results have then critically been addressed with respect to the physical and mathematical assumptions underlying their derivation, within a consistent framework which verified the rigor of superposing circulatory and noncirculatory contributions in the light of the CFD results. 

Exhibiting the correct limit behaviours for a generic aerofoil, this study also confirms that linear aerodynamics based on potential flow is remarkably reliable in predicting unsteady airloads when simply tuned to recover nonlinear (steady) viscous effects and thus suitably convenient for practical applications within the preliminary MDO of flexible subsonic wings.

\section*{Appendix: Analytical Solution for Aerofoil in Unsteady Potential Flow}

Exact analytical solutions exist for aerofoil in potential incompressible flow, for both the steady \cite{Theodorsen2,Theodorsen3} and the unsteady \cite{Wagner,Theodorsen1,Kussner,Garrick,VonKarman,Sears} cases; the corresponding approximate solutions for subsonic compressible flow may then suitably be obtained via well-established tuning/scaling techniques \cite{Glauert2,Jones3} based on rigorous fluid dynamics similitude \cite{Anderson}.

\begin{figure}[h!]
\begin{center}
\includegraphics[width=67mm]{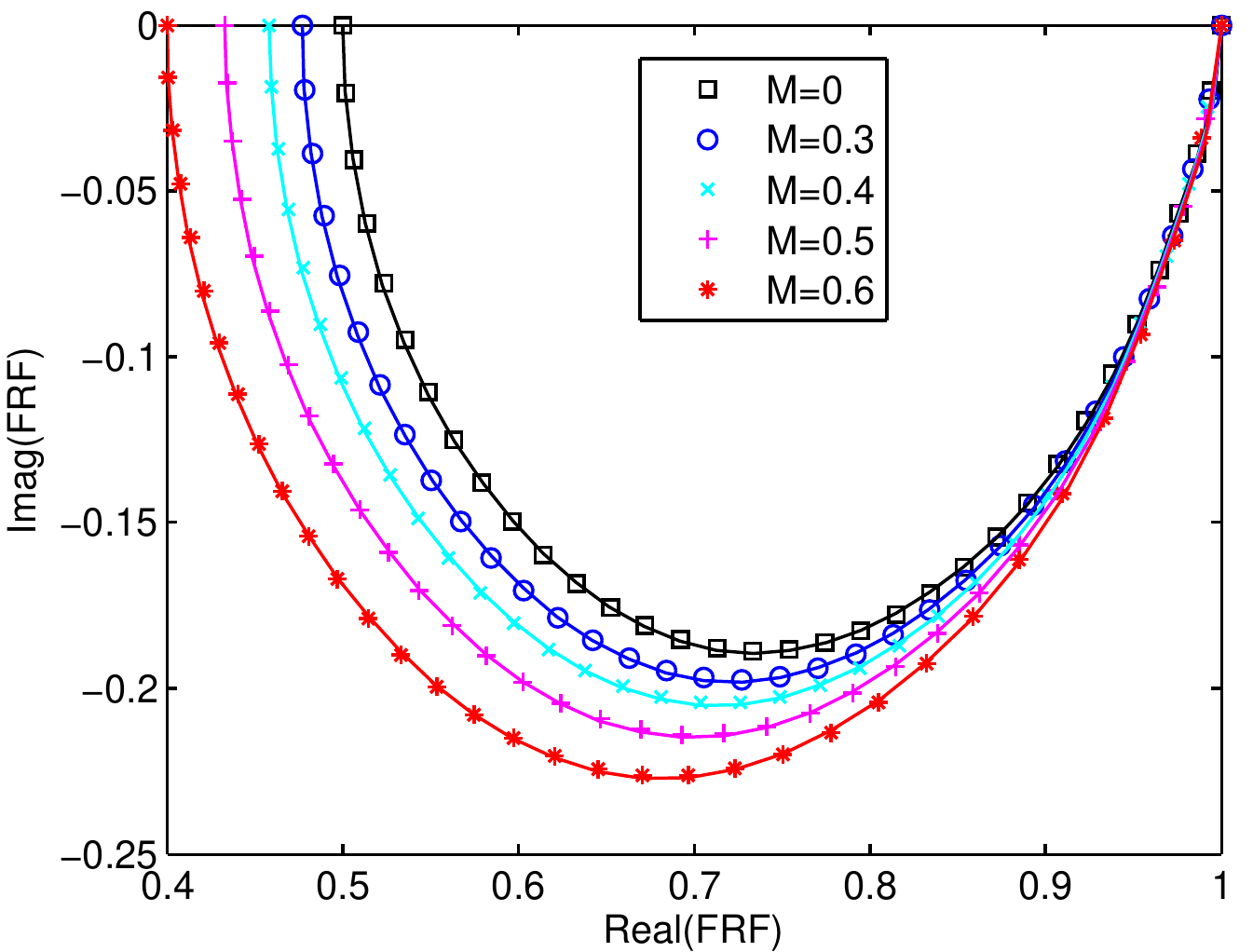}
\includegraphics[width=67mm]{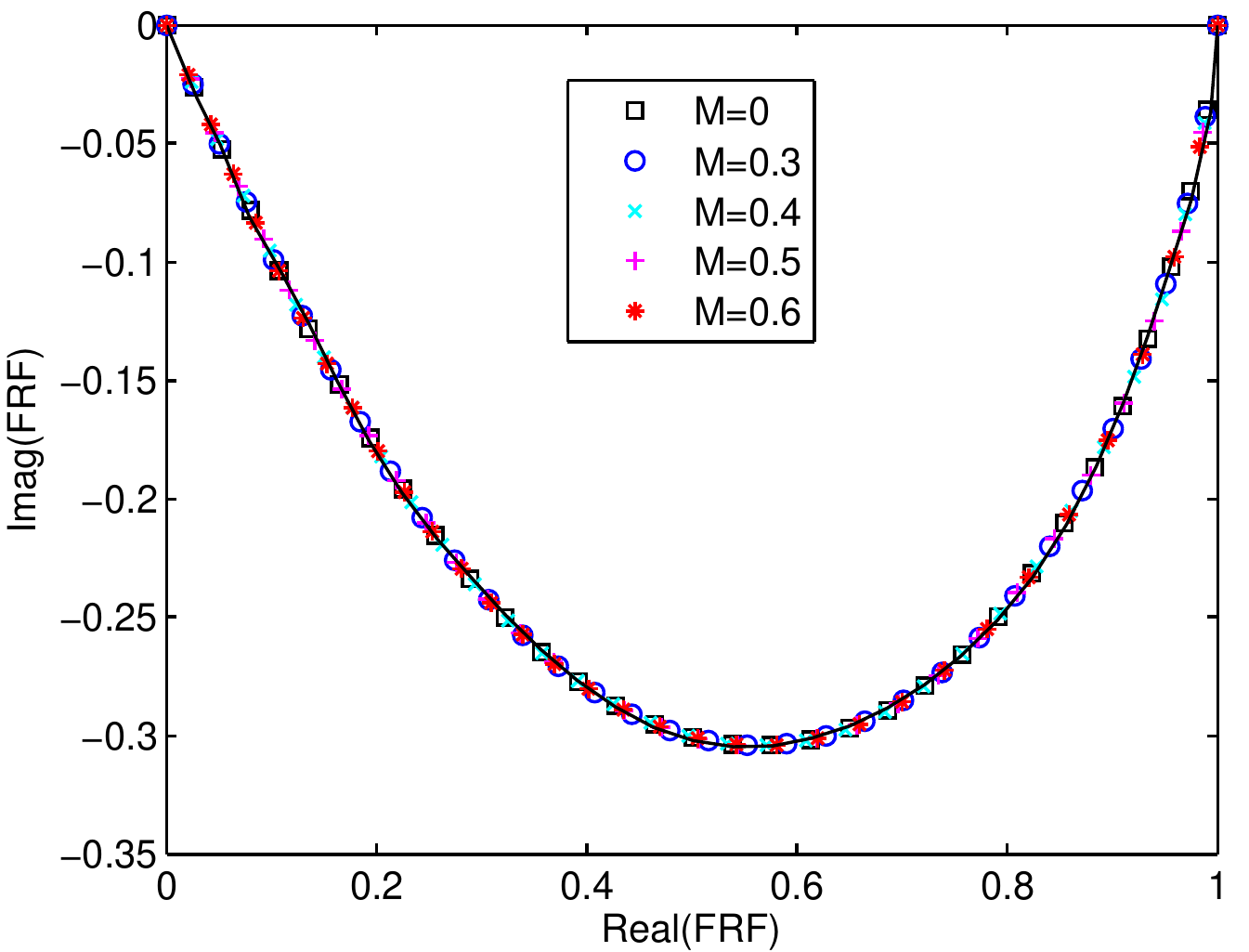}
\end{center}
\caption{Frequency response functions for unit AoA (left) and SEG (right); lines = approximate, symbols = exact.}
\label{fig:FRFs}
\end{figure}

\begin{figure}[h!]
\begin{center}
\includegraphics[width=67mm]{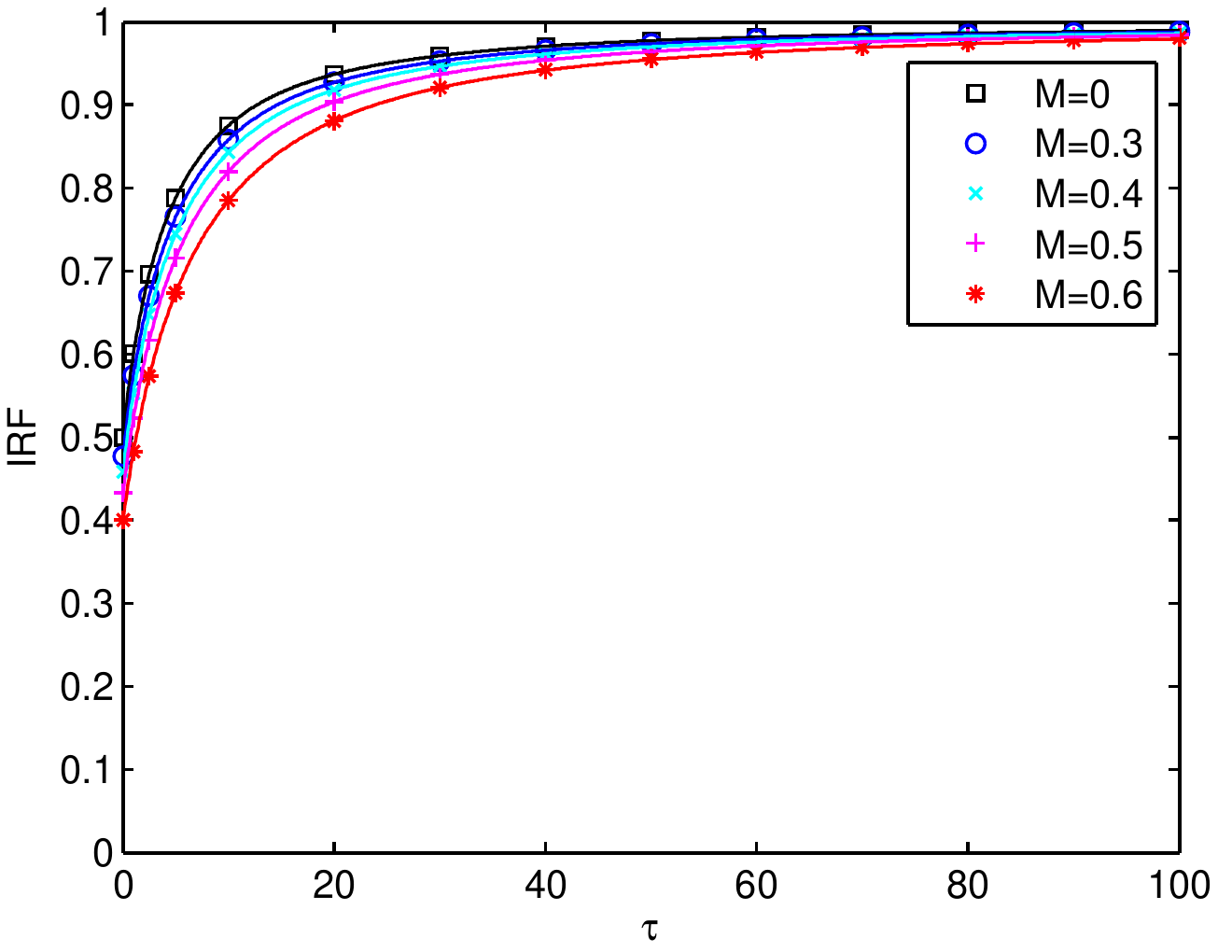}
\includegraphics[width=67mm]{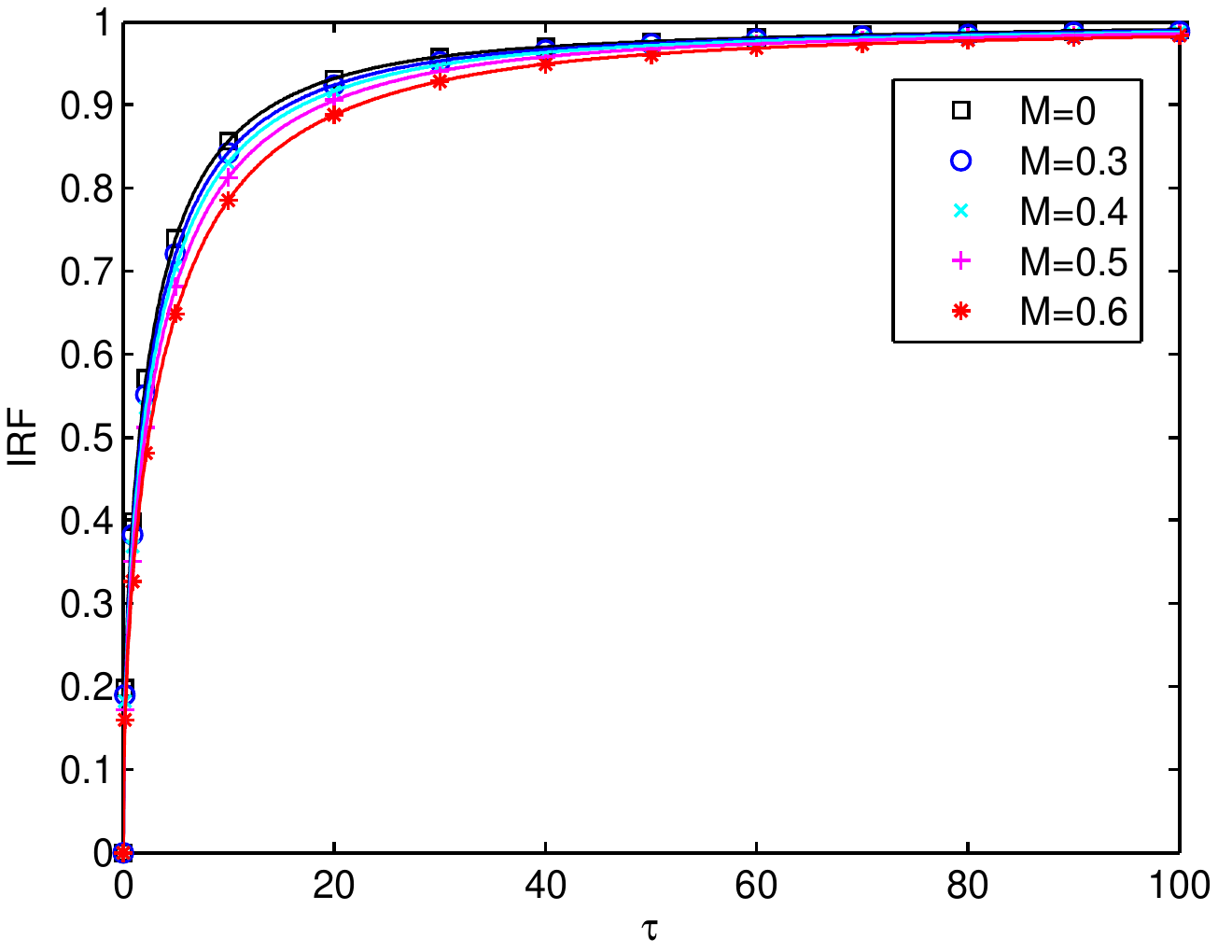}
\end{center}
\caption{Indicial response functions for unit AoA (left) and SEG (right); lines = approximate, symbols = exact.}
\label{fig:IFs}
\end{figure}

For flat aerofoil in incompressible potential flow, the exact lift-deficiency coefficients due to a unit step in the angle of attack and a unit sharp-edged gust are calculated by Wagner \cite{Wagner} and Kussner \cite{Kussner} in the reduced-time domain as:
\begin{equation}
\breve{C}_L^W = \int^{\infty}_{-\infty} \breve{C}_T \left(\frac{e^{\textrm{i}k\tau}}{\textrm{i}k}\right)dk,   \qquad    
\breve{C}_L^K = \int^{\infty}_{-\infty} \breve{C}_S \left(\frac{e^{\textrm{i}k\tau}}{\textrm{i}k}\right)dk,
\end{equation}
respectively, and exhibit limit behaviours given by \cite{Garrick,Sears}:
\begin{eqnarray}
\lim_{\tau \rightarrow 0} \breve{C}_L^W = \pi \left(1+\frac{\tau}{4}\right),   \qquad   \lim_{\tau \rightarrow 0} \breve{C}_L^K = 2\sqrt{2\tau} \left(1-\frac{\tau}{12}\right),   \\
\lim_{\tau \rightarrow \infty} \breve{C}_L^W = 2\pi \left(1-\frac{1}{\tau}\right),   \qquad   \lim_{\tau \rightarrow \infty} \breve{C}_L^K = 2\pi \left(1-\frac{1}{\tau-1}\right),
\end{eqnarray}
whereas the Theodorsen's \cite{Theodorsen1} and Sears' \cite{VonKarman,Sears} functions in the reduced-frequency domain read:
\begin{equation}
\breve{C}_T = \frac{H^2_1}{H^2_1+\textrm{i}H^2_0},   \qquad    \breve{C}_S = \left[\breve{C}_T\left(B^1_0-\textrm{i}B^1_1 \right)+\textrm{i}B^1_1\right]e^{-\textrm{i}k}.
\end{equation}
Using the Prandtl-Glauert factor \cite{Glauert2,Jones3} to scale both the reduced frequency $k=\frac{c\omega}{2*U}$ as $k^*=\frac{k}{\beta^2}$ and the (unsteady) asymptotic limit as $\frac{\beta}{2}$, according to the present method both Theodorsen's and Sears' functions (see Figure \ref{fig:FRFs}) may be generalised for compressible subsonic flow as:
\begin{equation}
\breve{C}^*_T = \beta-1+\frac{\left(2-\beta\right) H^{2*}_1}{H^{2*}_1+\textrm{i}H^{2*}_0},   \qquad    
\breve{C}^*_S = \left[\left(\frac{1-\beta+\breve{C}^*_T}{2-\beta}\right) \left(B^{1*}_0-\textrm{i}B^{1*}_1\right)+\textrm{i}B^{1*}_1\right] e^{-\textrm{i}k^*},
\end{equation}
with $H^{2*}_n=H^2_n(k^*)$ and $B^{1*}_n=B^1_n(k^*)$, whereas Wagner's and Kussner's functions (see Figure \ref{fig:IFs}) may hence be generalised for compressible subsonic flow in the scaled reduced time $\tau^*=\beta^2 \tau$ as:
\begin{equation}
\breve{C}_L^\alpha = \int^{\infty}_{-\infty} \breve{C}^*_T \left(\frac{e^{\textrm{i}k^*\tau^*}}{\textrm{i}k^*}\right)dk^*,   \qquad    
\breve{C}_L^G = \int^{\infty}_{-\infty} \breve{C}^*_S \left(\frac{e^{\textrm{i}k^*\tau^*}}{\textrm{i}k^*}\right)dk^*,
\end{equation}
the incompressible functions being consistently found for $M=0$ and $\beta=1$. Note that all these indicial functions have circulatory nature and quantify the decaying effect of the wake's downwash on the aerofoil's lift build-up \cite{Jones2}.

For both a unit step in the angle of attack and a unit sharp-edged gust, the theoretical incompressible non-circulatory contribution is a Dirac $\delta$ function centred at the start of the perturbation \cite{Pike}; however, this impulsive contribution becomes finite in compressible flow, where the exact initial behaviour of the aerofoil's total lift and leading-edge pitching moment coefficients due to a unit step in the angle of attack and a unit sharp-edged gust are calculated via piston theory \cite{Lighthill,Ashley} as:
\begin{equation}
\lim_{\tau \rightarrow 0}  C_L^\alpha = \frac{4}{M}\left[ 1 - \left(\frac{1-M}{2M}\right) \tau \right],   \qquad   \lim_{\tau \rightarrow 0} C_L^G = \frac{2\tau}{\sqrt{M}},   
\end{equation}

\begin{equation}
\lim_{\tau \rightarrow 0}  C_M^\alpha = -\frac{2}{M}\left[ 1 - \left(\frac{1-M}{2M}\right) \tau - \left(\frac{2-M}{8M}\right) \tau^2 \right],   \qquad   \lim_{\tau \rightarrow 0} C_M^G = -\frac{\left(1+M\right)\tau^2}{4M\sqrt{M}},
\end{equation}
respectively, which hold rigorously for $\tau \le \frac{2M}{1+M}$. In the former case, Lomax \cite{Lomax2} extended the analytical solution up to $\tau \le \frac{2M}{1-M}$ in order to include the complex transitory region where acoustic waves interact and the noncirculatory part eventually joins the circulatory part; nevertheless, the final expression is very complex and of little practical use. Note that $\acute{\tau}=\frac{2M}{1+M}$ and $\grave{\tau}=\frac{2M}{1-M}$ are the reduced times taken by the outgoing and incoming pressure waves to travel the aerofoil chord at their characteristic speed $a(1+M)$ and $a(1-M)$, respectively: at those reduced times, a significant change in the slope of the indicial function occurs as the pressure waves leave the aerofoil; finally, acoustic phenomena and gust penetration combine for $\tau \le 2$ (i.e., the reduced time taken by the gust to travel the aerofoil chord at the reference flow speed $U$).

\subsubsection*{Optimal Rational Approximation of Theodorsen's and Sears' Functions for incompressible flow}

The best analytical approximation of Theodorsen's and Sears' lift-deficiency functions for flat aerofoil in unsteady incompressible potential flow was sought in the (reduced) frequency $k$ domain, where the two functions are analytically defined as complex parametric curves of finite length and eventually expressed as a series of rational functions \cite{Roger}, namely:
\begin{eqnarray}
\breve{C}_T = 1 - \sum_{j=1}^{n_W} \frac{\breve{A}_j^W \textrm{i}k}{\breve{B}_j^W+\textrm{i}k},   \qquad   \sum_{j=1}^{n_W} \breve{A}_j^W = \frac{1}{2},   \\  
\breve{C}_S = 1 - \sum_{j=1}^{n_K} \frac{\breve{A}_j^K \textrm{i}k}{\breve{B}_j^K+\textrm{i}k},   \qquad   \sum_{j=1}^{n_K} \breve{A}_j^K = 1,
\end{eqnarray}
with both gains $\breve{A}_j$ and poles $\breve{B}_j$ acting as optimization variables; this was preferred to fitting their real and imaginary parts separately, since the latter have infinite extension and an adequate selection of their sample points would hence be more difficult \cite{Berci1}. Due to Laplace transform \cite{Dowell}, the corresponding optimal approximations of Wagner's and Kussner's lift-deficiency coefficients, respectively, may then consistently be expressed as a series of exponential functions in the (reduced) time $\tau$ domain \cite{Dunn}, namely: 
\begin{equation}
\frac{\breve{C}_L^W}{2\pi} = 1 - \sum_{j=1}^{n_W} \breve{A}_j^W e^{-\breve{B}_j^W \tau},   \qquad 
\frac{\breve{C}_L^K}{2\pi} = 1 - \sum_{j=1}^{n_K} \breve{A}_j^K e^{-\breve{B}_j^K \tau},
\end{equation}
with both amplitudes $\breve{A}_j$ and time constants $\breve{B}_j$ eventually specified in Table \ref{tab:2D}. Note that all these functions assume small (linear) flow perturbations and are rigorously applicable for thin aerofoils at low angle of attack; corrections are typically required otherwise \cite{Wieseman,Brink,Palacios,Dimitrov,Sucipto,Berci3}.

Constrained nonlinear optimization \cite{Holmes,Vanderplaats} was adopted to perform the curve-fitting effectively, where the Normalised Root Mean Squared Error (NRMSE) is the objective function to be minimized and the Normalised Maximum Absolute Error (NMAE) is monitored as quality measure \cite{Berci1,Tiffany}, namely: 
\begin{equation}
\textrm{NRMSE}= \frac{\left\| \tilde{F}(\textbf{s})-F(\textbf{s}) \right\|}{\sqrt{N}l},   \qquad   
\textrm{NMAE} = \frac{\textrm{max} \left| \tilde{F}(\textbf{s})-F(\textbf{s}) \right|}{l},
\end{equation}
where $\textbf{s}$ are the $N$ sample locations, $\tilde{F}(s)$ is the approximation function, whereas $F(s)$ and $l$ are the exact function and its length in the complex plane, respectively. 

In order to avoid focusing on certain regions a priori, a uniform distribution of $N=100$ sample points $F$ along both Theodorsen's and Sears' functions in the complex plane was first obtained by solving a nonlinear optimization problem where an equal distance between each couple of consecutive samples is sought \cite{Berci1}, with the (reduced) frequency acting as the free parameter $\textbf{s}$; Sequential Quadratic Programming \cite{Vanderplaats} (SQP) was used as the optimization algorithm. Figure \ref{fig:compaFRF} shows both Theodorsen's and Sears' functions along with their proposed approximations; the initial guess for the optimal parameters was based on Jones' results \cite{Jones1}. The same optimal results were efficiently obtained using either the unconstrained Nelder-Mead Simplex Method \cite{Nelder} (NMSM, where the absolute value of the poles was considered for a meaningful implementation of the optimization problem) or a Genetic Algorithm \cite{Affenzeller} (GA, where the initial population was taken as the SQP solution in order to boost convergence) as well, confirming consistency in the results and robustness of the proposed approach. No clustering of the optimal poles \cite{Tewari} was found and the global optimum was likely reached in all cases.

Note that more rational terms are necessary to approximate Sears' function with the same level of accuracy desired to approximate Theodorsen's function (e.g., both NRMSE and NMAE are below $1\%$ in the present case), due to the linear-like behaviour of the former in the complex plane at high reduced frequencies, where rational terms are not well suited; the same is then true for to Kussner's and Wagner's functions, respectively. Finally, it is worth stressing that any error in approximating Theodorsen's and Sears' functions translates into an error in approximating Wagner's and Kussner's functions, respectively; in particular, a good approximation of the formers at high reduced frequencies (i.e., highly-unsteady flow) translates into a good approximation of the latter at low reduced time (i.e., transient response), while a good accuracy at low reduced frequencies (i.e., quasi-steady flow) translates into a good accuracy at high reduced time (i.e., asymptotic response) \cite{Berci1}. Many approximations were readily available in the literature  \cite{Vepa1,Vepa2,Roger,Dunn,Dowell,Desmarais,Beddoes,Venkatesan,Peterson,Eversman,Tewari,Karpel,Poirel,Cotoi,Sedaghat,Brunton} but the present ones grant the best agreement along the entire exact curves with the least number of poles and possess the correct limit values.

Using the Prandtl-Glauert factor \cite{Glauert2,Jones3} to scale both the reduced frequency as $k^*=\frac{k}{\beta^2}$ and the reduced time as $\tau^*=\beta^2 \tau$, as well as the (steady) asymptotic limit as $\frac{2\pi}{\beta}$, the proposed approximations may then be generalised for compressible subsonic flow and are shown in Figures \ref{fig:FRFs} and \ref{fig:IFs}.

\section*{References}

\bibliography{berci_correct}

\begin{thebibliography}{100}
\newcommand{\enquote}[1]{``#1''}

\bibitem{Alexandrov}
Alexandrov, N. and Hussaini, M., {\em Multidisciplinary Design Optimization:
  State of the Art\/}, Proceedings in Applied Mathematics Series, SIAM, 1997.

\bibitem{Kesseler}
Kesseler, E. and Guenov, M., {\em Advances in Collaborative Civil Aeronautical
  Multidisciplinary Design Optimization\/}, Progress in Astronautics and
  Aeronautics Series, AIAA, 2010.

\bibitem{Kier1}
Kier, T., \enquote{Comparison of Unsteady Aerodynamic Modelling Methodologies
  with Respect to Flight Loads Analysis,} {\em AIAA-2005-6027\/}, 2005.

\bibitem{Kier2}
Kier, T. and Looye, G., \enquote{Unifying Manoeuvre and Gust Loads Analysis
  Models,} {\em IFASD-2009-106\/}, 2009.

\bibitem{Cavagna1}
Cavagna, L., Travaglini, L., and Ricci, S., \enquote{Gust Loads Assessment: a
  Multi-Fidelity Approach,} {\em IFASD-2011-000\/}, 2011.

\bibitem{Livne}
Livne, E., \enquote{The Future of Aircraft Aeroelasticity,} {\em Journal of
  Aircraft\/}, Vol.~40, No.~6, 2003, pp.~1066--1092.

\bibitem{Tobak}
Tobak, M., \enquote{On the Use of the Indicial-Function Concept in the Analysis
  of Unsteady Motions of Wings and Wing-Tail Combinations,} {\em NACA 1188\/},
  1954.

\bibitem{Pike}
Pike, E., \enquote{Manual on Aeroelasticity (Volume II, Chapter VI),} {\em
  AGARD 578\/}, 1971.

\bibitem{Graham}
Graham, G., \enquote{Aeroelastic Reciprocity: An Indicial Response
  Formulation,} {\em AFOSR-TR-95\/}, 1995.

\bibitem{Leishman1}
Leishman, J., {\em Principles of Helicopter Aerodynamics\/}, Cambridge
  Aerospace Series, Cambridge University Press, 2006.

\bibitem{Marzocca}
Marzocca, P., e.~a., \enquote{Development of An Indicial Function Approach for
  the Two-Dimensional Incompressible/Compressible Aerodynamic Load Modelling,}
  {\em Proceedings of the Institution of Mechanical Engineers, Part G: Journal
  of Aerospace Engineering\/}, Vol.~221, No.~3, 2007, pp.~453--463.

\bibitem{Quarteroni1}
Quarteroni, A. and Rozza, G., {\em Reduced Order Methods for Modeling and
  Computational Reduction\/}, Springer International Publishing, 2014.

\bibitem{Raveh}
Raveh, D., \enquote{Reduced-Order Models for Nonlinear Unsteady Aerodynamics,}
  {\em AIAA Journal\/}, Vol.~39, No.~8, 2001, pp.~1417--1429.

\bibitem{Gennaretti}
Gennaretti, M. and Mastroddi, F., \enquote{Study of Reduced-Order Models for
  Gust-Response Analysis of Flexible Fixed Wings,} {\em Journal of Aircraft\/},
  Vol.~41, No.~2, 2004, pp.~304--313.

\bibitem{Albano}
Albano, E. and Rodden, W., \enquote{A Doublet-Lattice Method for Calculating
  the Lift Distribution of Oscillating Surfaces in Subsonic Flows,} {\em AIAA
  Journal\/}, Vol.~7, No.~2, 1969, pp.~279--285.

\bibitem{Rodden}
Rodden, W., {\em Theoretical and Computational Aeroelasticity\/}, Crest Pub.,
  2011.

\bibitem{Wright}
Wright, J. and Cooper, J., {\em Introduction to Aircraft Aeroelasticity and
  Loads\/}, AIAA Education Series, AIAA, 2007.

\bibitem{Fung}
Fung, Y., {\em An Introduction to the Theory of Aeroelasticity\/}, Dover, 1993.

\bibitem{Bisplinghoff}
Bisplinghoff, R., Ashley, H., and Halfman, R., {\em Aeroelasticity\/}, Dover,
  1996.

\bibitem{Prandtl}
Prandtl, L., \enquote{Applications of Modern Hydrodynamics to Aeronautics,}
  {\em NACA TR-116\/}, 1921.

\bibitem{Glauert1}
Glauert, H., {\em The Elements of Aerofoil and Airscrew Theory\/}, Cambridge
  University Press, 1926.

\bibitem{Glauert2}
Glauert, H., \enquote{The Effect of Compressibility on the Lift of an
  Aerofoil,} {\em Proceedings of the Royal Society of London. Series A\/},
  Vol.~118, No. 779, 1928, pp.~113--119.

\bibitem{Wagner}
Wagner, H., \enquote{{\"U}ber die Entstehung des Dynamischen Auftriebes von
  Tragfl{\"u}geln,} {\em ZAMM-Journal of Applied Mathematics and
  Mechanics/Zeitschrift f{\"u}r Angewandte Mathematik und Mechanik\/}, Vol.~5,
  No.~1, 1925, pp.~17--35.

\bibitem{Theodorsen1}
Theodorsen, T., \enquote{General Theory of Aerodynamic Instability and the
  Mechanism of Flutter,} {\em NACA 496\/}, 1935.

\bibitem{Kussner}
K{\"u}ssner, H., \enquote{Zusammenfassender Bericht {\"u}ber den
  instation{\"a}ren Auftrieb von Fl{\"u}geln,} {\em Luftfahrtforschung\/},
  Vol.~13, No.~12, 1936, pp.~410--424.

\bibitem{Garrick}
Garrick, L., \enquote{On some Reciprocal Relations in the Theory of
  Nonstationary Flows,} {\em NACA 629\/}, 1938.

\bibitem{VonKarman}
Von~Karman, T. and Sears, W., \enquote{Airfoil Theory for Non-Uniform Motion,}
  {\em Journal of the Aeronautical Sciences\/}, Vol.~5, No.~10, 1938,
  pp.~379--390.

\bibitem{Sears}
Sears, W., \enquote{Operational Methods in the Theory of Airfoils in
  Non-Uniform Motion,} {\em Journal of the Franklin Institute\/}, Vol.~230,
  No.~1, 1940, pp.~95--111.

\bibitem{Jones1}
Jones, R., \enquote{The Unsteady Lift of a Wing of Finite Aspect Ratio,} {\em
  NACA 681\/}, 1940.

\bibitem{Heaslet}
Heaslet, M. and Spreiter, J., \enquote{Reciprocity Relations in Aerodynamics,}
  {\em NACA 1119\/}, 1953.

\bibitem{Basu}
Basu, B. and Hancock, G., \enquote{The Unsteady Motion of a Two-Dimensional
  Aerofoil in Incompressible Inviscid Flow,} {\em Journal of Fluid
  Mechanics\/}, Vol.~87, 1978, pp.~159--178.

\bibitem{Jones2}
Jones, R., \enquote{Classical Aerodynamic Theory,} {\em NASA RP-1050\/}, 1979.

\bibitem{Peters1}
Peters, D., Karunamoorthy, S., and Cao, W.-M., \enquote{Finite State Induced
  Flow Models (Part I),} {\em Journal of Aircraft\/}, Vol.~32, No.~2, 1995,
  pp.~313--322.

\bibitem{Kayran}
Kayran, A., \enquote{Kussner's Function in the Sharp Edged Gust Problem - A
  Correction,} {\em Journal of Aircraft\/}, Vol.~43, No.~5, 2006,
  pp.~1596--1599.

\bibitem{Peters2}
Peters, D., Hsieh, M.-C., and Torrero, A., \enquote{A State-Space Airloads
  Theory for Flexible Airfoils,} {\em Journal of the American Helicopter
  Society\/}, Vol.~52, No.~4, 2007, pp.~329--342.

\bibitem{Peters3}
Peters, D., \enquote{Two-Dimensional Incompressible Unsteady Airfoil Theory -
  An Overview,} {\em Journal of Fluids and Structures\/}, Vol.~24, No.~3, 2008,
  pp.~295--312.

\bibitem{Gaunaa}
Gaunaa, M., \enquote{Indicial Response Function for Finite-Thickness Airfoils,
  a Semi-Empirical Approach,} {\em AIAA-2011-542\/}, 2011.

\bibitem{Liu}
Liu, T., e.~a., \enquote{Unsteady Thin-Airfoil Theory Revisited: Application of
  a Simple Lift Formula,} {\em AIAA Journal\/}, Vol.~53, No.~6, 2015,
  pp.~1492--1502.

\bibitem{Possio}
Possio, C., \enquote{L'azione Aerodinamica sul Profilo Oscillante in un Fluido
  Compressibile a Velocità Iposonora,} {\em L'Aerotecnica\/}, Vol.~18, No.~4,
  1938, pp.~441--458.

\bibitem{Frazer}
Frazer, R., \enquote{Possio's Subsonic Derivative Theory and its Application to
  Flexural-Torsional Wing Flutter (Part I),} {\em ARC-RM-2553\/}, 1951.

\bibitem{Lomax1}
Lomax, H. and Heaslet, M., \enquote{The Indicial Lift and Pitching Moment for a
  Sinking or Pitching Two-Dimensional Wing Flying at Subsonic or Supersonic
  Speeds,} {\em NACA TN-2403\/}, 1951.

\bibitem{Mazelsky1}
Mazelsky, B., \enquote{Numerical Determination of Indicial Lift of a
  Two-Dimensional Sinking Airfoil at Subsonic Mach Numbers from Oscillatory
  Lift Coefficients with Calculations for Mach Number 0.7,} {\em NACA
  TN-2562\/}, 1951.

\bibitem{Mazelsky2}
Mazelsky, B., \enquote{Determination of Indicial Lift and Moment of a
  Two-Dimensional Pitching Airfoil at Subsonic Mach Numbers from Oscillatory
  Coefficients with Numerical Calculations for a Mach Number of 0.7,} {\em NACA
  TN-2613\/}, 1952.

\bibitem{Mazelsky3}
Mazelsky, B. and Drischler, J., \enquote{Numerical Determination of Indicial
  Lift and Moment Functions for a Two-Dimensional Sinking and Pitching Airfoil
  at Mach Numbers 0.5 and 0.6,} {\em NACA TN-2739\/}, 1952.

\bibitem{Lomax2}
Lomax, H., Fuller, F., and Sluder, L., \enquote{Two- and Three-Dimensional
  Unsteady Lift Problems in High-Speed Flight,} {\em NACA 1077\/}, 1952.

\bibitem{Jones3}
Jones, W., \enquote{Oscillating Wings in Compressible Subsonic Flow,} {\em ARC
  RM-2855\/}, 1957.

\bibitem{Leishman2}
Leishman, J., \enquote{Indicial Lift Approximations for Two-Dimensional
  Subsonic Flow as Obtained from Oscillatory Measurements,} {\em Journal of
  Aircraft\/}, Vol.~30, No.~3, 1993, pp.~340--351.

\bibitem{Leishman3}
Leishman, J., \enquote{Subsonic Unsteady Aerodynamics Caused by Gusts Using the
  Indicial Method,} {\em Journal of Aircraft\/}, Vol.~33, No.~5, 1996,
  pp.~869--879.

\bibitem{Balakrishnan}
Balakrishnan, A., \enquote{Possio Integral Equation of Aeroelasticity Theory,}
  {\em Journal of Aerospace Engineering\/}, Vol.~16, No.~4, 2003, pp.~139--154.

\bibitem{Polyakov}
Polyakov, P.~L., \enquote{Solvability of the Generalized Possio Equation in 2D
  Subsonic Aeroelasticity,} {\em Computational Methods and Function Theory\/},
  Vol.~7, No.~1, 2007, pp.~55--76.

\bibitem{Mateescu}
Mateescu, D., \enquote{Theoretical Solutions for Unsteady Compressible Subsonic
  Flows Past Oscillating Rigid and Flexible Airfoils,} {\em Mathematics in
  Engineering, Science and Aerospace\/}, Vol.~2, No.~1, 2011, pp.~1--27.

\bibitem{Gulcat}
Gulcat, U., {\em Fundamentals of Modern Unsteady Aerodynamics\/}, Springer,
  2011.

\bibitem{Chung}
Chung, T., {\em Computational Fluid Dynamics\/}, Cambridge Press, 2002.

\bibitem{Morino}
Morino, L., \enquote{A General Theory of Unsteady Compressible Potential
  Aerodynamics,} {\em NASA CR-2464\/}, 1974.

\bibitem{Katz}
Katz, J. and Plotkin, A., {\em Low Speed Aerodynamics\/}, Cambridge University
  Press, 2001.

\bibitem{McNamara}
McNamara, J.J., e.~a., \enquote{Approximate modeling of unsteady aerodynamics
  for hypersonic aeroelasticity,} {\em Journal of Aircraft\/}, Vol.~47, No.~6,
  2010, pp.~1932--1945.

\bibitem{Parameswaran}
Parameswaran, V. and Baeder, J.~D., \enquote{Indicial Aerodynamics in
  Compressible Flow-Direct Computational Fluid Dynamic Calculations,} {\em
  Journal of Aircraft\/}, Vol.~34, No.~1, 1997, pp.~131--133.

\bibitem{Silva}
Silva, W., \enquote{Discrete-Time Linear and Nonlinear Aerodynamic Impulse
  Responses for Efficient Use of CFD Analyses,} {\em PhD Thesis, College of
  William and Mary\/}, 1997.

\bibitem{Cavagna2}
Cavagna, L., e.~a., \enquote{Efficient Application of CFD Aeroelastic Methods
  Using Commercial Software,} {\em IFASD-2005-000\/}, 2005.

\bibitem{Romanelli}
Romanelli, G., Serioli, E., and Mantegazza, P., \enquote{A "Free" Approach to
  Computational Aeroelasticity,} {\em AIAA-2010-176\/}, 2010.

\bibitem{DaRonch}
Da~Ronch, A., Ghoreyshi, M., and Badcock, K., \enquote{On the Generation of
  Flight Dynamics Aerodynamic Tables by Computational Fluid Dynamics,} {\em
  Progress in Aerospace Sciences\/}, Vol.~47, 2011, pp.~597--620.

\bibitem{Ghoreyshi1}
Ghoreyshi, M., Jirasek, A., and Cummings, R., \enquote{Computational
  Investigation into the Use of Response Functions for Aerodynamic-Load
  Modeling,} {\em AIAA Journal\/}, Vol.~50, No.~6, 2012, pp.~1314--1327.

\bibitem{Jansson}
Jansson, N. and Eller, D., \enquote{Efficient Laplace-Domain Aerodynamics for
  Load Analyses,} {\em IFASD-2013-000\/}, 2013.

\bibitem{Ghoreyshi2}
Ghoreyshi, M., Jirasek, A., and Cummings, R., \enquote{Reduced Order Unsteady
  Aerodynamic Modeling for Stability and Control Analysis Using Computational
  Fluid Dynamics,} {\em Progress in Aerospace Sciences\/}, Vol.~71, 2014,
  pp.~167--217.

\bibitem{Cizmas}
Cizmas, P. and Gargoloff, J., \enquote{Mesh Generation and Deformation
  Algorithm for Aeroelasticity Simulations,} {\em Journal of Aircraft\/},
  Vol.~45, No.~3, 2008, pp.~1062--1066.

\bibitem{Farhat1}
Farhat, C., Lesoinne, M., and LeTallec, P., \enquote{Load and Motion Transfer
  Algorithms for Fluid/Structure Interaction Problems with Non-Matching
  Discrete Interfaces: Momentum and Energy Conservation, Optimal Discretization
  and Application to Aeroelasticity,} {\em Computer Methods in Applied
  Mechanics and Engineering\/}, Vol.~157, No.~1, 1998, pp.~95--114.

\bibitem{Farhat2}
Farhat, C. and Lakshminarayan, V., \enquote{An ALE Formulation of Embedded
  Boundary Methods for Tracking Boundary Layers in Turbulent Fluid-Structure
  Interaction Problems,} {\em Journal of Computational Physics\/}, Vol.~263,
  2014, pp.~53--70.

\bibitem{Anderson}
Anderson, J., {\em Fundamentals of Aerodynamics\/}, McGraw-Hill, 2007.

\bibitem{Pope}
Pope, S., {\em Turbulent Flows\/}, Cambridge University Press, 2000.

\bibitem{Jacobs}
Jacobs, E., Ward, K., and Pinkerton, R., \enquote{The Characteristics of 78
  Related Airfoil Sections from Tests in the Variable-Density Wind Tunnel,}
  {\em NACA 460\/}, 1935.

\bibitem{Abbott1}
Abbott, I. and von Doenhoff, A., \enquote{Summary of Airfoil Data,} {\em NACA
  824\/}, 1945.

\bibitem{Ladson}
Ladson, C. e.~a., \enquote{Computer Program To Obtain Ordinates for NACA
  Airfoils,} {\em NASA TM-4741\/}, 1996.

\bibitem{Quarteroni2}
Quarteroni, A., Sacco, R., and Saleri, F., {\em Numerical Mathematics\/},
  Springer-Verlag, 2000.

\bibitem{Vepa1}
Vepa, R., \enquote{On the Use of Pade' Approximants to Represent Unsteady
  Aerodynamic Loads for Arbitrarily Small Motions of Wings,} {\em
  AIAA-76-17\/}, 1976.

\bibitem{Vepa2}
Vepa, R., \enquote{Finite State Modeling of Aeroelastic Systems,} {\em NASA
  CR-2779\/}, 1977.

\bibitem{Roger}
Roger, K., \enquote{Airplane Math Modeling Methods for Active Control Design,}
  {\em AGARD-CP-228\/}, 1977.

\bibitem{Dunn}
Dunn, H., \enquote{An Analytical Technique for Approximating Unsteady
  Aerodynamics in the Time Domain,} {\em NASA TP-1738\/}, 1980.

\bibitem{Dowell}
Dowell, E., \enquote{A Simple Method for Converting Frequency Domain
  Aerodynamics to the Time Domain,} {\em NASA TM-81844\/}, 1980.

\bibitem{Desmarais}
Desmarais, R., \enquote{A Continued Fraction Representation for Theodorsen's
  Circulation Function,} {\em NASA TM-81838\/}, 1980.

\bibitem{Beddoes}
Beddoes, T., \enquote{Practical Computation of Unsteady Lift,} {\em Vertica\/},
  Vol.~8, No.~1, 1984, pp.~55--71.

\bibitem{Venkatesan}
Venkatesan, C. and Friedmann, P., \enquote{New Approach to Finite-State
  Modeling of Unsteady Aerodynamics,} {\em AIAA Journal\/}, Vol.~24, No.~12,
  1986, pp.~1889--1897.

\bibitem{Peterson}
Peterson, L. and Crawley, E., \enquote{Improved Exponential Time Series
  Approximations of Unsteady Aerodynamic Operators,} {\em Journal of
  Aircraft\/}, Vol.~25, No.~2, 1988, pp.~121--127.

\bibitem{Eversman}
Eversman, W. and Tewari, A., \enquote{Modified Exponential Series Approximation
  for the Theodorsen Function,} {\em Journal of Aircraft\/}, Vol.~28, No.~9,
  1991, pp.~553--557.

\bibitem{Tewari}
Tewari, A. and Brink-Spalink, J., \enquote{Multiple Pole Rational-Function
  Approximations for Unsteady Aerodynamics,} {\em Journal of Aircraft\/},
  Vol.~30, No.~3, 1993, pp.~426--428.

\bibitem{Karpel}
Karpel, M. and Strul, E., \enquote{Minimum-State Unsteady Aerodynamic
  Approximations with Flexible Constraints,} {\em Journal of Aircraft\/},
  Vol.~33, No.~6, 1996, pp.~1190--1196.

\bibitem{Poirel}
Poirel, D., \enquote{Random Dynamics of a Structurally Nonlinear Airfoil in
  Turbulent Flow,} {\em PhD Thesis, McGill University\/}, 2001.

\bibitem{Cotoi}
Cotoi, I. and Botez, R., \enquote{Method of Unsteady Aerodynamic Forces
  Approximation for Aeroservoelastic Interactions,} {\em Journal of Guidance,
  Control, and Dynamics\/}, Vol.~25, No.~5, 2002, pp.~985--987.

\bibitem{Sedaghat}
Sedaghat, A., e.~a., \enquote{Curve Fitting Approach for Transonic Flutter
  Prediction,} {\em The Aeronautical Journal\/}, Vol.~107, 2003, pp.~565--572.

\bibitem{Brunton}
Brunton, S. and Rowley, C., \enquote{Empirical State-Space Representations for
  Theodorsen's Lift Model,} {\em Journal of Fluids and Structures\/}, Vol.~38,
  No.~1, 2013, pp.~174--186.

\bibitem{Berci1}
Berci, M., \enquote{Optimal Approximations of Indicial Aerodynamics,} {\em
  Proceedings of the 1st OPTi\/}, 2014, pp.~1241--1277.

\bibitem{Berci2}
Berci, M., \enquote{Semi-Analytical Reduced-Order Models for the Unsteady
  Aerodynamic Loads of Subsonic Wings,} {\em Proceedings of the 5th COMPDYN\/},
  2015, pp.~1353--1378.

\bibitem{Wieseman}
Wieseman, C., \enquote{Methodology for Matching Experimental and Computational
  Aerodynamic Data,} {\em NASA TM-100592\/}, 1988.

\bibitem{Brink}
Brink-Spalink, J. and Bruns, J., \enquote{Correction of Unsteady Aerodynamic
  Influence Coefficients Using Experimental or CFD Data,} {\em
  AIAA-2000-1489\/}, 2000.

\bibitem{Palacios}
Palacios, R., e.~a., \enquote{Assessment of Strategies for Correcting Linear
  Unsteady Aerodynamics Using CFD or Test Results,} {\em IFASD-2001-074\/},
  2001.

\bibitem{Dimitrov}
Dimitrov, D. and Thormann, R., \enquote{DLM-Correction Method for Aerodynamic
  Gust Response Prediction,} {\em IFASD-2013-000\/}, 2013.

\bibitem{Sucipto}
Sucipto, T., Berci, M., and Krier, J., \enquote{Gust Response of a Flexible
  Typical Section via High- and (Tuned) Low-Fidelity Simulations,} {\em
  Computers \& Structures\/}, Vol.~122, 2013, pp.~202--216.

\bibitem{Berci3}
Berci, M., e.~a., \enquote{Dynamic Response of Typical Section Using
  Variable-Fidelity Fluid Dynamics and Gust-Modeling Approaches - With
  Correction Methods,} {\em Journal of Aerospace Engineering\/}, Vol.~27,
  No.~5, 2014, pp.~1--20.

\bibitem{Righi}
Righi, M., Koch, J., and Berci, M., \enquote{Subsonic Indicial Aerodynamics for
  Unsteady Loads Calculation via Numerical and Analytical Methods: a
  Preliminary Assessment,} {\em AIAA-2015-3170\/}, 2015.

\bibitem{Eliasson1}
Eliasson, P. and Weinerfelt, P., \enquote{Recent Applications of the Flow
  Solver Edge,} {\em Proceedings of the 7th Asian CFD Conference\/}, 2007.

\bibitem{EDGE}
Eliasson, P., \enquote{EDGE, a Navier-Stokes Solver for Unstructured Grids,}
  {\em Scientific Report FOI-R-0298-SE\/}, 2001.

\bibitem{Schlichting}
Schlichting, H. and Gersten, K., {\em Boundary Layer Theory\/}, Springer, 2000.

\bibitem{Wallin1}
Wallin, S. and Johansson, A., \enquote{A New Explicit Algebraic Reynolds Stress
  Turbulence Model for 3D flow,} {\em 11th Symposium on Turbulent Shear
  Flows\/}, 1997.

\bibitem{Wallin2}
Wallin, S. and Johansson, A., \enquote{An Explicit Algebraic Reynolds Stress
  Model for Incompressible and Compressible Turbulent Flows,} {\em Journal of
  Fluid Mechanics\/}, Vol.~403, 2000, pp.~89--132.

\bibitem{Jameson1}
Jameson, A., Schmidt, W., and Turkel, E., \enquote{Numerical Solutions of the
  Euler Equations by Finite Volume Methods Using Runge-Kutta Time-Stepping
  Schemes,} {\em AIAA-1981-1259\/}, 1981.

\bibitem{Roe}
Roe, P., \enquote{Approximate Riemann Solvers, Parameter Vectors, and
  Difference Schemes,} {\em Journal of Computational Physics\/}, Vol.~43,
  No.~2, 1981, pp.~357--372.

\bibitem{Jameson2}
Jameson, A., \enquote{Time Dependent Calculations Using Multigrid - with
  Applications to Unsteady Flows Past Airfoils and Wings,} {\em
  AIAA-1991-1596\/}, 1991.

\bibitem{Eliasson2}
Eliasson, P. and Weinerfelt, P., \enquote{High-Order Implicit Time Integration
  for Unsteady Turbulent Flow Simulations,} {\em Computers \& Fluids\/},
  Vol.~112, 2015, pp.~35--49.

\bibitem{Peng}
Peng, S. and Eliasson, P., \enquote{Unsteady Simulation for the Flow Over a
  Three-Element High-Lift Configuration at Stall,} {\em AIAA-2008-414\/}, 2008.

\bibitem{Eliasson3}
Eliasson, P., Weinerfeldt, P., and Nordström, J., \enquote{Application of a
  Line-Implicit Scheme on Stretched Unstructured Grids,} {\em
  AIAA-2009-0163\/}, 2009.

\bibitem{Rumsey}
Rumsey, C.~L., \enquote{Recent Developments on the Turbulence Modeling Resource
  Website,} {\em AIAA Paper 2015-2927\/}, 2015.

\bibitem{Lighthill}
Lighthill, M., \enquote{Oscillating Airfoils at High Mach Numbers,} {\em
  Journal of the Aeronautical Sciences\/}, Vol.~20, No.~6, 1953, pp.~402--406.

\bibitem{Ashley}
Ashley, H. and Zartarian, G., \enquote{Piston Theory-A New Aerodynamic Tool for
  the Aeroelastician,} {\em Journal of the Aeronautical Sciences\/}, Vol.~23,
  No.~12, 1956, pp.~1109--1118.

\bibitem{Chiang}
Chiang, H.-W. and Fleeter, S., \enquote{Prediction of loaded airfoil unsteady
  aerodynamic gust response by a locally analytical method,} {\em Mathematical
  and Computer Modelling\/}, Vol.~11, 1988, pp.~871--876.

\bibitem{Hoblit}
Hoblit, F.~M., \enquote{Gust Loads on Aircraft: Concepts and Applications,}
  {\em AIAA Education Series, AIAA\/}, 1988.

\bibitem{Holmes}
Holmes, R., {\em A Course on Optimization and Best Approximation\/}, Lecture
  Notes in Mathematics, Springer, 1972.

\bibitem{Vanderplaats}
Vanderplaats, G., {\em Numerical Optimization Techniques for Engineering
  Design: with Applications\/}, Series in Mechanical Engineering, McGraw Hill,
  1984.

\bibitem{Tiffany}
Tiffany, S. and Adams, W., \enquote{Nonlinear Programming Extensions to
  Rational Function Approximation Methods for Unsteady Aerodynamic Force,} {\em
  NASA TP-2776\/}, 1988.

\bibitem{Leishman4}
Leishman, J. and Nguyen, K., \enquote{State-Space Representation of Unsteady
  Airfoil Behavior,} {\em AIAA Journal\/}, Vol.~28, No.~5, 1990, pp.~836--844.

\bibitem{Meyer}
Meyer, M. and Matthies, H., \enquote{State-Space Representation of Instationary
  Two-Dimensional Airfoil Aerodynamics,} {\em Journal of Wind Engineering and
  Industrial Aerodynamics\/}, Vol.~92, No.~3, 2004, pp.~263--274.

\bibitem{Motta}
Motta, V., Guardone, A., and Quaranta, G., \enquote{Influence of Airfoil
  Thickness on Unsteady Aerodynamic Loads on Pitching Airfoils,} {\em Journal
  of Fluid Mechanics\/}, Vol.~774, 2015, pp.~460--487.

\bibitem{Theodorsen2}
Theodorsen, T., \enquote{Theory of Wing Sections of Arbitrary Shape,} {\em NACA
  411\/}, 1933.

\bibitem{Theodorsen3}
Theodorsen, T. and Garrick, I., \enquote{General Potential Theory of Arbitrary
  Wing Section,} {\em NACA 452\/}, 1934.

\bibitem{Kutta}
Kutta, W., \enquote{Auftriebskr{\"a}fte in str{\"o}menden Fl{\"u}ssigkeiten,}
  {\em Illustrierte Aeronautische Mitteilungen\/}, Vol.~6, No. 133, 1902,
  pp.~133--135.

\bibitem{Joukowski}
Joukowski, N., \enquote{Sur les Tourbillons Adjoints,} {\em Travaux de la
  Section Physique de la Societ\'e Imp\'eriale des Amis des Sciences
  Naturales\/}, Vol.~13, No.~2, 1906, pp.~261--284.

\bibitem{Imai}
Imai, I., Kaji, I., and Umeda, K., \enquote{Mapping Functions of the NACA
  Airfoils into the Unit Circle,} {\em Journal of the Faculty of Science,
  Hokkaido University (Series 2)\/}, Vol.~3, No.~8, 1950, pp.~265--304.

\bibitem{Karamcheti}
Karamcheti, K., {\em Principles of Ideal-Fluid Aerodynamics\/}, Wiley, 1967.

\bibitem{Barger}
Barger, R., \enquote{Adaptation of the Theodorsen Theory to the Representation
  of an Airfoil as a Combination of a Lifting Line and a Thickness
  Distribution,} {\em NASA TN-D-8117\/}, 1975.

\bibitem{Mateescu2}
Mateescu, D. and Abdo, M., \enquote{Efficient Second-Order Analytical Solutions
  for Airfoils in Subsonic Flows,} {\em Aerospace Science and Technology\/},
  Vol.~9, No.~2, 2005, pp.~101--115.

\bibitem{Spreiter}
Spreiter, J. and Alksne, A., \enquote{Thin Airfoil Theory Based on Approximate
  Solution of the Transonic Flow Equation,} {\em NACA 1359\/}, 1958.

\bibitem{Bendiksen}
Bendiksen, O., \enquote{Review of Unsteady Transonic Aerodynamics: Theory and
  Applications,} {\em Progress in Aerospace Sciences\/}, Vol.~47, No.~2, 2011,
  pp.~135--167.

\bibitem{Nelder}
Nelder, J. and Mead, R., \enquote{A Simplex Method for Function Minimization,}
  {\em Computer Journal\/}, Vol.~7, No.~4, 1965, pp.~308--313.

\bibitem{Affenzeller}
Affenzeller, M., e.~a., {\em Genetic Algorithms and Genetic Programming: Modern
  Concepts and Practical Applications\/}, Chapman \& Hall, 2009.

\end{thebibliography}
\bibliographystyle{aiaa}

\end{document}